%% file: article_main.tex
\documentclass[useAMS,usenatbib]{mnras}

\usepackage{graphicx}
\usepackage{epstopdf,color,verbatim}
\usepackage{threeparttable}
\usepackage[]{txfonts}
\usepackage[dvipsnames]{xcolor}
\usepackage{xspace}
\usepackage{url}

\newcommand{\smilei}{{\sc Smilei}\xspace}
\newcommand{\shockapic}{{\sc Shockapic}\xspace}

\usepackage[normalem]{ulem}

\DeclareFontFamily{U}{mathx}{\hyphenchar\font45}
\DeclareFontShape{U}{mathx}{m}{n}{
      <5> <6> <7> <8> <9> <10>
      <10.95> <12> <14.4> <17.28> <20.74> <24.88>
      mathx10
      }{}
\DeclareSymbolFont{mathx}{U}{mathx}{m}{n}
\DeclareFontSubstitution{U}{mathx}{m}{n}
\DeclareMathAccent{\widebar}{0}{mathx}{"73}

\title[Relativistic shocks in magnetized $e^{\pm}$ plasma]
{Perpendicular relativistic shocks in magnetized pair plasma}

\author[I. Plotnikov, A. Grassi and M. Grech]
{Illya Plotnikov$^{1,2}$ \thanks{E-mail: illyap@astro.princeton.edu}, Anna Grassi$^{3,4,5}$, and Mickael Grech$^6$\\
$^1$  Institut  de  Recherche  en  Astrophysique  et  Plan\'etologie,  University of  Toulouse, CNRS, 9  avenue  Colonel  Roche,  BP  44346 - 31028,  Toulouse,  France \\
$^2$ Department of Astrophysical Sciences, Peyton Hall, Princeton University, Princeton, NJ 08544, USA\\
$^3$ LULI, UPMC Universit\'e Paris 06: Sorbonne Universit\'es, CNRS, Ecole Polytechnique, CEA, Universit\'e Paris-Saclay, F-75252 Paris Cedex 05, France\\
$^4$ Dipartimento di Fisica Enrico Fermi, Universit\`a di Pisa, Largo Bruno Pontecorvo 3, I-56127 Pisa, Italy\\
$^5$    Istituto Nazionale di Ottica, Consiglio Nazionale delle Ricerche (CNR/INO), u.o.s. Adriano Gozzini, I-56127 Pisa, Italy\\
$^6$    LULI, CNRS, Ecole Polytechnique, CEA, Universit\'e Paris-Saclay, UPMC Universit\'e Paris 06: Sorbonne Universit\'es, F-91128 Palaiseau Cedex, France\\}

\begin{document}

\date{Received / Accepted}
\pagerange{\pageref{firstpage}--\pageref{lastpage}} \pubyear{2017}

\maketitle

\begin{abstract}

Perpendicular relativistic ($\gamma_0=10$) shocks in magnetized pair plasmas are investigated using two dimensional Particle-in-Cell simulations. 
A systematic survey, from unmagnetized to strongly magnetized shocks, is presented accurately capturing the transition from Weibel-mediated to magnetic-reflection-shaped shocks.
This transition is found to occur for upstream flow magnetizations $10^{-3}<\sigma<10^{-2}$ at which a strong perpendicular net current is observed in the precursor, driving the so-called current-filamentation instability. 
The global structure of the shock and shock formation time are discussed. The MHD shock jump conditions are found in good agreement with the numerical results, except for $10^{-4} < \sigma < 10^{-2}$ where a deviation up to 10\% is observed. 
The particle precursor length converges toward the Larmor radius of particles injected in the upstream magnetic field at intermediate magnetizations. For $\sigma>10^{-2}$, it leaves place to a purely electromagnetic precursor following from the strong emission of electromagnetic waves at the shock front.
Particle acceleration is found to be efficient in weakly magnetized perpendicular shocks in agreement with previous works,
and is fully suppressed for $\sigma > 10^{-2}$.
Diffusive Shock Acceleration is observed only in weakly magnetized shocks, while a dominant contribution of Shock Drift Acceleration is evidenced at intermediate magnetizations.
The spatial diffusion coefficients are extracted from the simulations allowing for a deeper insight into the self-consistent particle kinematics and scale with the square of the particle energy in weakly magnetized shocks.
These results have implications for particle acceleration in the internal shocks of AGN jets and in the termination shocks of Pulsar Wind Nebulae. 
\end{abstract}


\input{01_introduction} 
\input{02_setup}

\input{03_jump_conditions}
\input{04_structure}

\input{05_formation_timescales}
\input{06_downstream_fdist}
\input{07_particle_kinematics}
\input{08_discussion}

 \section*{Acknowledgments}
The authors acknowledge fruitful discussions with Martin Lemoine, Guy Pelletier, Beno\^it Cerutti, Lorenzo Sironi and Anatoly Spitkovsky 
as well as Laurent Gremillet and Mark Dieckmann for insightful remarks on PIC codes. 
The authors also thank the \smilei development team for technical support. 
Financial support from Grant No. ANR-11-IDEX-0004-02 Plas@Par is acknowledged. 
AG also acknowledges support from the Université Franco-Italienne through the Vinci program (Grant No. C2-133).
This work was granted access to the HPC resources of CALMIP supercomputing centre at Universit\'e de Toulouse under the allocation 2016-p1504, and from GENCI-TGCC (Grant 2017-x2016057678)

\bibliographystyle{mnras}
\bibliography{allbiblio_new}

\appendix
\input{appendix_codes_compare}

\input{appendix_jumpMHD}


\end{document}

%% file: 01_introduction.tex
\section{Introduction}
\label{sect:into}

Relativistic shocks in pair plasmas have been identified as one of the major sources of high-energy radiation and particle acceleration in several classes of astrophysical objects that involve tremendous energy releases over short timescales \citep{KirkDuffy}, such as in compact objects environments. At the present time, only remote sensing of these environments is possible, making theoretical investigations and numerical simulations central to the study of collisionless shocks. In parallel, laser-plasma facilities are on the verge of producing electron-ion collisionless shocks in the laboratory~\citep[e.g.][]{2015NatPh..11..173H, ross2017}, and production of collisionless shocks in pair plasmas may be possible in a near future~\citep[e.g.][]{chen2015,lobet2015}. These approaches give complementary insights into the physics of relativistic collisionless shocks, the amplification or generation of magnetic fields in magnetized or initially unmagnetized plasmas,  and the related production of accelerated particles.

In the case of initially unmagnetized electron-ion plasmas, the basic theoretical structure of collisionless shocks was described first by \citet{1963JNuE....5...43M, 1966RvPP....4...23S} and then generalized to the relativistic case by \citet{1999ApJ...526..697M}.
Here the magnetic field is generated by the Weibel-filamentation instability \citep{Weibel59,Fried59} that channels the beam kinetic energy into small skin depth scale magnetic fields and dissipate most of the flow kinetic energy into thermal energy. 
This model has been confirmed by Particle-In-Cell (PIC) simulations, where the dominant role of Weibel-filamentation instability and its ability to generate sub-equipartition magnetic fields has been highlighted for counter-propagating relativistic beams \citep{1999ApJ...526..697M,Silva03, Nishikawa05, Kato07}.

In the case of magnetized plasmas, the relevant parameter to describe the relativistic pair shock structure is the magnetization parameter~\citet{1992ApJ...391...73G}:
\begin{equation}
\label{eq:sigma_param}
\sigma = {B_0^2 \over 8\pi N_0 \gamma_0 m c^2} ,
\end{equation} 
where $m=m_{e^-}=m_{e^+}$ is the electron/positron mass, $c$ the speed of light, $B_0$ and $N_0$  are the magnetic field and particle number density of the upstream plasma and $\gamma_0$ is its Lorentz factor [CGS Gaussian units are used throughout this work, unless specified otherwise]. In the present study these quantities are defined in the downstream rest frame for convenience. This choice does not affect the generality of our results since $\sigma$ is invariant by Lorentz transformation along a direction transverse to $\mathbf{B}_0$ \citep{KirkDuffy}.
Theoretical studies of plasma instabilities in weakly magnetized relativistic shocks \citep{2009ApJ...699..990B, 2010MNRAS.402..321L, 2011MNRAS.417.1148L} have shown that the Weibel instability can fully develop for $\sigma \simeq \xi_{\rm \tiny CR}/\gamma^{\vert u\,2}_{\rm sh}< 10^{-4}$, where $\xi_{\rm \tiny CR}\sim 0.1$ is the fraction of the incoming energy transferred to the supra-thermal particles and $\gamma^{\vert u}_{\rm sh}$ is the shock front Lorentz factor as seen from the upstream rest-frame.

On the other hand, in strongly magnetized plasmas, the physics changes significantly as the external field is dynamically not negligible. The basic shock structure in this regime was described by \citet{AlsopArons88, Langdon88, 1992ApJ...391...73G}. The compression of the external magnetic field builds up a soliton-like structure at the shock front. This structure sustains a magnetic barrier that provides the local dissipation mechanism through Maser Synchrotron Instability \citep[MSI, see e.g.,][]{1991PhFlB...3..818H} and particles reflection at the shock front. 

Particle-In-Cell (PIC) simulations by \citet{SSA13} indicate that the critical magnetization for which the shock structure changes is $\sigma_{\rm crit} \simeq 3 \times10^{-3}$ for pair plasmas, independently of the shock Lorentz factor (in the ultra-relativistic limit).  Shocks with $\sigma<\sigma_{\rm crit}$ exhibit filamentary structures characteristic of the Weibel instability while for larger $\sigma$, the shock is shaped by the compression of the external magnetic field. This raised the question of what happens for mildly magnetized shocks $(10^{-4}<\sigma<10^{-2})$ where the Weibel instability cannot fully develop and the MSI is not strong enough to sustain the shock front. 
In this work we will discuss the role of a net perpendicular current building up in the overlapping region and driving the Current Filamentation Instability (CFI)  \citep{2014MNRAS.440.1365L}.

As a consequence of the shock propagation, different acceleration mechanisms can take place depending on the nature of the shock, 
i.e. the structure of the electromagnetic fields at the shock front. 
In the case of weak magnetization ($\sigma<10^{-4}$), the so-called Diffusive Shock Acceleration \citep[DSA,][]{1977DoSSR.234.1306K, Bell78}, 
based on the first-order Fermi mechanism, is believed to be the dominant mechanism \citep[e.g.,][]{Peacock81, 2001MNRAS.328..393A}. 
In the case of non-relativistic shocks the energy gain for each Fermi-cycle (upstream $\to$ downstream $\to$ upstream) is modest 
$\Delta E \sim \left(\upsilon^{\vert u}_{\rm sh}/c\right) E \ll 1$, with $\upsilon^{\vert u}_{\rm sh}$ the shock front velocity in the upstream frame 
and $E$ the initial particle energy. 
Whereas in the relativistic limit, particles gain a large amount of energy in the first Fermi-cycle, as $\Delta E \simeq E \gamma^{\vert u\,2}_{\rm sh}$ \citep{Vietri95}, where $\gamma^{\vert u}_{\rm sh}=\left(1-\upsilon^{\vert u\,2}_{\rm sh}/c^2\right)^{-1/2} \gg 1$. However, it was demonstrated that, for all subsequent cycles, the energy gain is reduced to $\Delta E \simeq E $ because of the strong anisotropy of the upstream particle momentum distribution \citep{1999MNRAS.305L...6G}.

In the presence of an external magnetic field, strong compression and Lorentz transformation effects on the downstream magnetic field may inhibit DSA in relativistic shocks.
Indeed, even if the upstream fluid is turbulent on large scales compared to the particle Larmor radius $R_L$, which could be the case in the Interstellar Medium, 
the strong transverse magnetic field in downstream medium prevents any efficient particle cross-field scattering, as demonstrated by \citet{2006ApJ...645L.129L, 2006ApJ...650.1020N}. The magnetic turbulence then acts as an ordered field and particles are tied to a given field line. These studies suggested that a strong self-generated micro-turbulence on scales much lower than $R_L$ is required to unlock the phase-space particle trapping and allow particles to scatter on both sides of the shock and gain energy.

This picture was successfully confirmed for unmagnetized shocks by self-consistent PIC simulations \citep{2008ApJ...682L...5S, 2009ApJ...695L.189M, 2011ApJ...739L..42H}. For long enough simulations, a non-thermal tail in the particle distribution function was observed, demonstrating that the Fermi process develops self-consistently in time \citep{2008ApJ...682L...5S}. The accelerated particle distribution is theoretically predicted to follow a power-law with index $s={\rm d}\log N/{\rm d}\log E \simeq 2.22$ in ultra-relativistic shocks \citep{2000ApJ...542..235K, 2003ApJ...589L..73L, 2005PhRvL..94k1102K}, in rather good agreement with the index $s\simeq 2.4$ found in PIC simulations \citep[e.g.,][]{2008ApJ...682L...5S, 2009ApJ...698.1523S}. Moreover, the magnetic turbulence is found to be on the skin depth scale and of a quasi-static nature in the downstream fluid rest-frame, as well as to decay by phase mixing \citep{2008ApJ...674..378C, 2015JPlPh..81a4501L}. This turbulence is seeded on larger and larger spatial scales as, in the absence of the external magnetic field, the precursor is allowed to extend with no space limitation \citep{2009ApJ...693L.127K}.

The particle acceleration efficiency from unmagnetized to moderately magnetized regimes, up to $\sigma \simeq 10^{-2}$ in perpendicular shocks\footnote{We note that ultra-relativistic shocks are generically quasi-perpendicular for a random upstream magnetic field orientation \citep{1992ApJ...391...73G}. This is due to the shock compression and Lorentz transformation from the lab frame to the shock front rest-frame. Hence, in this study, only perpendicular shocks are considered. The study of the acceleration efficiency for different field orientations can be found in \citet{2009ApJ...698.1523S}.}, was studied by \citet{SSA13} using 2D and 3D long-term PIC simulations. These authors found that, in weakly magnetized shocks, DSA is efficient, with the maximum particle energy scaling as $E_{\rm max} \propto \sigma^{-1/4}$, and a fraction of energy transmitted to the supra-thermal particles $\xi_{\rm \tiny CR} \sim 10\%$, while for $\sigma \gtrsim 3 \times 10^{-3}$ DSA is inhibited.

As the magnetization increases, a very different particle acceleration mechanism occurs in magnetized oblique and transverse relativistic shocks. It is the Shock Drift Acceleration \citep[SDA;][]{1990ApJ...353...66B, 2009ApJ...698.1523S} for which particles gain energy from the motional electric field, and are found to drift along the shock front surface. 
This mechanism is found to be faster than DSA, but does not allow to reach considerably high energies as it requires the particle to remain close to the shock front. An important contribution of SDA to particle acceleration in mildly magnetized shocks will actually be demonstrated in this work.

In addition, the study of the spatial diffusion coefficients is most interesting. Indeed, non-linear Monte-Carlo simulations \citep[e.g.,][]{2016MNRAS.456.3090E} have demonstrated the dependence of the particle acceleration efficiency on the spatial diffusion coefficients in relativistic shocks.
Up to now, however, only an indirect demonstration of the scaling of the diffusion coefficient as $D \propto E^2$, was given by \citet{2012ApJ...755...68S, SSA13}, where the maximum energy of the particles has been shown to grow in time as $E_{\rm max} \propto \sqrt{t}$. This work will provide a direct demonstration of the $D \propto E^2$ in weakly magnetized shocks by means of a detailed analysis of single particle trajectories in PIC simulations. It will also be shown that this dependence is affected by the external magnetic field as particles scattering on turbulence is in competition with regular gyration, as expected from the theoretical study of \citet{2011A&A...532A..68P}.

The present study provides a detailed analysis of the shock formation, structure and late stage evolution across a wide range of $\sigma$, from very weakly magnetized shocks $\sigma \leq 10^{-5}$ relevant to shocks in the Interstellar Medium, to strongly magnetized shocks $\sigma=10$ relevant to shocks in Pulsar Wind Nebulae. 
In particular, this work extends the study of \citet{1992ApJ...391...73G} and \citet{SSA13} by performing two-dimensional PIC simulations with a more systematic grid of $\sigma$, and using two different PIC codes. These simulations provide a unique way to study the relevant instabilities, particle diffusion and acceleration processes in-depth and self-consistently. 

In Section~\ref{sect:setup}, we first present the numerical tools, methods as well as our simulation setup and parameters. 
Section~\ref{sect:jump} presents an analysis of the shock jump conditions, and the results of PIC simulations are compared
 to ideal Magnetohydrodynamics (MHD) predictions.
In Section~\ref{sect:global_struct}, the global structure of the shock is discussed in terms of density, magnetic fields and particle phase-space distributions. 
Particular attention is paid to the nature of the precursor, the mechanisms leading to shock formation and their dependence on $\sigma$. 
The presence of a perpendicular current in the shock precursor is demonstrated (Sec.~\ref{subsect:perp_current}), and we highlight that it could play a role in shock formation for intermediate magnetizations. In particular, it will be shown that the amplitude of this current increases with $\sigma$, and that its contribution is most important for the mildly magnetized cases $10^{-4} < \sigma <10^{-2}$.
The characteristic time and processes mediating shock formation are then discussed in Sec.~\ref{sect:formation}. An operating definition
for the time of shock formation is introduced, based on the temporal evolution of both the density compression and level of anisotropy in the
downstream region. Depending on the magnetization, it is found that either compression or isotropization is established first.

The next part of the paper is then devoted to a detailed investigation of particle acceleration in the shock.
First, Section~\ref{sect:fdist} provides a study, as a function of the magnetization, of the evolution of the accelerated particle distribution functions and of the acceleration efficiency. Then, Section~\ref{sect:track_particles} deals with the kinetic aspects of particle transport and acceleration in the shock using self-consistent particle kinematics. The spatial diffusion coefficient in the shock propagation direction, $D_{\perp}$, is extracted directly by analyzing single particle trajectories. The scaling $D_\perp \propto \gamma^2$ is found to hold for weakly magnetized shocks for high-energy particles while at larger magnetizations, $\sigma \gtrsim 10^{-3}$, $D_\perp$ is almost constant. The mechanisms responsible for particle acceleration are also discussed. DSA is found to be  efficient at small magnetizations, while SDA  gives an important contribution to particle acceleration in mildly magnetized purely perpendicular shocks, up to $\sigma\sim 10^{-3}$, where particle energy $\gamma \sim 40 \gamma_{\rm sh}$ can be reached on very short timescales.

Finally, we discuss the main findings and give our conclusions in Sec.~\ref{sect:discuss}. 

%% file: 02_setup.tex
\section{Simulation methods and setup}
\label{sect:setup}

\subsection{Simulation codes}
The simulations presented in this paper have been performed using two different PIC codes. 
Both codes are multi-dimensional in space (only the 2D version is used in this study) and 3D in momentum. Our aim in using the two codes is solely to ensure that the physics at hand does not depend on the numerical schemes and their implementation, both very different in each code. Throughout this work a very good agreement was found when comparing the two codes, as also be presented in more details in Appendix~\ref{appendix:comparison_codes}.

The first code is the open-source code \smilei\footnote{\url{http://www.maisondelasimulation.fr/smilei}} developed in a collaborative framework by physicists and HPC experts \citep{derouillat2018}. 
It is a fully electromagnetic, relativistic code that solves Maxwell's equations using the Finite-Difference-Time-Domain (FDTD) method and a staggered Yee-type mesh~\citep{taflove2005}. 
In the simulations presented here, macro-particles (with second order shape-functions) are advanced using the standard Boris algorithm \citep{1991ppcs.book.....B}, and particle currents are deposed onto the grid using the charge conserving Esirkepov’s algorithm \citep{Esirkepov01}. \smilei benefits from a state-of-the-art, hybrid MPI-OpenMP, parallelization and a modern dynamic load balancing strategy. 

Simulations have also been performed with the PIC code \shockapic, developed by one of the authors (IP). 
It is a pseudo-spectral time domain code [PSTD; as defined by \citet{Liu97PSTD, 2013JCoPh.243..260V}] that solves Fourier-transformed Maxwell's equations in $k$-space, while time advancing of the electromagnetic fields is done using an explicit finite difference method.
Particle advancing is done using the Boris algorithm, and currents are deposited onto the grid using second order particle shape-functions.
The spectral Boris correction on the electric field \citep{2013JCoPh.243..260V} is applied to ensure charge conservation at each time step. 

The essential difference in between the two codes is the method they use to solve Maxwell's equations. PSTD methods have the advantage to provide a more accurate representation of the electromagnetic waves dispersion relation. 
In particular, the light-wave dispersion relation in an unmagnetized plasma, $\omega^2=\omega_{p}^2+k^2 c^2$, is well-captured at high-$k$ modes in PSTD codes, while an important deviation occurs in FDTD codes \citep{1991ppcs.book.....B}.
FDTD solvers indeed introduce a spurious dispersion resulting in a numerical speed of light lower than $c$~\citep[see e.g.][]{nuter2014}. 
This aspect is important when one deals with relativistic flows drifting with velocity arbitrarily close to $c$. 
It produces a spurious grid-Cherenkov instability responsible for Cherenkov-like radiation, which dramatically heats the upstream plasma. 
In FDTD codes the instability is practically very difficult to avoid in long-term simulations as it involves coupling to low-$k$ modes.
In \smilei's simulations, using a 3-pass digital filter on currents together with a temporal Friedmann filter (with the control parameter $\theta= 0.3$) on the electric field \citep{Greenwood04} allowed us to prevent spurious effects for the simulation parameters discussed in this work (see Sec.~\ref{sec:geom}).
The instability is much easier to suppress in spectral codes using a simple spectral filtering of the highest $k$-modes.
In \shockapic's simulations, using a spectral cut on all field components proved sufficient. 
A 2-pass digital filter was however used to reduce the coarse-grain effect as only 2 particles-per-cell were used for each species. 
Note that the digital filtering and spectral cut affect only the smallest wavelengths such that the relevant physics is not affected.

Let us note that the `price to pay' for using spectral codes is that their real space representation might be questioned due to the non-local character of Fourier transforms (global stencil) and an imposed periodicity on fields. 
The former rises some questions on whether spectral codes respect the causality constraint. 
The latter imposes a careful dealing with boundary conditions when particle reflective walls are imposed in the simulation box. 
Yet, as will be demonstrated in this paper, both codes give very similar results for this study.

Finally, spectral codes rely on global communications, the Fourier transform requiring to `know' the fields everywhere on the grid,
whereas FDTD codes only require local calculations.
This makes spectral codes particularly difficult to efficiently parallelize on a large number of processors\footnote{Let us note that techniques for efficient parallelization of spectral codes have been recently proposed \citep{2013JCoPh.243..260V}.}. For this reason the simulations performed with \smilei are much longer than those done with \shockapic.

\subsection{Simulation geometry}\label{sec:geom}

\begin{figure}
\centering
\includegraphics[width=0.49\textwidth]{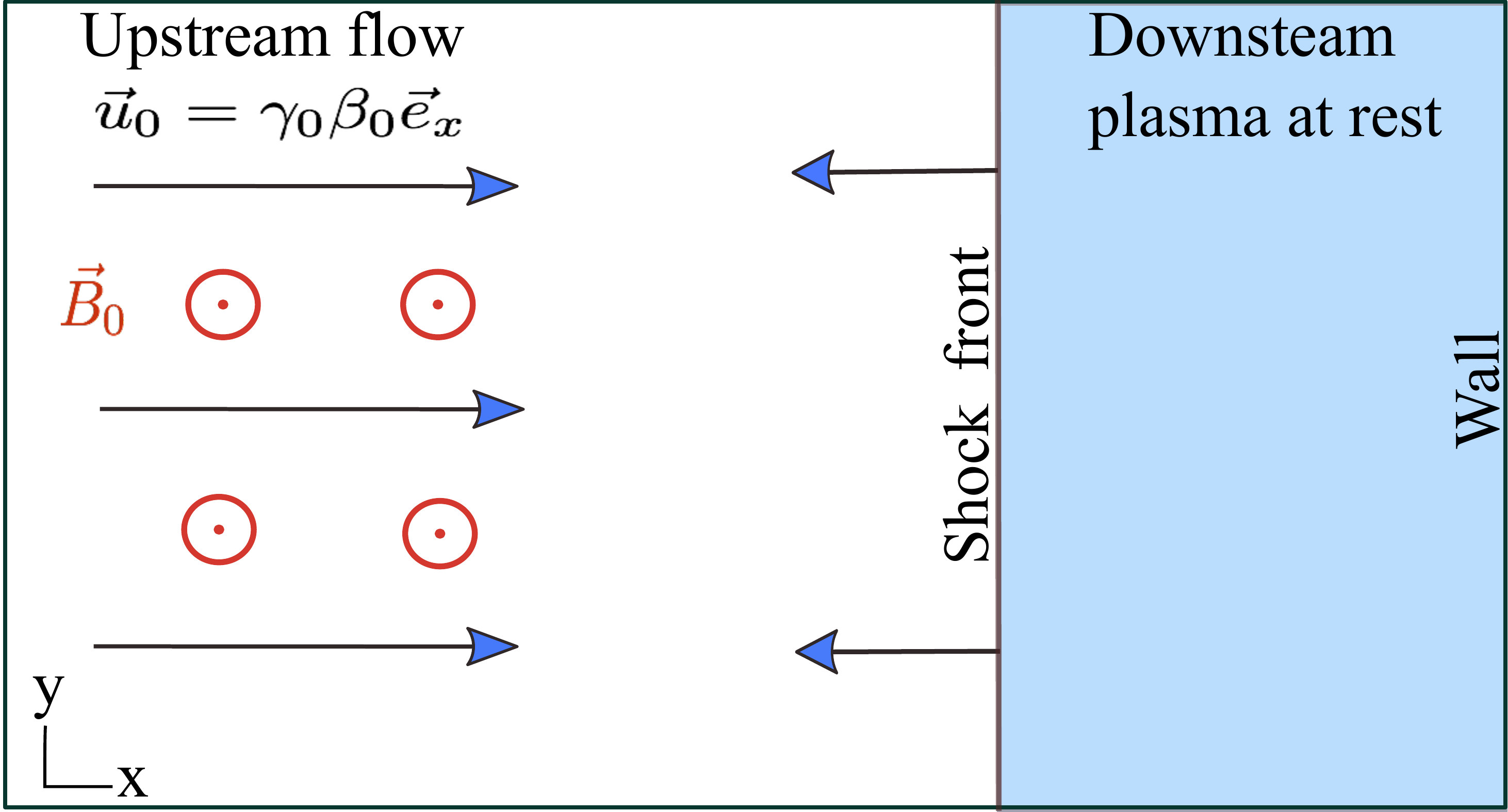}
\caption{Simulation set-up. The $(x,y)$ simulation plane contains an upstream pair plasma drifting in the positive $x$-direction. 
This plasma carries a frozen-in magnetic field $\mathbf{B}_0$ in the $z$-direction perpendicular to the simulation plane
and an electric field $\mathbf{E}_0$ in the $y-$direction to ensure an initial equilibrium condition. 
The shock front, once formed, propagates from right to left, and the downstream plasma has no average speed.}
\label{fig:scheme}%
\end{figure}

All simulations are performed in a two-dimensional Cartesian geometry in the $x$-$y$ plane. 
Initially, the simulation box is filled with a cold ($T_0 \ll m c^2$) plasma drifting along the positive $x$-direction with a relativistic velocity ${\bf v}_0=v_0\mathbf{e}_x$ [corresponding to a Lorentz factor $\gamma_0=(1-v_0^2/c^2)^{-1/2}$]. 
An external uniform magnetic field $\mathbf{B}_0$ is imposed in the out-of-plane $z$-direction. 
This choice is motivated by the fact that the out-of-plane field allows to recover the main ingredients of tri-dimensional particle dynamics such as scattering and acceleration process, as demonstrated by \citet{SSA13}. 
To provide an initial equilibrium condition, an electric field $\mathbf{E}_0={-\mathbf{p}_0/(mc\gamma_0) \times \mathbf{B}_0}$, where $\mathbf{p}_0=mc\gamma_0\beta_0 \mathbf{e}_x$ is the momentum of the upstream flow, is initialized as well.
The flow is reflected on a conducting wall at the right edge of the simulation domain. 
Particles are elastically reflected at that wall. 
The interaction between the incoming and wall-reflected flows builds up a shock that propagates in $-x$ direction.
This setup is presented schematically in Fig.~\ref{fig:scheme}. 
In this way the shocked (downstream) plasma has no average speed as it corresponds to the center-of-mass frame of the system composed of the incoming and wall-reflected flow. Hence, the frame of the simulations is the downstream rest frame.

All discussions (quantities) will henceforth be made (defined) in this simulation/downstream frame, unless specified otherwise.

\subsection{Units, physical and numerical parameters} 

In both codes times are expressed in units of the inverse of the non-relativistic electron plasma frequency $\omega_{\rm pe}=\left[ 4 \pi N_0 e^2 / m \right]^{1/2}$, where $-e$ is the electron charge, and $N_0$ is the electron (and positron) initial density. 
Velocities are expressed in units of the speed of light $c$. 
Hence, the space unit is the non-relativistic electron skin depth $\delta_{\rm e}= c/ \omega_{\rm pe}$. Note that the total plasma frequency is $\omega_{\rm p}=\sqrt{2}\omega_{\rm pe}$ as both constituents have the same mass, and fields are in units of $mc\omega_{\rm pe}/e$.
The cyclotron frequency in the external magnetic field is $\omega_{\rm ce}= e B_0/(m c)$.
These quantities are related to their relativistic counterpart using $\omega_{\rm pe,rel}=\omega_{\rm pe}/\sqrt{\gamma_0}$, $\delta_{\rm e,rel}= \delta_{\rm e} \sqrt{\gamma_0}$ and $\omega_{\rm ce, rel}= \omega_{\rm ce}/\gamma_0$, where $\gamma_0$ is the Lorentz factor of the upstream flow. 
In the present work, we use non-relativistic notations unless stated otherwise. 
We note however that the reference Larmor radius (i.e. the Larmor radius of an electron/positron with initial velocity $\pm v_0\mathbf{e}_x$ in the external magnetic field ${\bf B}_0$) is always defined as $R_{L,0}=\gamma_0 c/\omega_{\rm ce}$.
Note also that, the magnetization parameter $\sigma$, defined by Eq.~\ref{eq:sigma_param}, can also be written as
\begin{equation}
\sigma={1\over 2} \left( {\omega_{\rm ce,rel}  \over \omega_{\rm pe,rel}} \right)^2\, .
\end{equation}

\begin{table*}
\begin{center}
\begin{tabular}{|c|c|c|c|c|c|c|c|c|}
\hline 
PIC code & $\Delta t \omega_{\rm pe}$ & $T_{\rm sim} \omega_{\rm pe}$ & $\Delta x/\delta_{\rm e}$ & $L_{\perp}/\delta_{\rm e}$ & $N_{\rm ppc}$ & $\sigma_{\rm min}$ & $\sigma_{\rm max}$ & $N_{\rm sim}$ \\ 
\hline 
\smilei & 0.125 & 2000 & 0.25* & 256 & 20 & $5\times 10^{-6}$ & 5 & 24 \\ 
\shockapic & 0.1 & 300 & 0.33 & 85 & 2 & $10^{-5}$ & $10^{-1}$ & 20 \\ 
\hline 
\end{tabular} 
\end{center}
\caption[]{Parameters used in the two simulation series performed with \smilei and \shockapic: $\Delta t$ is the time-step,  $T_{\rm sim}$ is the simulation time, $\Delta x$ is the cell size (in both $x$- and $y$-directions), $L_\perp$ is the transverse ($y$) width of the simulation box, $N_{\rm ppc}$ is the number of particles-per-cell for each species. $N_{\rm sim}$ denotes the number of simulations in each series varying 
the magnetization from $\sigma_{\rm min}$ to $\sigma_{\rm max}$ over a logarithmic scale. (*) For \smilei simulations at $\sigma \ge 1$, the cell-size in both directions was reduced to $\delta_{\rm e}/16$.}
\label{table:sim_params}
\end{table*}
 
All simulations have been performed with a streaming plasma using $\gamma_0=10$ and initial temperature $T_0=10^{-5} m c^2$ (in the rest frame of the flow).
The main numerical parameters of the simulations are presented in the Table~\ref{table:sim_params}.
With \smilei, we performed 20 simulations varying $\sigma$ from $5\times 10^{-6}$ to $5\times10^{-2}$ on an equally spaced logarithmic scale
and four supplementary simulations with $\sigma=\{0.1, 0.3, 1, 5\}$. 
With \shockapic, we carried out 20 simulations with $\sigma \in [10^{-5}, 10^{-1}]$ on an equally spaced logarithmic scale. 

The transverse box size was $256 \delta_{\rm e}$ for \smilei and $85 \delta_{\rm e}$ for \shockapic, chosen in order to observe at least a dozen of filaments with typical size equal to the relativistic skin depth $\delta_{\rm e, rel}$. 
The cell size was $\Delta x = 0.25 \delta_{\rm e}$ for \smilei's simulations ($0.33 \delta_{\rm e}$ for \shockapic's), that in units of the relativistic skin depth give $\Delta x = 0.08 \delta_{\rm e,rel}$ ($0.1\delta_{\rm e,rel}$) ensuring that all relativistic scales are properly resolved. For \smilei simulations with large magnetizations ($\sigma \ge 1$), the spatial resolution was set to $\Delta x = \delta_{\rm e}/16$.

%% file: 03_jump_conditions.tex
\section{Shock jump conditions}
\label{sect:jump}
In this Section we derive the shock jump conditions from our PIC simulations and compare the results with the ideal MHD prediction. 
The derivation by means of relativistic ideal MHD conservation laws of the theoretical jump conditions, between the upstream incoming relativistic flow and the downstream plasma, is carried out in Appendix~\ref{appendix:jump}. In this derivation, no assumption is made on the upstream flow Lorentz factor (no ultra-relativistic limit). We only consider the strong shock limit for which the upstream flow internal  pressure (both magnetic and thermal) is neglected. This approximation is fully justified here, since the upstream plasma is cold with $T_0=10^{-5} m c^2$ and the magnetic pressure can be neglected up to $\sigma \sim \gamma_0^2 = 100$.

\begin{figure} \centering
\includegraphics[width=0.98\columnwidth]{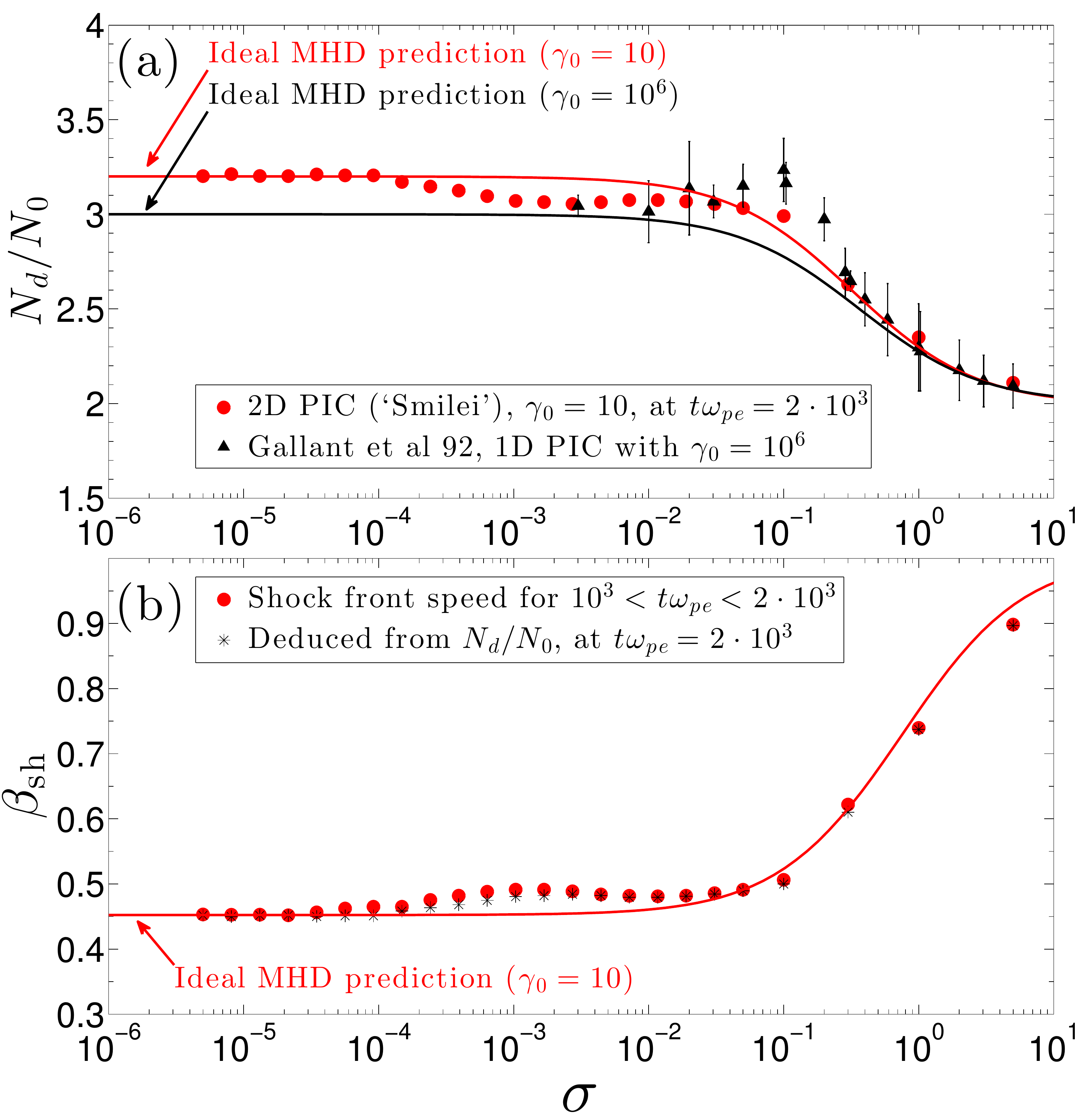}
\caption{Variation of the density compression ratio $N_d/N_0$ (panel a) and of the shock front speed (panel b) as a function of the upstream flow magnetization $\sigma$. Red circles correspond to the values extracted from \smilei simulations at $t\omega_{\rm pe}=2000$. Values plotted using black triangles with error bars are taken from the 1D simulation results of \citet{1992ApJ...391...73G}. Black star symbols in panel (b) correspond to the shock front speed derived from the measured density compression ratio using Eq.~\ref{eq:jump}. Solid lines follow the ideal MHD expectation, derived in the Appendix~\ref{appendix:jump}. Red and black lines use $\gamma_0=10$ (used in the present study) and $10^6$ [used in \citet{1992ApJ...391...73G}], respectively.} \label{fig:jump_dens}%
\end{figure}

Figure~\ref{fig:jump_dens}(a) shows the density jump $N_d/N_0$ as a function of the magnetization. 
$N_d$ is the electron density in the post-shock downstream medium measured in \smilei simulations 
at $t \omega_{\rm pe}=2\times10^3$ (red circles). 
The theoretically predicted density jump is shown for comparison (red line). 
It corresponds to Eq.~\ref{eq:jump}, adopting the appropriate adiabatic index for the downstream relativistic 
2D plasma $\Gamma_{\rm ad}=3/2$. 
In addition, we show the values extracted from 1D simulations by \citet{1992ApJ...391...73G} (black triangles with error bars) along with
the corresponding ideal MHD prediction for $\gamma_0=10^6$ (black solid line).

For $\sigma<3\times10^{-4}$, the compression ratio keeps its unmagnetized limit $N_d/N_0=1+(\gamma_0+1)/\left[ \gamma_0 (\Gamma_{\rm ad}-1) \right]= 3.2$ and the simulations results are in very good agreement with the ideal MHD predictions. 
A gradual deviation occurs for intermediate magnetizations, $10^{-4} < \sigma < 10^{-2}$, where the measured compression ratio can be up to the 10\% lower than the theoretical expectations. Note that measurements of the compression ratio at shorter times, $t\omega_{\rm pe} = 500$, show a smaller discrepancy, not exceeding 3\%.
A good agreement with the ideal MHD predictions is recovered for large magnetizations, $\sigma > 10^{-2}$, and only a small discrepancy is observed for the case $\sigma=0.1$ for which the compression ratio is slightly larger than expected.

Correspondingly, in Fig.~\ref{fig:jump_dens}(b), we show the shock front speed $\beta_{\rm sh}=v_{\rm sh}/c$ obtained from \smilei simulations. 
Defining the shock front position $x_{\rm sh}$ as the position where the transverse averaged density increases from the upstream value
\footnote{Note that this density value is reached at the shock front even in the highly magnetized cases, for which the predicted density jump is smaller than $2.8 N_0$, thanks to the density overshoot building up at the shock front, see Sec.~\ref{sect:global_struct} and Fig.~\ref{fig:densities}.}
 $N_0$ to $N(x_{\rm sh})=2.8 N_0$, we measured $x_{\rm sh}(t)$ every $2\omega_{\rm pe}^{-1}$ and we deduced the front velocity by a linear fit: $x_{\rm sh}(t) = -\beta_{\rm sh} t+x_0$, considering a constant front velocity (red circles). 
Also shown (black stars) is the front speed deduced from the theoretical relation $\beta_{\rm sh}=\beta_0(N_d/N_0-1)$, see Appendix~\ref{appendix:jump}, using the measured final state density jump shown in Fig.~\ref{fig:jump_dens}(a). 
The two measures are consistent. At low magnetization, the shock speed is in good agreement with the unmagnetized limit $\beta_{\rm sh} (\sigma=0)=(\Gamma_{\rm ad}-1)(\gamma_0-1)/ (\gamma_0\beta_0)=0.452$.
A deviation for intermediate magnetizations ${10^{-4} < \sigma < 10^{-2}}$ is observed, consistent with Fig.~\ref{fig:jump_dens}(a). 
In this interval of $\sigma$, the shock propagates faster than expected from the ideal MHD.
At high magnetizations, we recover the standard result of \citet{1984ApJ...283..694K}, that is the shock speed tends asymptotically to $c$ for $\sigma \gg 1$.

We tested the convergence of a limited number of simulations performed  with the code \shockapic up to $t\omega_{\rm pe} = 1200$ and the obtained results are in good agreement with \smilei simulations. Observing the same effect with the two codes supports the idea that the observed deviations are indeed physical and not a result of the numerical filtering.

As mentioned above, the standard ideal MHD does not account for kinetic effects that may take place at the shock front. 
In particular, \citet{1992ApJ...391...73G} has observed in 1D PIC simulations a departure from MHD jump conditions at $\sigma \simeq 0.1$, which they attributed to the emission of electromagnetic waves at the shock front. 
They further demonstrated that it is possible to account for this dissipative mechanism by including an additional term into the equation for the momentum and energy conservation across the shock front. 
As clear from Fig.~\ref{fig:jump_dens}(a) the deviation observed in their 1D simulations (black triangles) is more pronounced than ours, obtained in a 2D geometry, for $\sigma = 0.1$. 
Indeed, in 2D, the situation can be different as transverse effects may be important. 
Our simulations suggest that in a 2D configuration the cyclotron loop, responsible for the electromagnetic wave emission, loses in coherence because of the presence of transverse instabilities, such as Weibel-filamentation or CFI, as already suggested by \citet{1992ApJ...391...73G}. 
Consequently, the amplitude of the emitted electromagnetic waves is lower in 2D than in 1D and their contribution to the energy balance is reduced.
This results in the reduced [as compared to \citet{1992ApJ...391...73G}] departure from MHD predictions in our 2D PIC simulations for $\sigma \simeq 10^{-1}$.

The deviation that we observe in the intermediate range ${10^{-4} < \sigma < 10^{-2}}$ is more puzzling as it goes in the opposite direction: that is the density compression is somewhat lower in the PIC simulations than predicted by the MHD jump condition. 
Moreover, no strong electromagnetic precursor is observed there. Simulations however show that the downstream region immediately behind the shock front is very turbulent. Including the downstream wave turbulence terms in the conservation equations, in the same way as done in \citet{1992ApJ...391...73G}, we can recover the observed 10\% deviation in the jump conditions if about one tenth of the incoming kinetic energy is converted in the downstream magnetic wave turbulence. This latter term produces the supplement of downstream magnetic pressure that is necessary to increase the shock front speed. This, however, does not explain why the downstream wave turbulence is only important in the intermediate range of magnetization while a very good agreement with ideal MHD is found for either $\sigma<10^{-4}$ or $\sigma>10^{-2}$. 

%% file: 04_structure.tex
\section{Global shock structure}
\label{sect:global_struct}

In the absence of binary particle collisions the dissipation required to build up the shock front is given by collective plasma phenomena. 
For weakly magnetized shocks the transition region between the unperturbed upstream and the hot downstream plasmas is dominated by beam-type instabilities in the region where the energetic returning particle precursor pervades the upstream flow. 
With growing $\sigma$ the penetration length shortens, a cyclotron loop at the shock-front gains in coherence and the dissipation is gradually assumed by a coherent large-amplitude electromagnetic wave emission from the leading edge of the cyclotron ring at the shock front. 
Once the transition region is settled, the incoming cold flow is heated and relaxes to the downstream state on a spatial scale related to the relativistic skin depth, typically $\sim 10 \delta_{\rm e,rel}$, in weakly magnetized shocks and to the particle Larmor radius $\sim R_{L,0}$ in highly magnetized shocks.

Let us note that, as relativistic pair plasmas are considered here, the potential barrier at the shock front is of an electromagnetic nature.
This contrasts with shocks in electron-ion plasmas where important electrostatic effects are present \citep{2010MNRAS.402..321L}. 

\subsection{Density structure}\label{sect:global_struct:density}
\begin{figure}
\centering
 \includegraphics[width=0.48\textwidth]{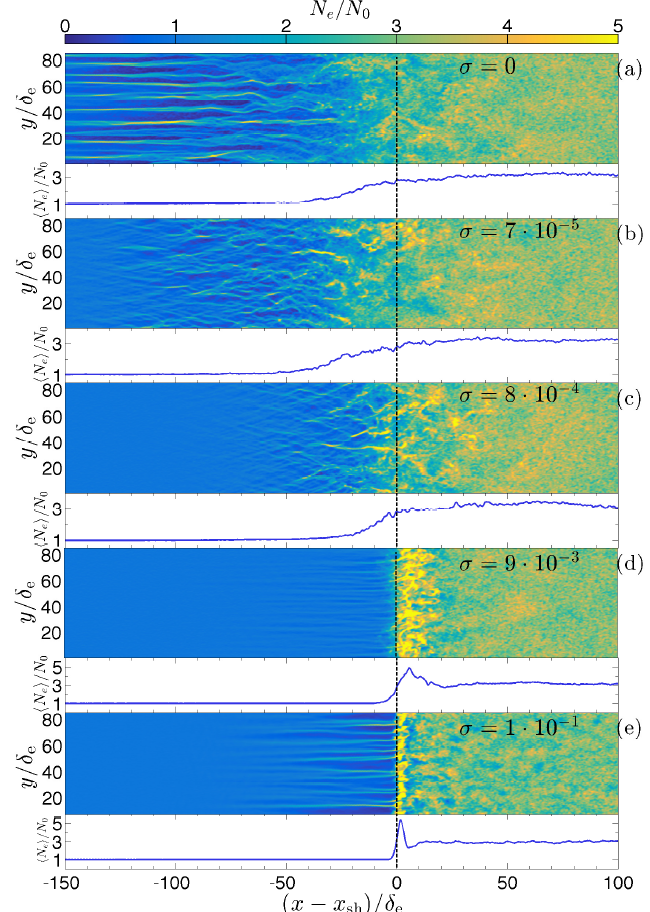}
\caption{Electron density maps in the simulation plane at $t \omega_{\rm pe}=300$ from the spectral \shockapic code. Five different magnetizations are presented from top to bottom panels: (a) $\sigma=0$, (b) $\sigma=7\times 10^{-5}$, (c) $\sigma=8 \times 10^{-4}$, (d) $\sigma=9\times10^{-3}$, and (e) $\sigma=10^{-1}$. The $x$-axis denotes the position relative to the shock front $x_{\rm sh}$. The latter is defined here as the position where the density 
is equal to $2.8 N_0$.}
\label{fig:densities}
\end{figure}

In Figure~\ref{fig:densities} we present the 2D density maps from the simulations performed with the code \shockapic for five different magnetizations at $t\omega_{\rm pe}=300$, when the shock is well-formed but has not yet evolved over long time scales.
Due to the complete symmetry between the two species we present only the electron density maps. 
The shock front is located at $x=x_{\rm sh}$ (black dashed line) and propagates from right to left.
In panel (a), the case of an unmagnetized shock is presented, $\sigma=0$. 
The precursor region, $x-x_{\rm sh}<0$, is filled with Weibel-generated filaments, elongated in the direction parallel to the shock-normal. 
These filaments have initially a transverse width $\sim \delta_{\rm e,rel}=\sqrt{\gamma_0}\delta_{\rm e}$.
They grow in size when approaching the shock front where they mix up and are disrupted. 
The shock front where the density increases from $N_0$ to its downstream value $N_d \simeq 3.2 N_0$, has a width of $60$-$70\delta_{\rm e}$. 
Panels (b) and (c) depict the shock structure for $\sigma = 7\times 10^{-5}$ and $8\times 10^{-4}$, respectively. 
The filamentary structures in the precursor are shorter than in the unmagnetized case and the filaments are slightly oblique. 
In addition, the shock width is narrower than in panel (a). 
For these low magnetization, the downstream asymptotic density value is roughly the same as in panel (a). 

Panel (d), for which $\sigma = 9\times 10^{-3}$, presents a case where the magnetic field in the downstream plasma becomes dynamically non-negligible.  
The particle precursor is very short and the shock is shaped by the compression of the external magnetic field. 
In this case, the so-called magnetic reflection dominates the shock transition and a characteristic overshoot at the shock front can be seen between $0< x - x_{\rm sh} < 20\delta_{\rm e}$ where a density hump is formed. 
The filamentary structure in the precursor, seen for $x-x_{\rm sh}<0$, is due to the reaction of the incoming background plasma to the electromagnetic large-amplitude waves emitted from the shock front \citep[also evidenced by][]{2017ApJ...840...52I}. The presence of these non-Weibel filamentary structures is even more prominent in the $\sigma=0.1$ case (panel e) for which the emitted wave intensity is larger than for the case $\sigma=9 \times 10^{-3}$, and we observed this kind of filamentation for any $\sigma > 10^{-3}$. 
We attribute these filamentary structures to the self-focusing and filamentation of the high-amplitude electromagnetic wave when in propagates through the upstream plasma, 
as introduced theoretically in, e.g., \citet{1974PhRvL..33..209M}. Our results in panels (d-e) illustrate this effect in a pair plasma.

\subsection{Magnetic field and phase space}
\begin{figure*}
\centering
\includegraphics[width=0.95\textwidth]{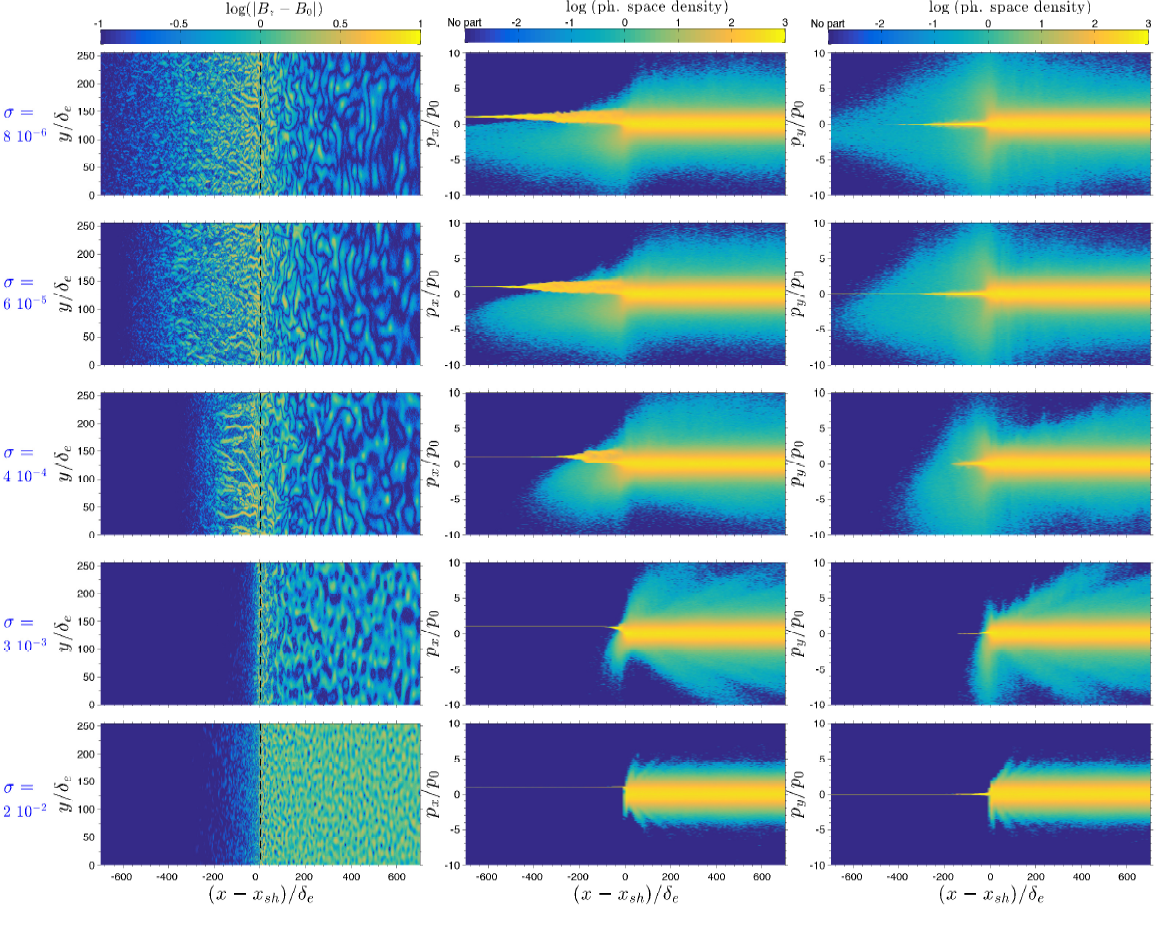}
\caption{Magnetic field and flow structure around the shock at $t \omega_{\rm pe}=2000$ (\smilei code) for five representative magnetizations, from top to bottom: $\sigma=8\times 10^{-6}, 6\times 10^{-5}, 4\times 10^{-4},3\times10^{-3}$, and $2\times10^{-2}$. \emph{Left column:} $\log (|B_z-B_0|)$ magnetic field 
in the simulation plane. \emph{Central column:}  electron $x$-$p_x$ phase space distribution. \emph{Right column:} electron $x$-$p_y$ phase space distribution.  }
\label{fig:bz_phasespace_struct}%
\end{figure*}

We now discuss the shock structure at longer times with respect to the one presented in the previous Sec.~\ref{sect:global_struct:density}. 
The simulations presented here have been performed with \smilei.  
Figure~\ref{fig:bz_phasespace_struct} shows at $t\omega_{\rm pe}=2000$ the magnetic field strength $\log[{e\vert B_z-B_0\vert} /(m c \omega_{\rm pe})]$ in the simulation plane (left column), the electron phase space density projected in the $x$-$p_x$ plane (central column) and in the $x$-$p_y$ (right column) where $p_{x,y}=m c\gamma\beta_{x,y}$.  
The results of simulations performed with five different values of $\sigma$ are presented from top to bottom: $\sigma=8\times 10^{-6}, 6\times 10^{-5}, 4\times 10^{-4},3\times10^{-3}, 2\times 10^{-2}$, respectively. \footnote{The magnetizations are different than presented in Fig.~\ref{fig:densities} as we here use the \smilei simulation series.} 
The phase spaces of positrons show the same main features, except for the inversion in the $p_y$ momentum as a result of the opposite direction of gyration in the $\mathbf{B}_0$ field.

Weibel turbulence is observed in the upstream region, $x-x_{\rm sh}<0$, for small magnetizations and up to $\sigma = 4 \times 10^{-4}$ and the field strength peaks at the shock front $x=x_{\rm sh}$. 
The filamentary structures are the result of the interaction between the cold incoming flow with $p_{x,0}=m c\gamma_0\beta_0$ and the hot, tenuous cloud of returning particles (precursor) that propagates in the opposite direction, as seen in the $x$-$p_x$ phase-space. 
In the downstream region, $x-x_{\rm sh}>0$, the plasma reaches a relativistic temperature, as expected from the jump conditions, and the magnetic field turbulence decays by phase mixing.
The length of this unstable region decreases with increasing $\sigma$ and for $\sigma \geq 10^{-2}$ the upstream magnetic field is large enough to completely suppress the precursor beam. 
A quantitative study of the precursor length is presented later in Sec.~\ref{sect:global_struct:precursor}. 
For $\sigma>10^{-3}$ the Weibel-unstable precursor is replaced by large amplitude electromagnetic waves emitted from the shock front. 
The regular downstream shock-compressed component of the magnetic field $\mathbf{B}_{0,d}$ is more and more prominent with growing $\sigma$. 
The sharp transition between upstream and downstream is clearly seen for the $\sigma=2\times 10^{-2}$ case.

We note that we obtain qualitatively a structure very similar to the one presented by \citet{SSA13} (Figure~3 of their article), with a finer phase-space resolution here as 20 particles/cell/species were used instead of 2.

\subsection{Precursor length}\label{sect:global_struct:precursor}

\begin{figure}
\centering
\includegraphics[width=0.49\textwidth]{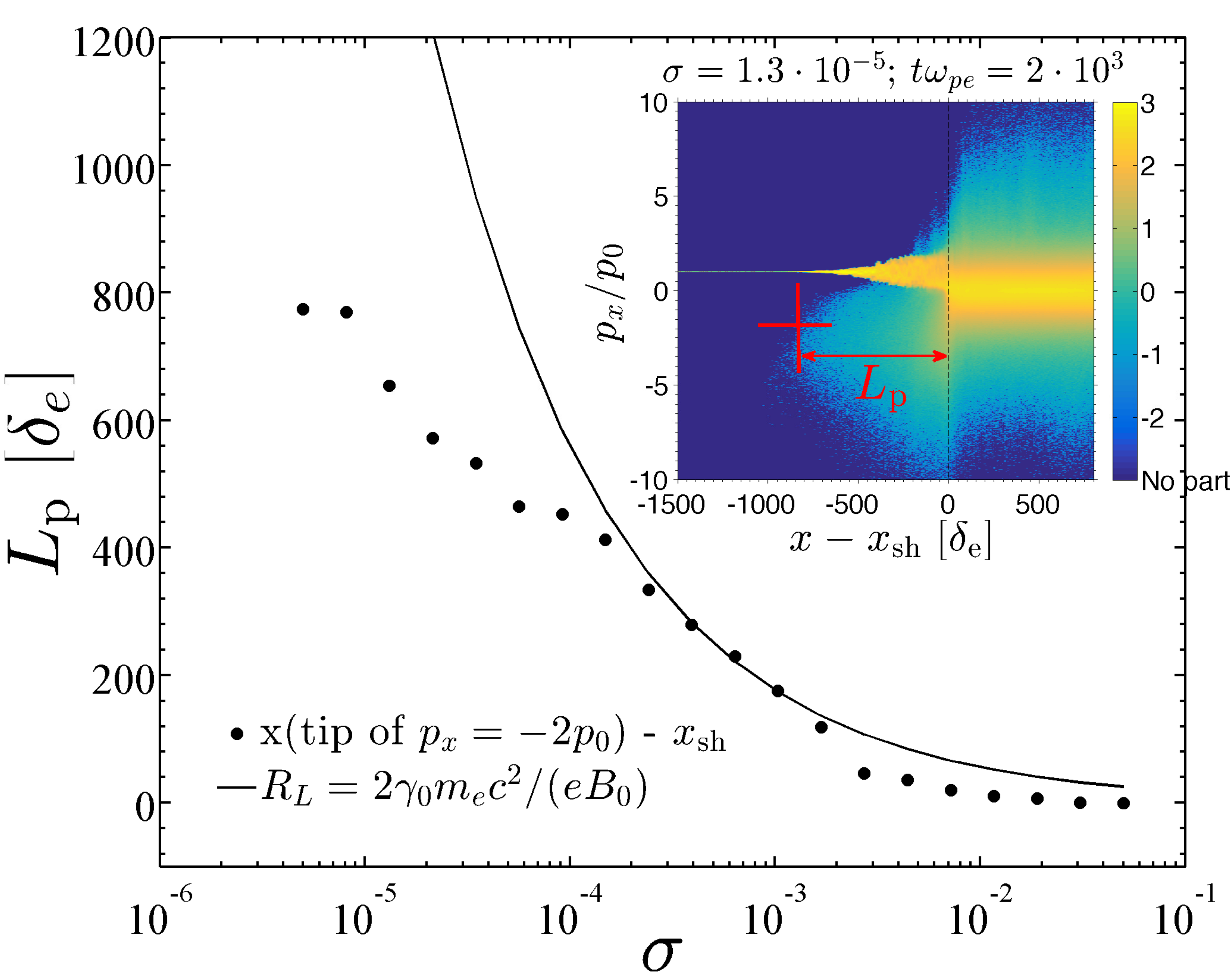}
\caption{Dependence of the precursor length $L_{\rm p}$ on the upstream flow magnetization $\sigma$. 
Filled circles present the value derived from \smilei simulations at $t \omega_{\rm pe}=2000$ and the solid line reports the Larmor radius of 
particles with a Lorentz factor $\gamma=2\gamma_0$ in the external $\bf{B}_0$ magnetic field. 
The \emph{subset} in the upper right corner presents a typical electron $x-p_x$ phase space density (here for $\sigma=1.3\times 10^{-5}$ at $t\omega_{\rm pe}=2000$). 
The red cross shows where the precursor tip is located and the corresponding precursor length $L_p$ is delimited by the red double arrow.}
\label{fig:Lp}%
\end{figure}

It was shown in Figures~\ref{fig:densities}-\ref{fig:bz_phasespace_struct} that the length of the
particle precursor $L_p$, composed of the hot beam of returning particles, decreases as the magnetization increases. 
Eventually, the particle precursor is completely suppressed for $\sigma \simeq 10^{-2}$, leaving
place to an electromagnetic wave precursor.  As long as one is interested in estimating the particle precursor
length (i.e. for $\sigma<2 \times 10^{-3}$ for our simulations, see Fig.~\ref{fig:Lp} and discussion below),
one should consider two competing mechanisms.

(i) The first mechanism that can limit the expansion of the particle precursor is the regular gyration of returning particles in the ordered upstream magnetic field $\mathbf{B_0}$.
The corresponding precursor length then scales as $L_p \approx (\gamma_{\rm inj}/\gamma_0) R_{L,0}$, where $\gamma_{\rm inj}$ is a representative Lorentz factor of the returning particles. 
This scaling was found by \citet{SSA13} for any $\sigma>10^{-5}$ shocks. 
These authors proposed that the characteristic energy of the returning particles corresponds to $\gamma_{\rm inj} \simeq 5 \gamma_0$. 
It is roughly the lower cut-off energy of the power-law part of the distribution function in the precursor region. It carries the bulk of the energy content in the non thermal particles if the power-law spectral index $s={\rm d}\log N/{\rm d}\log E>2$.  As will be evidenced at the end of this Section, in our simulations, this mechanism efficiently limits the particle precursor length for large enough magnetizations (typically $\sigma \gtrsim 10^{-4}$).

(ii) The second mechanism that can limit the expansion of the particle precursor relates to the returning particle scattering on the self-excited microturbulence.
This process should dominate over the regular gyration for weak enough magnetization, leading to particle precursor length $L_p$ of order of the diffusive length in the direction of the shock propagation. It was argued that this scaling may occur in very weak magnetization shocks with $\sigma<\xi_{\rm  \tiny CR}/\gamma^{\vert u\,2}_{\rm sh}$ \citep{2009MNRAS.393..587P, 2010MNRAS.402..321L},  where $\xi_{\rm \tiny CR}$ is the fraction of the incoming energy channeled to the supra-thermal particles and $\gamma^{\vert u}_{\rm sh}$ is the Lorentz factor of the shock front as seen from the upstream rest frame.

To measure the precursor length at the end of the simulation ($t\omega_{\rm pe}=2000$), we localize
the tip of the precursor as the position $x_{\rm tip}$ of the phase space element in the upstream 
region with $p_x=-2p_0$ at the largest distance from the shock front. 
The choice of $2p_0$ momentum is done as a compromise, the value of $L_p$ depending weakly on the
choice of the phase space element between $p_0$ and $5p_0$. 
The precursor length is then defined as $L_p = x_{\rm tip}-x_{\rm sh} $. 
Figure~\ref{fig:Lp} presents the particle precursor length $L_p$, measured at 
$t\omega_{\rm pe}=2000$, as a function of $\sigma$. 
The figure subset in the upper right corner presents the $x$-$p_x$ phase space density for the
case $\sigma=1.3\times 10^{-5}$. 
The red cross shows where the precursor tip is located. 
The measured precursor length, $L_p$, is compared to twice the Larmor radius of an electron with
Lorentz factor $\gamma_0$  in the upstream magnetic field (solid line), the diffusion length being to large to be reported
in this figure.

For low magnetizations, $\sigma<10^{-4}$, the precursor is still expanding at the end of the
simulations so that it did not yet reach its stationary size. At $t\omega_{\rm pe}=2000$ the precursor
size in our simulation is limited by the ballistic expansion of the hot particles moving 
at $\langle v_x \rangle \sim c/2$ to roughly $800\delta_{\rm e}$. Present simulations are
too short to allow distinguishing between the two discussed length scales for very weak magnetizations.

For intermediate magnetizations, $10^{-4}<\sigma<10^{-3}$, $L_p$ stabilizes at $L_p \simeq 2\,R_{L,0}$.
This confirms that the particle precursor extent is here limited by the Larmor gyration in the 
external magnetic field, as found by \citet{SSA13}.

Finally, for $\sigma>10^{-2}$, the particle precursor  disappears as the external magnetic field 
is strong enough to suppress it. 
Indeed, in Figure~\ref{fig:Lp}, for these magnetizations $L_p \ll R_{L,0}$, and it is very close to 0. 

\subsection{Perpendicular current in the precursor}
\label{subsect:perp_current}

It is still a matter of debate if the filamentary turbulence seen in Figure~\ref{fig:bz_phasespace_struct} for $\sigma> 10^{-5}$  results from the `pure' Weibel instability, 
since the latter may not be able to grow to sufficient strength in the precursor  \citep{2010MNRAS.402..321L, 2014MNRAS.440.1365L,2014IJMPS..2860193P}. 
For such magnetizations indeed, the precursor may be too narrow for the Weibel instability to fully develop.

Nevertheless, in the presence of an (ordered) perpendicular  magnetic field, the flow of the precursor hot electrons can be deviated in the negative $y$-direction.
This can be clearly observed in the right column of Figure~\ref{fig:bz_phasespace_struct}: 
with growing $\sigma$ the deviation of the `cloud' of the precursor electrons towards negative $p_y$ values becomes more pronounced. 
Conversely, positrons are deviated in the positive $y$-direction. 
This produces a net current $j_{y,\rm beam}$ in that region, which is in return efficiently compensated by the incoming plasma in order to ensure global neutrality. 
This situation is unstable with respect to the so-called Current Filamentation Instability (CFI) if the drift speed of the background plasma imposed by $j_{y,\rm beam}$ is larger than the upstream sound speed . Its nature is different from the standard Weibel instability, the latter being produced by a neutral beam along $x$. 
As studied theoretically by \citet{2014MNRAS.440.1365L} the role of the CFI in the magnetic field generation can be as important as the Weibel instability for 
$\xi_{\rm \tiny CR}/\gamma^{\vert u\,2}_{\rm sh}<\sigma \leq 10^{-2}$.

\begin{figure}
\centering
\includegraphics[width=0.49\textwidth]{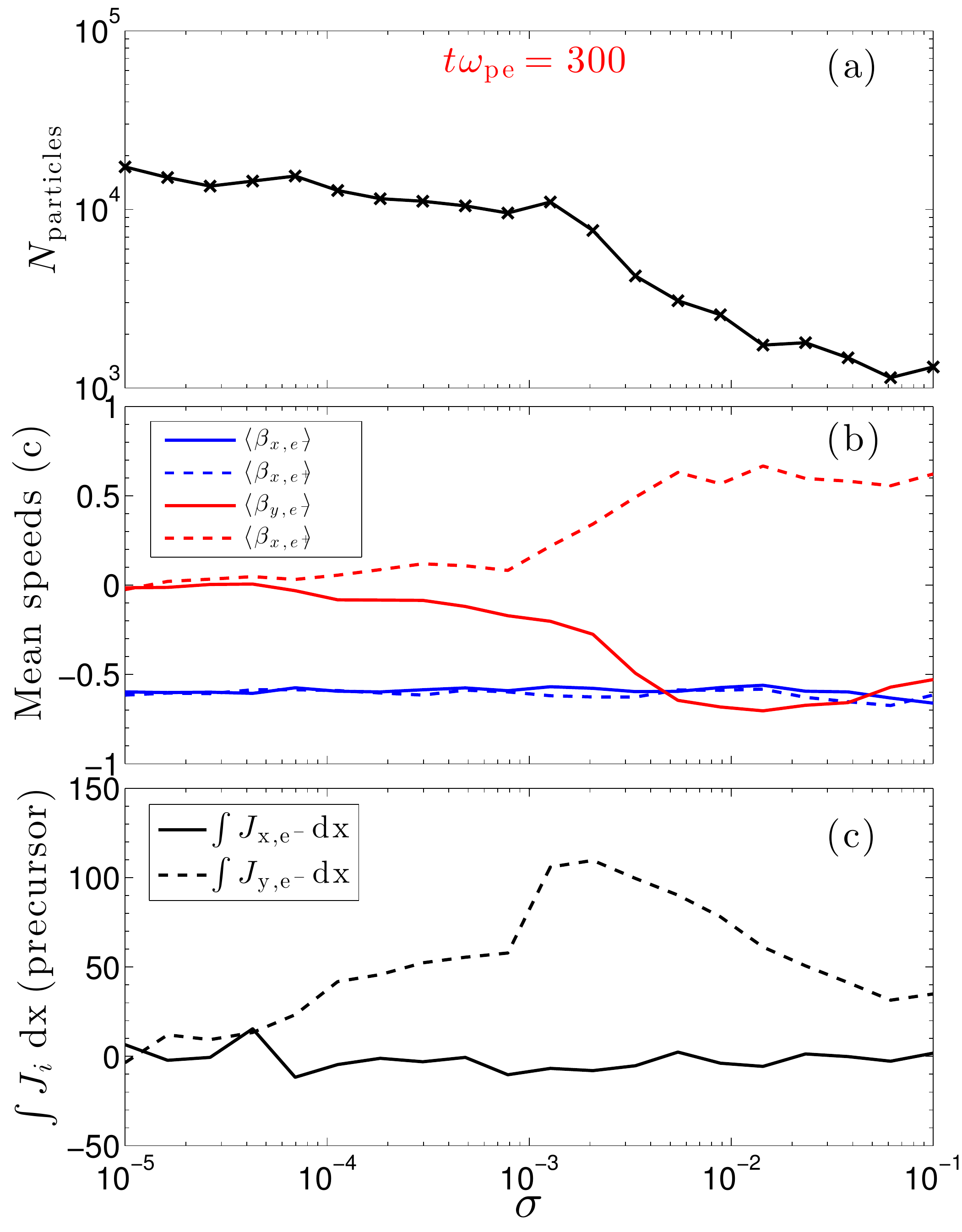}
\caption{Dependence on the magnetization of (a) the number of the returning beam macro-particles in the shock precursor, (b) the mean speeds of the returning beam electrons (dashed lines) and positrons (solid lines) in shock-normal $x$-direction (blue color) and perpendicular $y$-direction (red color), and (c) the integrals of the beam contribution to the $j_x$ current (solid line) and to the $j_y$ current (dashed line).  Done with the spectral code \shockapic at $t\omega_{\rm pe}=300$.}
\label{fig:beam_speeds}%
\end{figure}

The existence and amplitude of this perpendicular current in the precursor of relativistic shocks was never evidenced in PIC simulations. 
Indeed the total $j_y$ current, formed by the contributions of the returning beam particles and of the incoming cold flow particles, 
is very small as a result of the efficient compensation of the beam-produced current by the background plasma. 
We show here that, for a finite $\sigma$, the beam of outgoing particles in the precursor indeed produces a net perpendicular current. 
 
In order to measure the returning beam contribution to the current, it is necessary to isolate the beam particles in the precursor from the background incoming population. 
In our simulations, this is done by selecting as the precursor particles those with negative momentum along the flow $x$ direction, $p_x<0$.
In Figure~\ref{fig:beam_speeds} we plot as a function of $\sigma$ the number of macro-particles (panel a), the average velocity in the $x$ and $y$ directions (panel b), 
as well as the integral over the precursor length of the current of the returning beam particles (panel c).  
As seen in panel (a) the number of macro-particles in the beam is roughly constant up to $\sigma \sim 10^{-3}$, while it decreases for larger magnetizations, the particle precursor eventually disappearing (see previous Sec.\ref{sect:global_struct:precursor}). 
Clearly, electrons and positrons have the same outgoing average velocity $\simeq 0.6 c$ in the shock normal direction $-x$, as seen in panel b (blue lines). 
This implies that no net current $j_x$ will develop in the background plasma. 
The picture is different in the perpendicular $y$ direction. 
As expected, electrons and positrons go in opposite directions with roughly the same average velocity (red lines). 
The average $y$ velocity vanishes for $\sigma<10^{-5}$ and increases up to $|v_y|\simeq 0.6c$ for $\sigma\geq 10^{-3}$. 
Such a configuration produces a net $j_y$ current carried by the returning particles, as shown in panel (c). 
The parallel component is always close to 0, as expected, while the $j_y$ component increases gradually for $10^{-5}<\sigma<10^{-3}$. 
The maximum current is reached for $\sigma \simeq 2\times 10^{-3}$ and, for larger magnetizations $\int j_y {\rm d}x$ decreases as the precursor shortens rapidly. 
Based on these results we expect the influence of the perpendicular current, and in turns of the CFI on the shock structure to be most important for this level of magnetization, 
$\sigma \sim 10^{-3}$.

%% file: 05_formation_timescales.tex
\section{Characteristic time of shock formation}
\label{sect:formation}

In this Section, we investigate, as a function of the upstream magnetization, the characteristic time for shock formation.
Our definition of the shock formation time follows from the study of the temporal evolution of two complementary quantities: 
(i)~the compression factor of the downstream region, 
(ii)~a measure of the anisotropy in the $p_x$-$p_y$ phase-space of the downstream plasma.
Both quantities have been computed in each simulation considering a region in the overlapping plasma
 located close to, but no exactly at, the reflecting wall on the right border of the simulation box (see Sec.~\ref{sect:formation:time} for more details).
The compression factor is obtained simply as the ratio of the averaged plasma density in this region by the initial plasma density $N_0$.
The anisotropy is measured introducing the anisotropy parameter:
\begin{eqnarray}\label{eq:aiso}
a_{\rm iso} = \frac{\langle v_x\,p_x \rangle}{\langle v_y\,p_y \rangle} -1\,,
\end{eqnarray}
where $\langle v_i\,p_j\rangle$ is the pressure tensor averaged over the whole overlapping region,
so that an isotropized plasma corresponds to values $a_{\rm iso} \ll 1$.

As previously underlined, the main instability at the origin of shock formation depends on the upstream magnetization. 
In what follows, we first (Sec.~\ref{sect:formation:early}) discuss the early phase of shock formation that corresponds to the initial phase dominated by the instability,
focusing in particular on the modification of the $p_x$-$p_y$ particle distribution.
We then address the shock formation time (Sec.~\ref{sect:formation:time}), discussing the methods used to identify the time of shock formation then presenting our results of time formation as a function of the magnetization parameter.

\subsection{Instability and isotropization of the flows}\label{sect:formation:early}

\begin{figure*}
\centering
\includegraphics[width=0.8\textwidth]{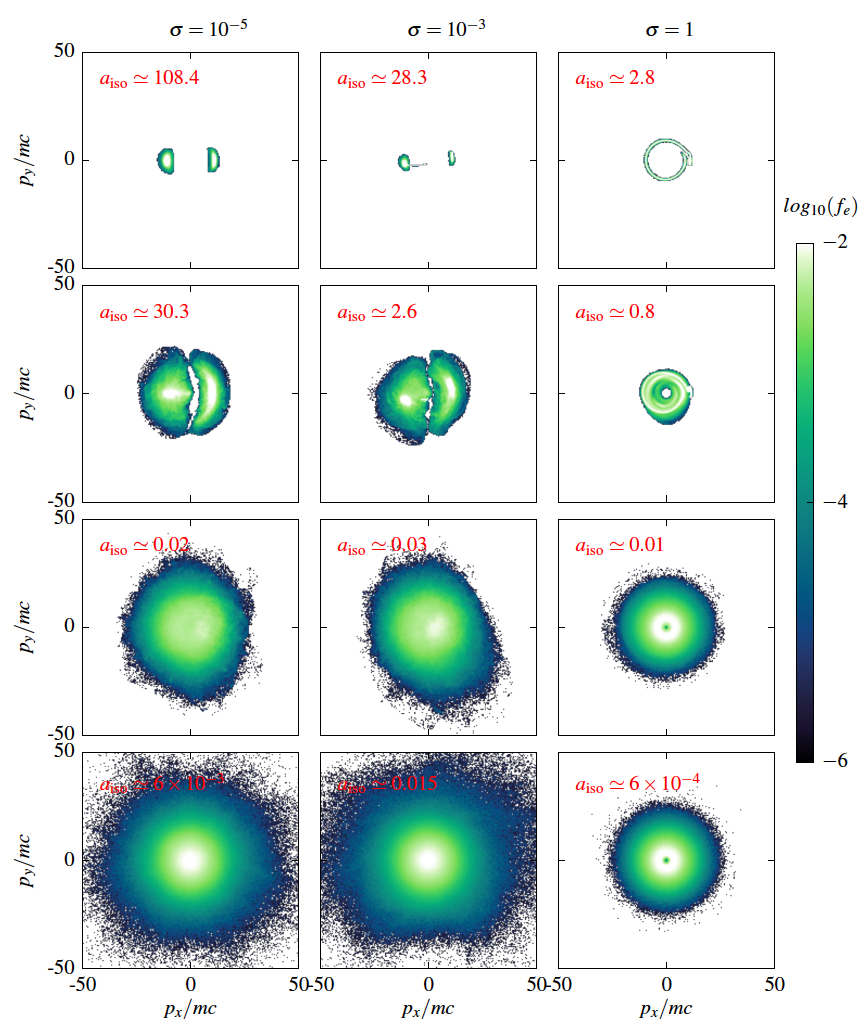}
\caption{Evolution of the $p_x-p_y$ phase space density in the downstreaming region $8 \delta_{\rm e}<\vert x-x_{\rm wall}\vert<28 \delta_{\rm e}$ for times from $t \omega_{\rm pe}=20$, $30$, $100$ and $500$ (form top to bottom) and for three magnetizations: $\sigma= 10^{-5}$ (left column), $\sigma=10^{-3}$ (central column), and $\sigma=1$ (right column). Done with \smilei. }
\label{fig:tform_pxpy}%
\end{figure*}

Before looking at the shock formation timescale, we illustrate in the initial phase of shock formation the difference between the weakly and highly magnetized plasmas. We follow the $p_x$-$p_y$ phase-space density  in the overlapping region, close to the right boundary of the simulation box $x_{\rm wall}$, in between $x_{\rm wall}-28\,\delta_{\rm e}$ and $x_{\rm wall}-8\,\delta_{\rm e}$. 
The time evolution is presented in Figure~\ref{fig:tform_pxpy} where we show the $p_x$-$p_y$ phase-space density for three representative magnetizations: 
$\sigma= 10^{-5}$ (left column), $\sigma=10^{-3}$ (central column) and $\sigma=1$ (right column). 
The evolution in time, for $t\omega_{\rm pe}=$20, 30, 90, and 500, is presented from top to bottom. 

For $\sigma \ll 1$ (left and middle columns), two counter streaming cold beams are present at $t \omega_{\rm pe}=20$. 
A slight shift in the $p_y$ direction is seen for the case $\sigma=10^{-3}$ as a result of the beam rotation in the external $B_0$ field, 
while both beams are centered in $p_y=0$ for $\sigma= 10^{-5}$. 
These beams quickly become unstable, and at time $t \omega_{\rm pe}=30$ the phase-space distribution of the two beams appear to be strongly modified, and in particular broader.
Yet two distinct beams can still be identified and the anisotropy parameter [Eq.~\ref{eq:aiso}] 
remains large ($a_{\rm iso}\simeq 30$ for $\sigma = 10^{-5}$ and $a_{\rm iso}\simeq 2.6$ for $\sigma = 10^{-3}$).
At $t \omega_{\rm pe}=90$ the distribution is almost fully isotropized, and the anisotropy parameter has strongly decreased down to $a_{\rm iso} \sim 0.02$ and $0.03$ for $\sigma = 10^{-5}$ and $10^{-3}$, respectively.
At much a larger time, $t \omega_{\rm pe} = 500$, the $p_x$-$p_y$ phase space assumes a Maxwell-Juttner-like distribution and thermalization has been reached.
Note that the temperatures measured in these simulations are found to be in excellent agreement with Rankine-Hugoniot predictions.

The evolution is different in the highly magnetized case $\sigma=1$ (right column of Fig.~\ref{fig:tform_pxpy}).
At $t \omega_{\rm pe}=20$ the plasma flows show a cold ring distribution, which quickly turns unstable with respect to the Maser Synchrotron Instability~\citep{1991PhFlB...3..818H}. This instability quickly heats up the flows and a much broader ring-like structure is observed at $t \omega_{\rm pe}=30$.
Yet, the anisotropy parameter for these two early times remains quite large, $a_{\rm iso}\simeq2.8$ and 0.8 for $t \omega_{\rm pe}=20$ and 30, respectively.
Isotropization is reached at time $t \omega_{\rm pe}=90$ for which $a_{\rm iso} \simeq 0.01$, and remains approximately stable at later times. 
It is worth noting that the ring distribution does not relax completely to a Maxwell-Juttner: the phase space region around $[p_x,p_y]=0$ remains depleted. 
The relaxation towards the Maxwell-Juttner distribution is even less efficient for larger $\sigma$. 

\subsection{Formation time}\label{sect:formation:time}
 
We now turn to extracting the shock formation time from our PIC simulations. 
To do so, one has to find an operating definition for the formation time, and the question arises as to what is the relevant quantity to look at. 

In~\citet{2014PhPl...21g2301B} (and following works), the authors compute the formation time from their PIC simulations by extracting, as a function of time, 
the shock front position, there defined as the position for which the plasma density reaches $3\,N_0$. At large time, this position is found to follow a straight line (the shock front velocity being constant). Expanding this line toward shorter times, they define the time of shock formation as the time at which the line crosses 
the $x=x_{\rm wall}$ position.
While this method allow them to compute formation times consistent with simple analytical estimates (as will be discussed later in this Section), this method proved to be unreliable for our study.

Instead, our definition of the time of shock formation follows from the study of the temporal evolution, 
in the overlapping plasma region, of both the compression factor and anisotropy parameter, Eq.~\ref{eq:aiso}.
Different averaging region have been considered, all of them extending up to $x_{\rm wall}-8\,\delta_{\rm e}$, 
but with different widths, $[5,10,15,20]~\delta_{\rm e}$.  

The {\it compression} time $T_{\rm dens}$ is obtained as the time at which this averaged density reaches 95\% of the maximum density of the downstream plasma obtained at late times (well after shock formation). 
Note that for either small ($\sigma \lesssim 10^{-4}$) or large ($\sigma \gtrsim 10^{-1}$) magnetizations, this threshold corresponds to
95\% of the downstream density theoretically predicted from Rankine-Hugoniot conditions .
For intermediate magnetizations, this threshold is somewhat lower than predicted as the late time compression factor obtained in our simulations
is for these magnetizations slightly lower then predicted [see Fig.~\ref{fig:jump_dens}a) and Sec.~\ref{sect:jump}].
Finally, the {\it isotropization} time $T_{\rm iso}$ is defined as the time such that $\vert a_{\rm iso}\vert(t \ge T_{\rm iso}) \le 0.04$.

\begin{figure}
\centering
\includegraphics[width=0.49\textwidth]{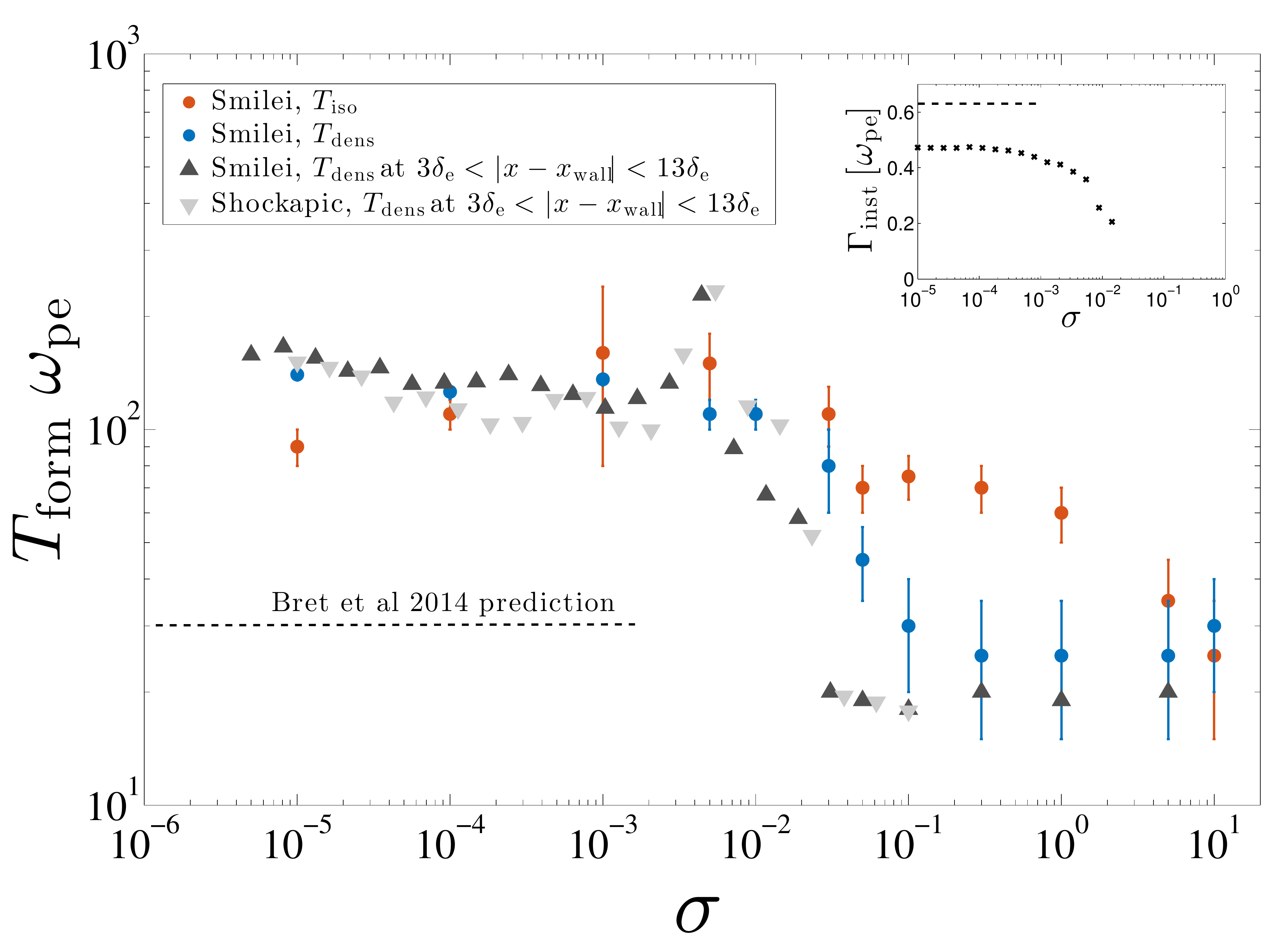}
\caption{Dependence on $\sigma$ of the measured  {\it isotropization} and {\it compression} times (orange and blue circles, respectively), extracted from selected \smilei simulations. Also shown are the {\it compression} times derived from complementary \smilei simulations (dark triangles) and \shockapic simulations (light-grey triangles), considering only the $3\delta_{\rm e}<|x-x_{\rm wall} |<13\delta_{\rm e}$ region. The dashed-line reports the prediction for 
the time of shock formation using the model presented in \citet{2014PhPl...21g2301B}. \textit{Inset:}  magnetic field growth-rate dependence on $\sigma$ during the linear phase of the instability. The averaged magnetic field energy was taken over the region $3\delta_{\rm e}<|x-x_{ \rm wall} |<13\delta_{\rm e}$. The dashed line follows the maximum theoretical growth rate in the unmagnetized limit $\Gamma_{\rm weib} = 2\omega_{\rm pe}/\sqrt{\gamma_0}$.}
\label{fig:tform}%
\end{figure}

Figure~\ref{fig:tform} reports, as a function of the upstream magnetization $\sigma$, the measured {\it compression} and {\it isotropization} times 
extracted from our simulations. The reported values are obtained considering different widths of the
averaging region. 
Error bars account for the spread around the mean value, the diagnostics output frequency 
and the measurement errors when analyzing the time series of the density compression ratio and isotropization parameter.
As shock formation implies both a compression consistent with Rankine-Hugoniot conditions and isotropization of the flow, the shock formation time $T_{\rm form}$ is
henceforth defined as the maximum of $T_{\rm dens}$ and $T_{\rm iso}$:
\begin{eqnarray}
T_{\rm form} \equiv {\rm max}\left\{ T_{\rm dens}, T_{\rm iso} \right\}\,.
\end{eqnarray}

For small magnetizations, $\sigma < 10^{-3}$, isotropization is reached at time $t \omega_{\rm pe} \simeq 100$. 
At that time however the compression factor in the overlapping region still remains below the value predicted using Rankine-Hugoniot conditions.
One has to wait $t \omega_{\rm pe} \simeq 145$ before the compression factor reaches 95\% of this maximum value.
Hence, for $\sigma < 10^{-3}$, the formation time is $T_{\rm form} \equiv T_{\rm dens} \simeq 145~\omega_{\rm pe}^{-1}$.
Additional simulations considering small magnetizations (triangles in Fig.~\ref{fig:tform}) show that the formation time is roughly constant for these magnetizations,
indicating that the external magnetic field is too weak to influence  shock formation.
Note also that both PIC codes give similar predictions for this time (dark up-triangles for \smilei's simulations, and light down-triangle for Shockapic's ones).

For such low magnetizations, one expects the Weibel instability to mediate the shock. This is confirmed in our simulations where the energy in the magnetic field $B_z$ 
is found to increase exponentially with a growth rate consistent, if slightly smaller, than the maximum growth rate $\Gamma_{\rm weib} = 2\omega_{\rm pe}/\sqrt{\gamma_0}$ for the Weibel instability driven by two cold, counterstreaming electron-positron plasmas (see insert in Fig.~\ref{fig:tform}).

A simple model for the time of shock formation in an unmagnetized plasma was proposed in \citet{2014PhPl...21g2301B}.
This model considers that shock formation is reached, in a two-dimensional geometry, at a time corresponding to twice the time $T_{\rm sat}$ necessary for the Weibel instability to saturate, which for our parameters would lead $T_{\rm sat} \simeq 15\,\omega_{\rm pe}^{-1}$. 
This predicted value does agree with the characteristic time for the Weibel instability to saturate $T_{\rm sat} \simeq 18-27\,\omega_{\rm pe}$ measured in reduced simulations (not shown) considering initially overlapping counterstreaming plasmas and periodic boundary conditions in both $x$ and $y$ directions with otherwise similar physical and numerical parameters. 
It is however much smaller than either the isotropization or compression times, and thus than the measured time of shock formation.
Indeed, our reduced simulations show that particles (electrons and positrons) are not fully stopped at the time at which the instability saturates 
[see also the discussion on the saturation mechanisms of the Weibel instability in \citet{grassi2017}]. 
As a result, in the full (shock formation) simulations, the overlapping region continues to expand slowly well after the time of saturation. This prompts us to suggest that the late-time merging phase and development of turbulent magnetic fields e.g. through kink-like instabilities are central to the late stages of formation of the collisionless shock and require additional time.

For intermediate magnetizations,  $2\times 10^{-3} < \sigma < 5 \times 10^{-2}$, there is a transitory regime
for which both compression and isotropization times decrease gradually. As demonstrated in Sec.~\ref{sect:global_struct}, in this transitory regime,
a slight departure from RH conditions was observed on the density compression factor, the precursor length was reduced down to twice the gyration 
radius of particles in the external magnetic field, and a net current in the $y$-direction was observed suggesting that the instability mediating shock formation
was more related to CFI rather than the standard, neutral, Weibel instability. This regime was also associated with the presence of an overshoot 
in the particle density at the tip of the reflected beam [see, e.g. Fig.~\ref{fig:densities}(d)]. 

Yet, for these intermediate magnetizations, the external magnetic field is not strong enough to trigger a rapid turnover of the reflected particles while the filamentation growth rate decreases rapidly (see the inset in Fig.~\ref{fig:tform}).
As a result, both compression and isotropization times in this regime decrease slowly as the magnetization increases.
Note also that,  for this range of magnetizations,  $2\times 10^{-3} < \sigma < 5 \times 10^{-2}$, both compression and isotropization times are found to be of the same order.

Finally, for high magnetizations, $\sigma \gtrsim 5 \times10^{-2}$, one enters the regime for which 
the dynamical effect of the external magnetic field becomes dominant
and neither Weibel nor CFI is observed (see Sec.~\ref{sect:global_struct}).
For such high magnetizations, both the compression and isotropization times are reduced so that the averaging region 
over which these times are computed is reduced\footnote{Reducing the averaging region for smaller magnetizations does not impact our results.} 
to $3 < \vert x-x_{\rm wall} \vert < 13$.
In this regime, the downstream density accumulation predicted by the RH conditions is reached very quickly, 
for times $\omega_{\rm pe} T_{\rm dens}  \sim (20 - 30)$.
This can be explained by the faster particle gyration in the compressed magnetic field, 
as well as by the reduced value of the density jump predicted by the RH conditions at high magnetizations.
This reduced compression factor makes the definition of the shock formation time at these high magnetization quite difficult to diagnose using the density compression argument only. 
As a result, even though density compression is quite fast, it takes more time to achieve isotropization of the flows, and one typically
obtains $\omega_{\rm pe} T_{\rm iso} \sim (50-70)$ for magnetizations $\sigma \lesssim 1$, 
and $\omega_{\rm pe} T_{\rm iso} \sim (20-40)$ for the two simulations performed at high-magnetizations $\sigma \gg 1$. 
In this regime of high magnetization, isotropization thus defines the time of shock formation: $T_{\rm form} \equiv T_{\rm iso} > T_{\rm dens}$. 
For times larger than $T_{\rm form}$, the downstream region is relaxed,
the shock is stabilized and propagates steadily with a constant speed consistent with RH conditions 
[see Sec.~\ref{sect:jump} and Fig.~\ref{fig:jump_dens}(b) in particular].

Yet, no clear dependency of either compression or
isotropization times on $\sigma$ is observed. This is somewhat intriguing as one could expect a $\sigma^{-1/2}$-dependence corresponding to the gyration time in the compressed magnetic field, or some different scaling relevant to the characteristic MSI growth.
Indeed, in magnetically dominated shocks
one needs to reconsider the relevant timescales since any filamentation instability is no
longer relevant. We can identify three of them:
\begin{enumerate}
\item  the relativistic particle gyro-period in the external $\mathbf{B}_0$ field:
$\tau_{g,0} = 2 \pi \gamma_0/\omega_{\rm ce} = 2\pi \gamma_0 m c/(e B_0)$.
It is the shortest timescale. During the first stages of the shock
formation a dense shell of plasma (overshoot) is formed on this timescale.
\item the Maser Synchrotron Instability saturation timescale. It corresponds to the
time on which the cold ring distribution in $p_x-p_y$ space collapses
towards a quasi-Maxwellian.
\item the timescale of convection/diffusion of the dense shells built up in the overlapping region during the early stage ($\propto \tau_{g,0}$).
These shells are responsible for a highly modulated density profile, and need time to fully mix before getting a stationary downstream density.
\end{enumerate}

In the simplest scenario of the shock formation by magnetic reflection,
only the first timescale is considered, leading to
$T_{\rm form} \sim \tau_{g,0} \propto \sigma^{-1/2}$.
In a more refined picture, full relaxation of the downstream plasma is
required, i.e., distribution function isotropization and convergence towards
Rankine-Hugoniot conditions. Then one needs to consider the two last
timescales. 

\begin{figure}
\centering
\includegraphics[width=0.49\textwidth]{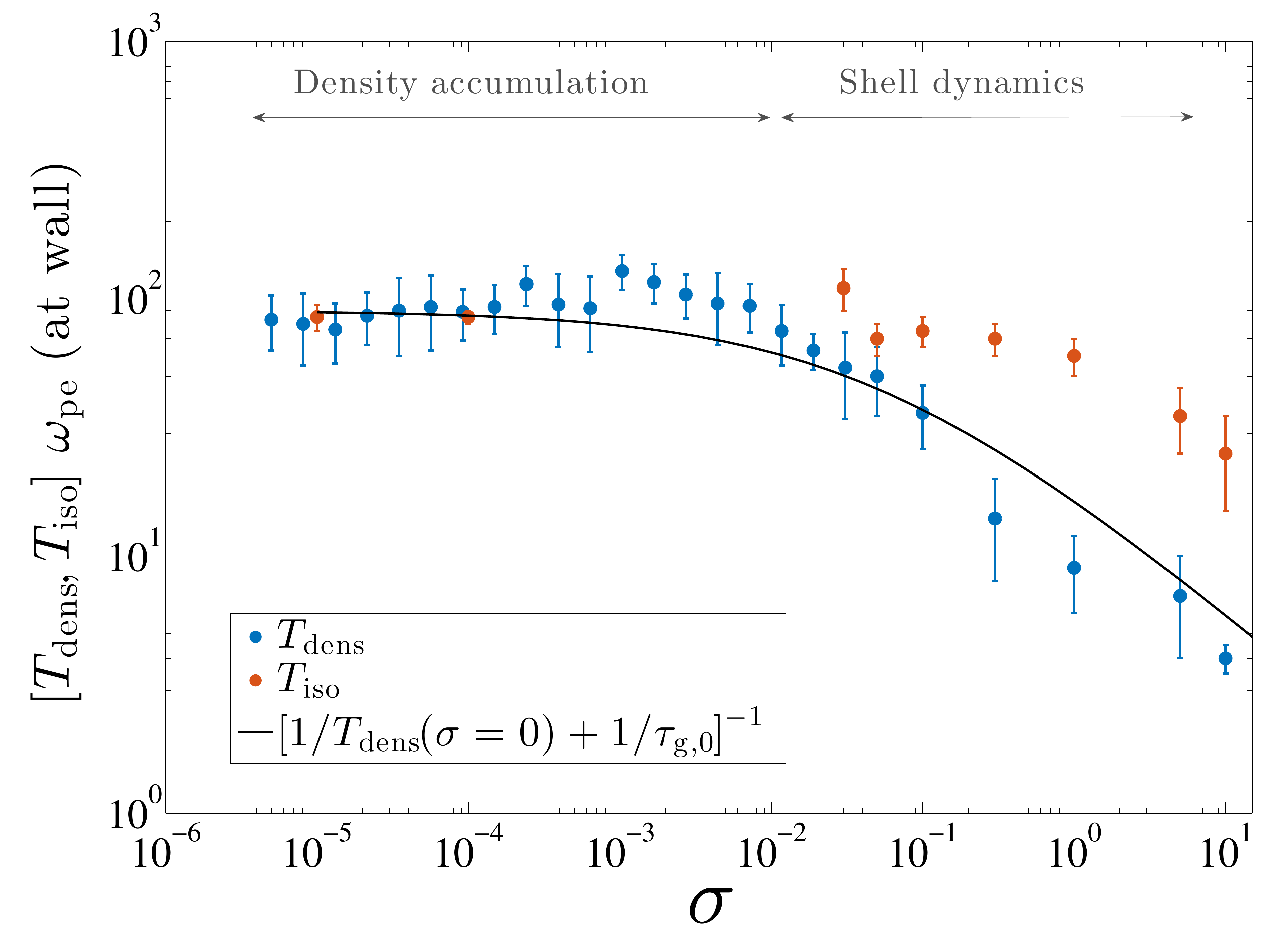}
\caption{Dependence on $\sigma$ of the density compression time and isotropization time measured at the exact position of the wall. Here $T_{\rm dens}$ is derived by considering the density accumulation up to $0.9N_d$ (for $\sigma<5\times 10^{-2}$) or the formation of the second density shell (for $\sigma>5\times 10^{-2}$).}
\label{fig:tform_wall}%
\end{figure}

We further assess this picture by recomputing the characteristic time of density compression and isotropization 
right at the reflecting wall ($x=x_{\rm wall}$) of the simulation. 
The results are reported in Fig.~\ref{fig:tform_wall}. The values of $T_{\rm dens}$ and $T_{\rm iso}$ remain globally unchanged for $\sigma <10^{-2}$.
Then, as $\sigma$ increases, the compression time is found to decrease much faster than the time of isotropization.
For $\sigma>10^{-2}$, the compression time measured at the reflecting wall corresponds to the time of formation of density overshoots (shells).
It is found to scale with $\sigma^{-1/2}$ (as indicated by the solid line for $\sigma \gg 1$ in Fig.~\ref{fig:tform_wall}), 
consistent with the idea that the density overshoot is formed on the $\tau_{g,0}$ timescale.
In addition, this compression time is much shorter than the time of isotropization, that has a much weaker dependency on $\sigma$ (it slightly decrease
from $80$ to $25 \omega_{\rm pe}^{-1}$ over the three decades $\sigma \in [10^{-2},10]$), and 
is related to the relaxation of the downstream plasma.

%% file: 06_downstream_fdist.tex
\section{Downstream particle distribution functions}
\label{sect:fdist}

We now turn to the particle energy spectra as measured behind the shock front, in the downstream plasma.
We first discuss the evolution in time, and for different magnetizations, of the thermal component 
of the distribution and the development of a high-energy tail (Sec.~\ref{sect:fdist:tail}).
We then discuss in more details the temporal evolution of the maximum particle energy (Sec.~\ref{sect:fdist:maxenergy}). 

\subsection{Thermal component and high energy tail}\label{sect:fdist:tail}

After the initial phase of shock formation the particle distribution relaxes to a downstream isotropic, quasi-thermal spectrum 
with, in the case of small enough magnetization, a high-energy non-thermal component eventually developing. 
The acceleration efficiency of the shocks as a function of the magnetization can then be measured by the extent of this tail and its high-energy cut-off. 

\begin{figure}
\centering
\includegraphics[width=0.49\textwidth]{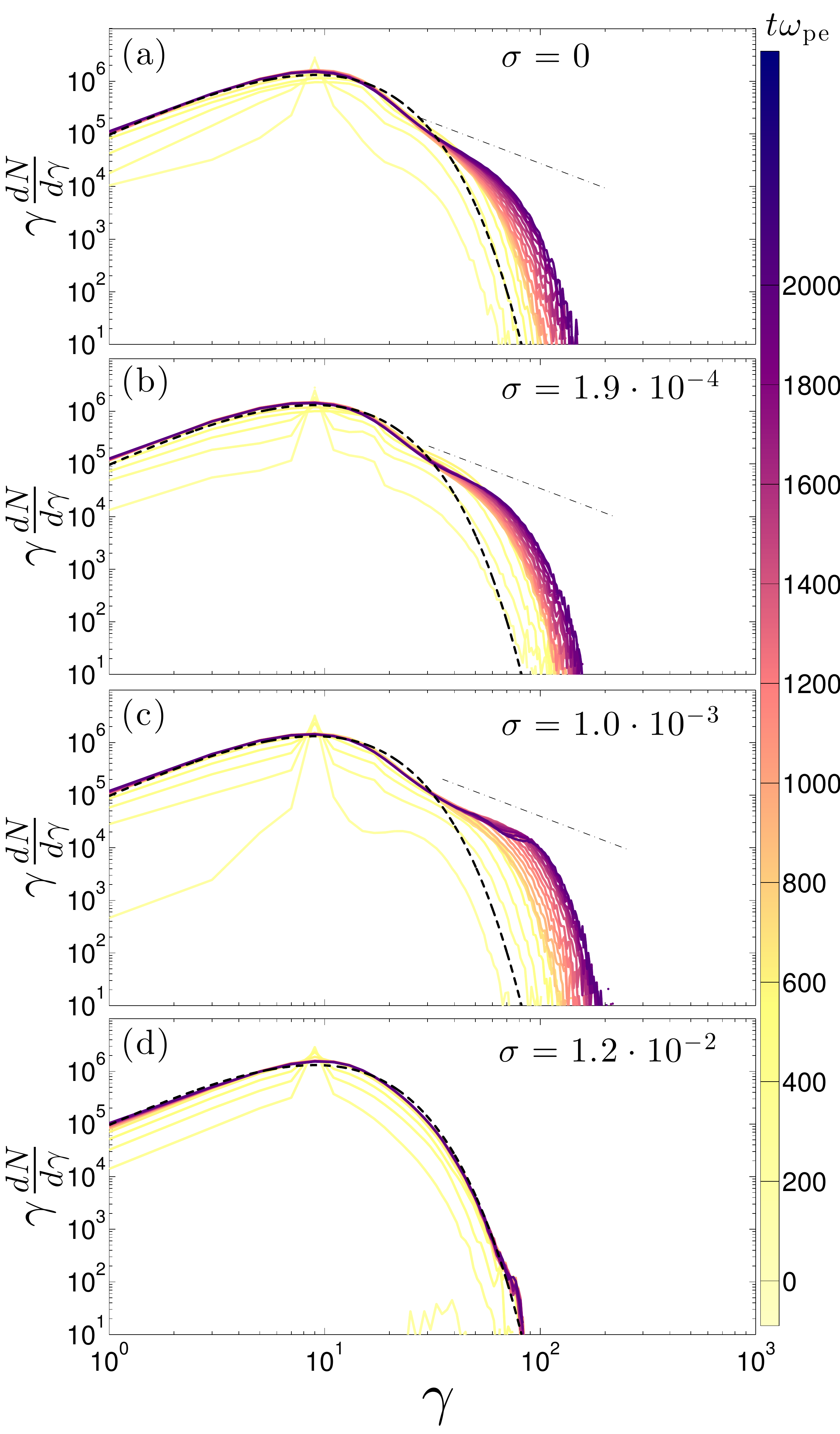}
\caption{Temporal evolution of the electron energy spectrum (\smilei code) in the $200\delta_e$-wide slice between $500\delta_e$ and $300\delta_e$ from the reflecting wall for four representative magnetizations: (a) $\sigma=0$, (b) $\sigma=1.9 \cdot 10^{-4}$, (c) $\sigma=10^{-3}$, and (d) $\sigma=1.2 \cdot 10^{-2}$. The lines of different color correspond to increasing simulation times from $<500 \omega_{\rm pe}^{-1}$ (yellow) to $2000 \omega_{\rm pe}^{-1}$ (violet), as indicated in the colorbar. The \textit{black dashed lines} show the energy spectra corresponding to a 2D relativistic Maxwell-J\"uttner distribution with  temperature expected from the ideal MHD jump conditions. The \textit{black dot-dashed lines} show the  power-law guide line $dN/d\gamma \propto \gamma^{-s}$ with $s=2.5$.}
\label{fig:fdist_downstream}%
\end{figure}

Figure~\ref{fig:fdist_downstream} presents the time evolution of the electron distribution function far downstream for four different magnetizations: 
$\sigma = 0$ in panel (a), $\sigma = 1.9\times 10^{-4}$ in panel (b), $\sigma = 10^{-3}$ in panel (c), and $\sigma = 1.2 \times 10^{-2}$ in panel (d). 
Because of the symmetry with positrons we do not present their distribution here. 
For all magnetizations, at early times $t\omega_{\rm pe}<200$, Figure~\ref{fig:fdist_downstream} illustrates the transition from the initial beam-like 
to the isotropic distribution during the shock formation. 
A thermal state with $dN/d\gamma \propto \gamma e^{-\gamma/\gamma_{\rm th}}$ is reached before $t\omega_{\rm pe}=500$, 
with $\gamma_{\rm th}=k_B T/(m c^2)$ and temperature $T$ in good agreement with the value expected from the MHD jump conditions (derived in Appendix~\ref{appendix:jump}). 
Indeed, for comparison we show (dashed lines) the expected 2D Maxwell-J\"uttner distribution function with $\gamma_{\rm th}$ deduced from Eq.~\ref{eq:Tdownstream}.

\begin{figure*}
\centering
\includegraphics[width=0.9\textwidth]{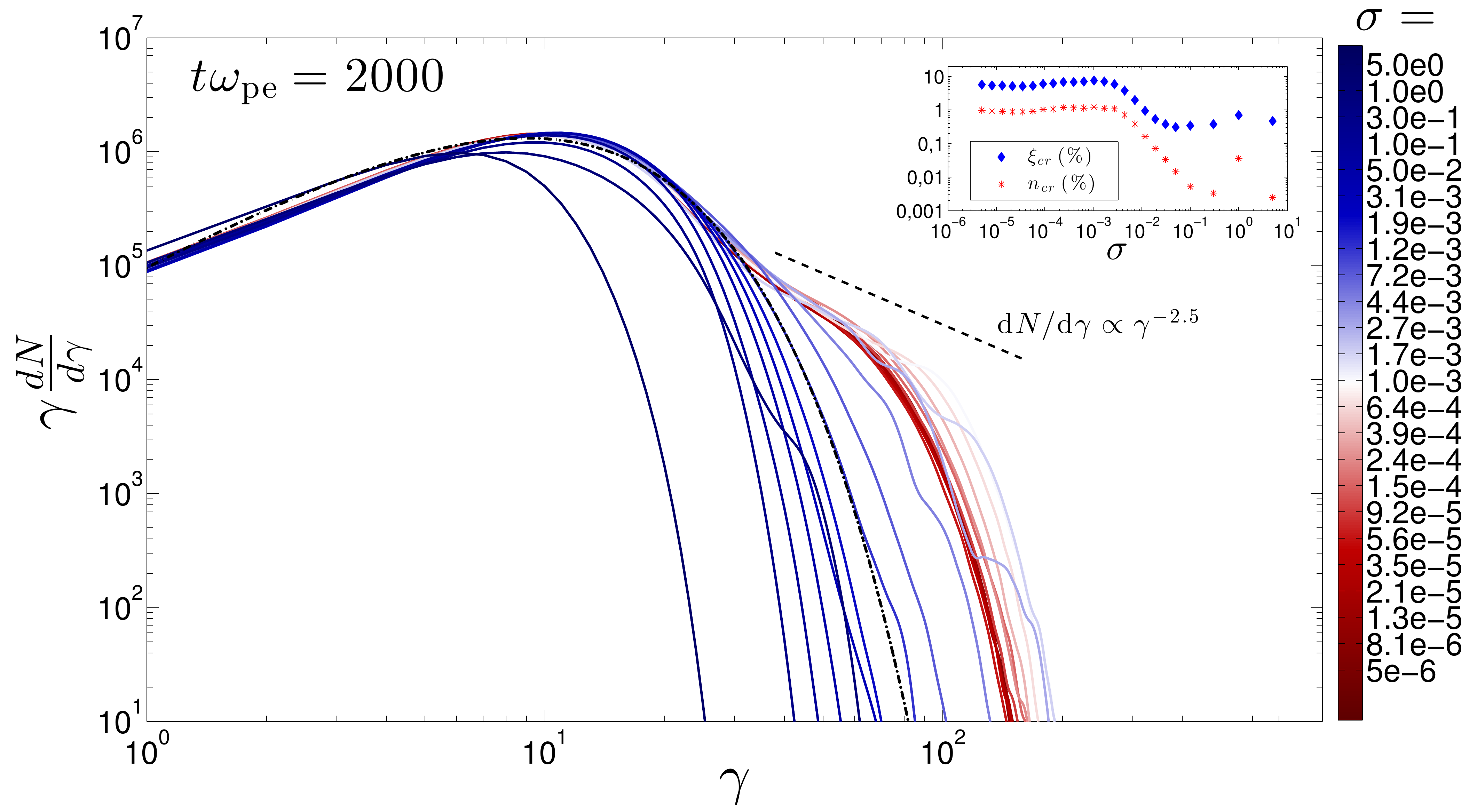}
\caption{Electron energy spectrum far downstream of the shock (\smilei code) at $t  \omega_{\rm pe}=2000$ for simulations with growing magnetization 
from $\sigma=5 \times 10^{-6}$ (deep red solid line) to $\sigma=5$ (deep blue solid line). The \emph{black dot-dashed line} corresponds to the energy
distribution corresponding to the 2D Maxwell-J\"uttner with the temperature $T_d=0.45 m c^2$ expected from MHD jump conditions in the unmagnetized case. The \emph{black dashed line} shows the power law guide line $dN/d\gamma \propto \gamma^{-s}$ with $s=2.5$. 
The \emph{subpanel} presents the dependence on $\sigma$ of the particle number fraction (red stars) and energy fraction (blue diamonds) contained in the power-law part of the distribution function.}
\label{fig:fdist_sigmas}%
\end{figure*}

For the three top panels ($\sigma \le 10^{-3}$), a non-thermal power-law part develops at large times and its high-energy cut-off extends with time. 
The dot-dashed guide-line in panels (a-c) follows the $dN/d\gamma \propto \gamma^{-2.5}$ scaling at non-thermal energies. 
The power-law index is roughly the same than found in previous PIC studies \citep{2008ApJ...682L...5S,SSA13}, and it is slightly larger than the theoretically expected 
slope for ultra-relativistic shocks $s \simeq 2.2$ \citep[e.g.,][]{2000ApJ...542..235K, 2001MNRAS.328..393A, 2003ApJ...589L..73L, 2005PhRvL..94k1102K}. 

No supra-thermal component is present in the $\sigma = 1.2 \times 10^{-2}$ case (panel d) and 
the 2D Maxwell-J\"uttner distribution provides here a very good fit of the observed distribution.

We finally note that close to the transition magnetization $\sigma \sim 10^{-3}$ the acceleration is faster: 
the maximum cut-off energy is larger than for the two lower magnetization cases, and so is the overall number of particles in the high-energy tail. 
This larger acceleration rate may be attributed to the smaller spatial diffusion coefficient $D_{\perp}$ in the shock normal direction $x$ also perpendicular to $\mathbf{B_0}$.
This reduces the residence time $t_{\rm res} \propto D_{\perp}/c^2$ on both  sides of the shock.  
This situation is known to be beneficial for the Diffusive Shock Acceleration (DSA) mechanism as, in the Fermi picture, the acceleration time is governed by 
the scattering time of particles on both sides of the shock front. 
Another explanation for this faster acceleration may be attributed to non-Fermi acceleration, in particular to the Shock Drift Acceleration (SDA) as
some particles may gain energy in the convective upstream electric field $|\mathbf{E}_0|=\mathbf{\beta}_0 |\mathbf{B}_0|$. 
This point will be discussed further in the next section.

In Figure~\ref{fig:fdist_sigmas} we present the downstream distribution functions at the final (\smilei) simulation time $t\omega_{\rm pe}=2\times10^3$
for various magnetizations $\sigma \in [10^{-5},5]$, the lines color coding corresponding to different $\sigma$ as given in the colorbar. 
Red shaded lines correspond to weak magnetization shocks $\sigma \ll 10^{-3}$ and blue shaded lines correspond to the high magnetizations $\sigma \gg 10^{-3}$. 
The dashed guide-line follows the $dN/d\gamma \propto \gamma^{-2.5}$ scaling as in the previous figure.

While a smooth high-energy power-law tail is observed for $\sigma<10^{-3}$, its shape becomes irregular for $10^{-3}\leq \sigma < 10^{-2}$ 
and the tail disappears for larger magnetizations. 
The width of the thermal part of the distribution function narrows at high magnetizations $\sigma>0.1$ as required by the MHD jump conditions (see Appendix~\ref{appendix:jump}). 
The expected temperature, given by Eq.~\ref{eq:Tdownstream}, is $T_d = 4.5 m c^2$ at $\sigma=10^{-3}$, and decreases gradually to $\simeq 3.7 m c^2$ for $\sigma=1$. 

The figure insert depicts the dependence on $\sigma$ of the energy ratio $\xi_{\rm CR}$ in the non thermal tail  by the total energy (blue diamonds) 
and of the non thermal particle number fraction $n_{\rm CR}$ (red stars). Both are constant for $\sigma <10^{-3}$ with $\xi_{\rm CR} \simeq 8$\% and $n_{\rm CR}\simeq 1$\%. 
A rapid decrease occurs between $\sigma=2\times 10^{-3}$ and $\sigma=10^{-1}$, where the supra-thermal part disappears. 
An irregular behavior is observed for $\sigma=1$. 

\subsection{Maximal energy evolution and saturation}\label{sect:fdist:maxenergy}

It was shown by \citet{2012ApJ...755...68S} and by \citet{SSA13} that tracking the temporal evolution of the particle maximum energy in the simulation unveils the properties of the spatial scattering of the accelerated particles. 
In the Bohm regime the diffusion coefficient evolves as $D_\perp \propto \gamma$ and consequently the particle maximal energy evolves linearly with time, $\gamma_{\rm max} \propto t$. 
This follows from the acceleration time dependence in relativistic shocks as $t_{\rm acc} \sim D_{\perp} /c^2$, derived assuming that the particle residence time is dominated by the downstream residence time. 
In contrast, in the small-angle scattering regime $D_\perp \propto \gamma^2$ and the particle maximal energy scales as $\gamma_{\rm max} \propto \sqrt{t}$. 
This characteristic growth with the square-root of time was demonstrated in relativistic pair shocks by \citet{SSA13}.

\begin{figure}
\centering
\includegraphics[width=0.49\textwidth]{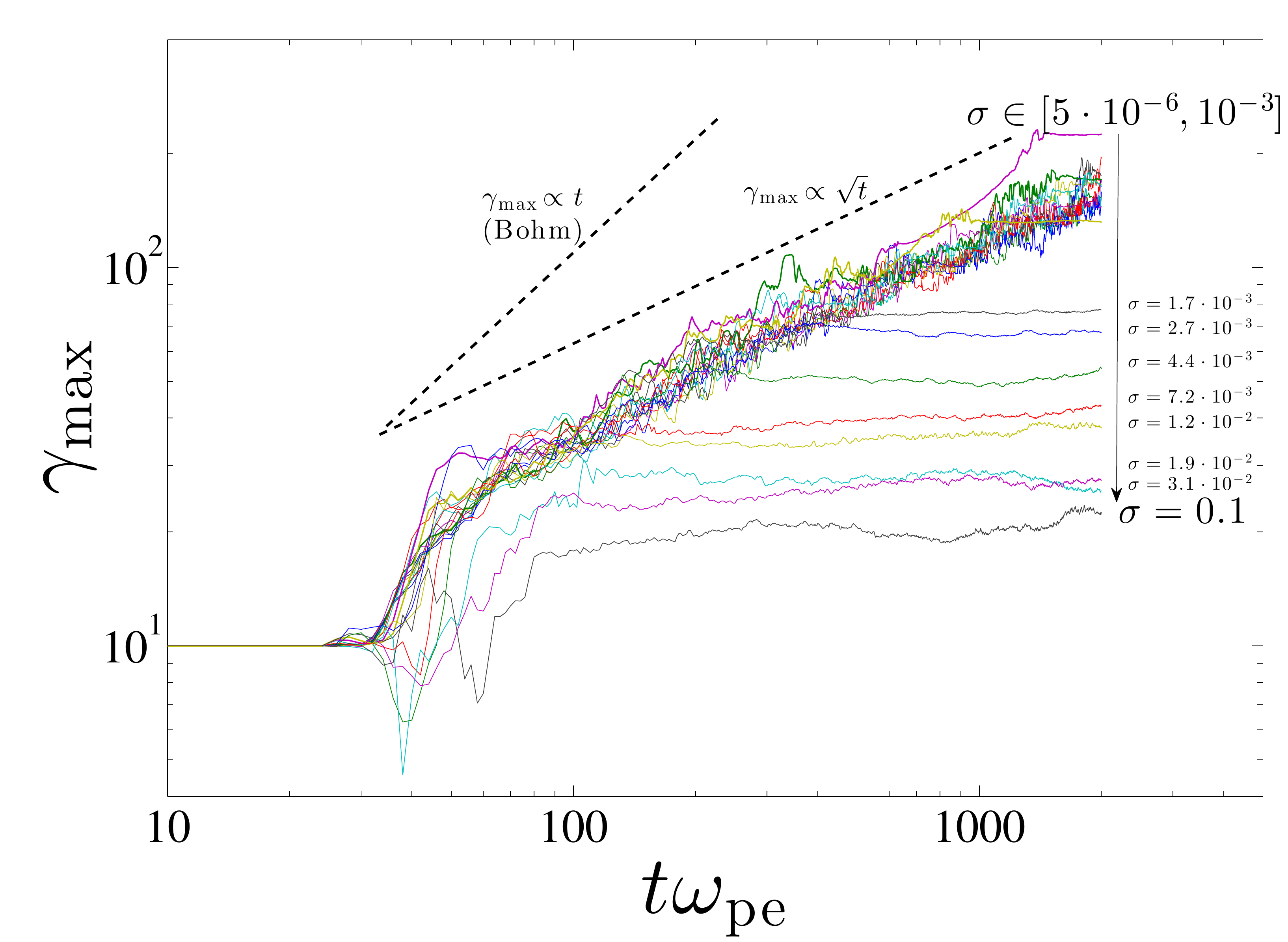}
\caption{Temporal evolution of the highest particle energy in different simulations with increasing magnetization (lines with different colors). The dashed line indicates the $\gamma_{\rm max} \propto \sqrt{t}$ scaling.}
\label{fig:gmax_timeseries}%
\end{figure}

Figure~\ref{fig:gmax_timeseries} presents the temporal evolution of the maximum Lorentz factor $\gamma_{\rm max}(t)$ of the particles in the simulations 
with magnetization $\sigma \in [5\times10^{-6},10^{-1}]$ (curves of different color). 
The dashed line represents the $\sqrt{t}$ scaling. After the initial shock-formation phase ($t\omega_{\rm pe}<100)$, 
the scaling $\gamma_{\rm max}\propto \sqrt{t}$ is satisfied in all simulations with $\sigma < 10^{-3}$ and at early times for $\sigma > 10^{-3}$, 
confirming the results of \citet{SSA13}.
This also provides an indirect confirmation of the small-scale nature of the self-generated turbulence and as a consequence of the small-angle scattering regime
\footnote{We note that in the Bohm regime the diffusion coefficient is $\propto \gamma$ and the maximal particle energy would evolve as $\gamma_{\rm max} \propto t$.
Hence, Bohm diffusion is ruled out here.}. 
For simulations with $\sigma \ge 10^{-3}$, the particle maximum energy is found to saturate after some time. Both the time at which saturation is reached, and the value of the particle energy at that time (saturation energy) are found to decrease as the upstream magnetization increases. Note that the final simulation time, $t \omega_{\rm pe}=2000$, allows to reach energy saturation only for magnetizations $\sigma > 10^{-3}$. 
Longer simulations would allow us to investigate the energy saturation for lower magnetizations.

\begin{figure}
\centering
\includegraphics[width=0.49\textwidth]{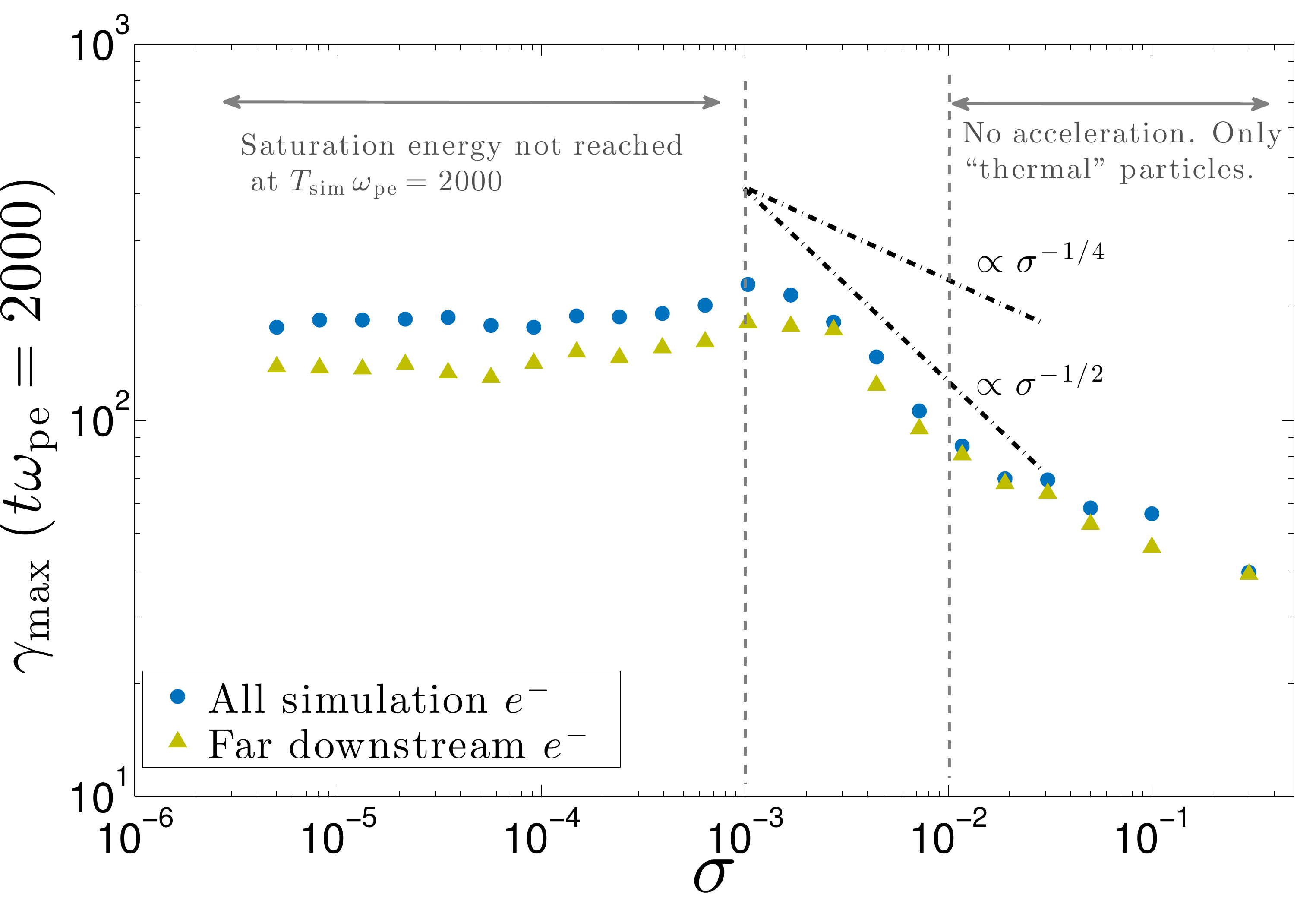}
\caption{Dependence on $\sigma$ of the maximum energy of particles at the final simulation time $t \omega_{\rm pe}=2000$ (\smilei code). Two different populations of particles are considered: (blue circles) all electrons in the simulation, (yellow triangles) only far downstream electrons. 
The \emph{dot-dashed lines} indicate the power-law scalings $\gamma_{\rm sat}\propto \sigma^{-1/4}$ and $\gamma_{\rm sat}\propto \sigma^{-1/2}$. }
\label{fig:gsat}%
\end{figure}

The particle maximal energy at the final simulation time $\gamma_{\rm max}(t \omega_{\rm pe}=2000)$ are presented in Figure~\ref{fig:gsat}
as a function of $\sigma$ considering either all electrons in the simulation (blue circles) or only electrons in the downstream far from the shock front $x-x_{\rm wall}=200 \delta_{\rm e}$ (yellow triangles). 

For low magnetizations ($\sigma<10^{-3}$), $\gamma_{\rm max}(t \omega_{\rm pe}=2000)$ has a constant value.
This is because the saturation time is not yet reached and particles are still being accelerated at the shock front.
As a result, in the far downstream region, electrons have lower energy respect to the ones close to the shock front, 
the latter are still being accelerated and need more time to escape and reach the far downstream. 

In the opposite case of high-magnetizations, for $\sigma>10^{-2}$, the saturation energy does not correspond to any acceleration process.
It simply reflects the high-energy cut-off of the thermal distribution.

More interesting is the case of intermediate magnetizations, $10^{-3}<\sigma<10^{-2}$, for which the particle maximal energy has reached saturation 
(as discussed earlier and shown in Fig.~\ref{fig:gmax_timeseries}).
The energy saturation level $\gamma_{\rm sat} \equiv \gamma_{\rm max}(t \omega_{\rm pe}=2000)$ is found to decrease with increasing $\sigma$. 
Note also that in this stationary regime, the saturation level is the same for both electron populations. 
The scaling seems to follow $\gamma_{\rm sat} \propto \sigma^{-1/2}$ with better confidence than the scaling $\gamma_{\rm sat} \propto \sigma^{-1/4}$ found by \citet{SSA13}. 
Yet, the latter can not be rejected here because the scaling is derived in a region ($10^{-3} < \sigma < 10^{-2}$) where the particle acceleration efficiency drops rapidly, 
while \citet{SSA13} investigated a wider range of magnetizations $10^{-4} < \sigma < 10^{-2}$ (their simulations being longer than the ones presented here).

 The saturation mechanism is an important point to discuss further. 
 However, this discussion is postponed to the last section of this article
 since it is required to derive the behavior of the spatial diffusion
 coefficient before. This latter is done in the section below.
 
%% file: 07_particle_kinematics.tex
\section{Particle kinematics}
\label{sect:track_particles}

Important insights into the particle acceleration physics can be provided by studying the particle trajectories. 
In this Section we study the statistical and individual behavior of the particles' dynamics. 
We select and follow test particles, injected into the simulation box at different positions and times, and with different initial Lorentz factors. 
These particles do not contribute to the current deposition on the grid, but they fully experience the time-dependent electromagnetic fields during the simulation. 
In each \smilei simulation, several populations of test-particles were injected: 
\begin{enumerate}
\item 8192 electrons located initially at $t\omega_{\rm pe} =0$ at the wall (right side of the simulation box) with the same  initial momentum and energy as the bulk particles, i.e. ${\bf p}_x=p_0 \vec{e}_x$ and $\gamma_i=\gamma_0$. These particles are initially located at the leading edge of the wall-reflected flow and participate in the shock formation and evolution.
\item 8192 positrons injected far upstream at $t\omega_{\rm pe} =0$ with the same distribution as the upstream drifting particles. These particles interact with a well-formed shock front.
\item 8192 electrons injected at $t\omega_{\rm pe} =500$ and at the right side of the simulation box with an initial momentum in $-x$ direction and Lorentz factor $\gamma_i=\gamma_0$. 
They experience the downstream decaying micro turbulence and do not interact with the shock front nor with the upstream medium. These particles are considered to study the spatial transport downstream.
\item Same population 3, but with an initial particle Lorentz factor $\gamma_i=10 \gamma_0$.  They are considered to study the particles spatial transport downstream at a different initial energy.
\item In four dedicated simulations with $\sigma=0,9\times 10^{-5}, 4\times10^{-4}$, and $10^{-3}$, same as population 3 but with initial particle Lorentz factors $\gamma_i=\{1.5, 2, 3, 5 \}\gamma_0$ in order to study the energy dependence of the spatial diffusion laws.
\end{enumerate}

\subsection{Spatial diffusion downstream}
It is a well-known result that the particle transport in a small-scale magnetic and isotropic turbulence is governed by small-angle scattering \citep{2009MNRAS.393..587P, Aloisio04}. 
We define the effective particle Larmor radius $R_L = \gamma m c^2 /(e \widebar{B})$, where $\widebar{B}=\sqrt{ B_0^2 + \langle \delta B^2 \rangle}$ is the total rms magnetic field strength. 
A particle crossing this magnetic turbulence is randomly deviated by a small angle $\delta \theta \sim \ell_c/R_L$ at each coherence length of the field $\ell_c$. 
It is common to define the parameter $\rho=R_L/\ell_c$ so that the small-angle scattering regime corresponds to $\rho>1$, generally satisfied for any supra-thermal particle in the downstream of the shock \citep[e.g.,][]{2010ApJ...710L..16K}. 

In the limit of vanishing $B_0$ (unmagnetized upstream plasma) it is straightforward to derive the isotropic spatial diffusion coefficient from order of magnitude considerations 
as $D = c R_L^2 /(3\ell_c) \propto \gamma^2$.

In the presence of an external (even weak such as $B_0^2 \ll \langle \delta B^2\rangle$) magnetic field, the spherical symmetry is broken and one needs to distinguish between the diffusion coefficients $D_{\parallel}$ in the direction parallel to the external magnetic field $\mathbf{B}_0$, and $D_{\perp}$ in the direction perpendicular to $\mathbf{B}_0$. 
A rigorous theory of the particle scattering under these conditions was presented and verified by Monte-Carlo simulations by \citet{2011A&A...532A..68P}. 
These authors found that the diffusion coefficient  $D_{\parallel}$ is the same as in the unmagnetized limit as far as $\rho>1$,
while the diffusion in the transverse direction is affected by the regular gyration imposed by the external field ${\bf B}_0$. 
It was found that the diffusion coefficient $D_{\perp}$ obeys a law similar to classical diffusion:
\begin{eqnarray}
\label{eq:dperp}
D_\perp &=& {\langle \Delta x^2 \rangle \over 2 \Delta t} = {c^2 \over 3} {\nu_{\rm s} \over \nu_{\rm s}^2 +\omega_{L,0}^2} ,
\end{eqnarray}
where $\omega_{L,0} = e B_{0,{\rm d}} /(\gamma_i m c)$ is the relativistic Larmor frequency of a particle in the homogeneous downstream magnetic field $B_{0,{\rm d}}$,
and $\nu_{\rm s} = e^2 \langle \delta B^2\rangle \ell_c/(\gamma^2 m^2 c^3)$ is the pitch-angle scattering frequency governed by the turbulent component. 
If $B_0^2 \ll \langle \delta B^2\rangle$ and the particle energy is not very high, $\nu_{\rm s} >\omega_{L,0}$ and the mean field has no influence on the particle scattering. 
As a result, the scaling $D_\perp \propto \gamma^2$ holds. 
When the particle energy increases up to the value for which $\rho = \sqrt{\langle \delta B^2\rangle / B_0^2}$, the regular gyration produces the same angular deviation than the random scattering, 
since $\nu_{\rm s}$ decreases faster then $\omega_{L,0}$. 
Consequently, the transverse diffusion coefficient saturates when $\omega_{L,0} \gg \nu_{\rm s}$ to a value $D_\perp \simeq c^2\nu_{\rm s}/ (3\omega_{L,0}^2)$. 

This saturation effect has been previously observed in Monte-Carlo simulations but it was never evidenced in self-consistent PIC simulations of relativistic shocks, where the presence of an approximately isotropic magnetic micro-turbulence is known to develop. 
Here, we directly derive the diffusion coefficient in the shock normal direction (perpendicular to the mean field $\mathbf{B_0}$) by following particle trajectories in the downstream medium.

\begin{figure}
\centering
\includegraphics[width=0.49\textwidth]{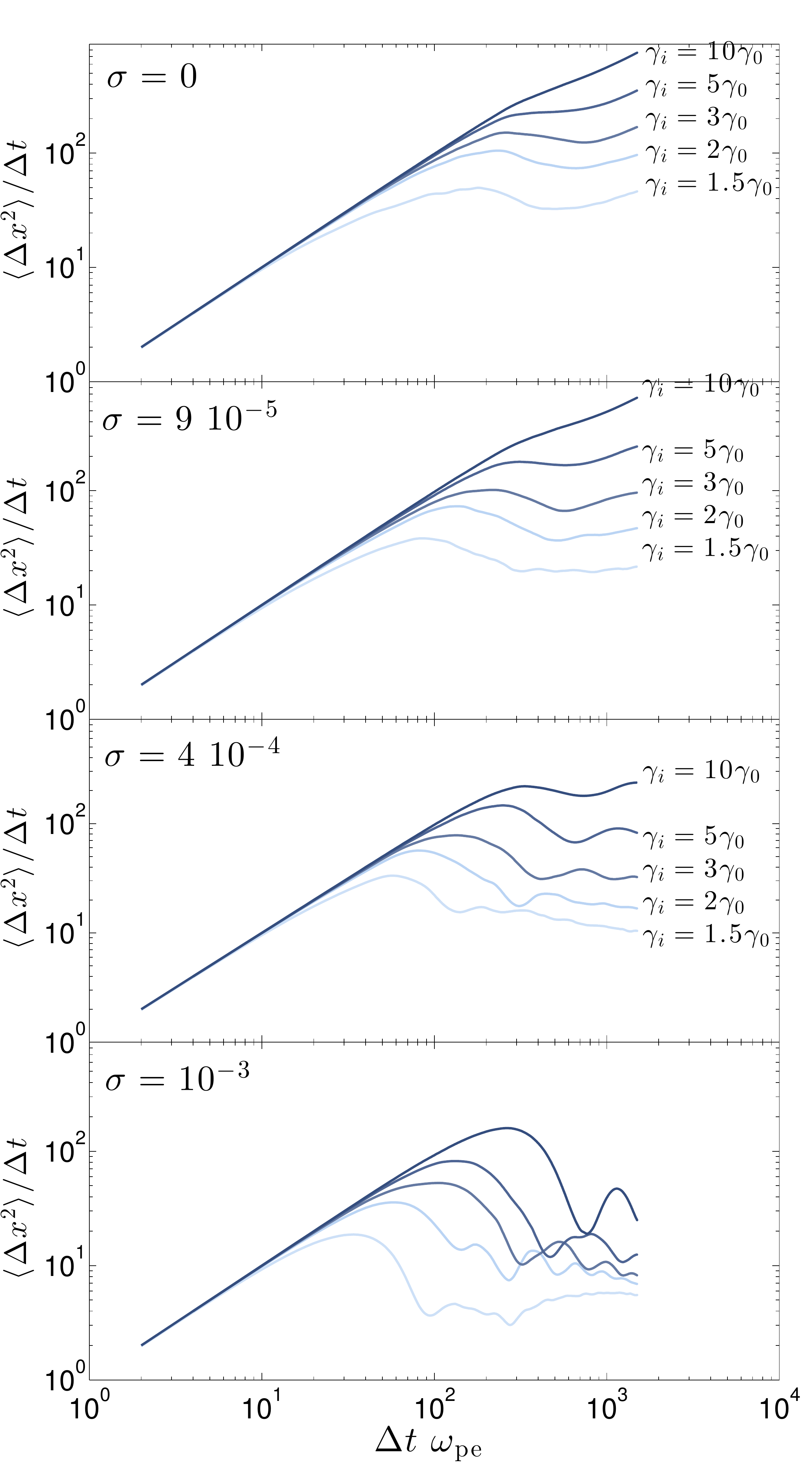}
\caption{Running diffusion coefficients $\langle \Delta x^2 \rangle /\Delta t$, in units of $c \delta_e$, for five initial particle energies ($\gamma_i/\gamma_0=\{1.5,2,3,5,10\}$) plotted using different shades of blue and for four magnetizations, from top to bottom: $\sigma=0$, $\sigma=9\times  10^{-5}$, $\sigma=4\times 10^{-4}$, and $\sigma= 10^{-3}$.}
\label{fig:dxx_time}%
\end{figure}

In Figure~\ref{fig:dxx_time} we present the evolution in time of the coefficient $D_\perp \equiv {\langle \Delta x^2 \rangle / 2\Delta t}$, 
where $\Delta x = x(t_0+\Delta t)-x(t_0)$ measures at a given time $t_0+\Delta t$ 
the distance travelled by a particle along the $x$ direction since its time of injection $t_0=500\omega_{\rm pe^{-1}}$, 
and angular brackets $\langle \rangle$ denote the average over all particles (populations 4 and 5 considered here).
We report the results of simulations performed with four representative magnetization $\sigma=0$, $9\times 10^{-5}$, $4\times 10^{-4}$, and $10^{-3}$, from top to bottom respectively. The five initial energies $\gamma_i/\gamma_0=\{1.5, 2, 3, 5, 10\}$ corresponds to lines with different shades of blue.

At early times, the behavior is ballistic: $\langle \Delta x^2 \rangle \propto \Delta t^2$. 
Particles typically follow a straight path, at constant velocity. 
When the diffusive process occurs ${\langle \Delta x^2 \rangle / \Delta t}$ reaches a constant value, i.e. $\Delta t$ becomes larger then the scattering time $\tau_{\rm s}=1/\nu_{\rm s}$. 
This is observed at large times in Fig.~\ref{fig:dxx_time}. 
Moreover, since the scattering time increases with the particle initial energy, so does the value of the plateau. 
In addition, the final value of $D_\perp$, decreases with increasing $\sigma$. 
Note that, in weakly magnetized cases $\sigma \ll 10^{-4}$, the diffusion does not reach an asymptotic value. 
This suggests that there may be a second saturation time-scale that is not reached at the end of these simulations\footnote{
Let us note that the long-term behavior is hard to constrain because (i) simulations are obviously not long enough, (ii) the downstream magnetic turbulence is decaying in time by phase mixing. This implies that after some time (and at large distance from the shock front) the level of turbulence decreases to the level of the thermal `noise'. PIC codes are known to produce a level of thermal fluctuations much larger than expected in realistic systems. This latter effect may strongly impact the particle scattering properties at later times. For this reason, we cannot clearly identify a different scattering time scale or change of particle scattering from micro turbulence to the thermal noise dominated regimes.}.

\begin{figure}
\centering
\includegraphics[width=0.49\textwidth]{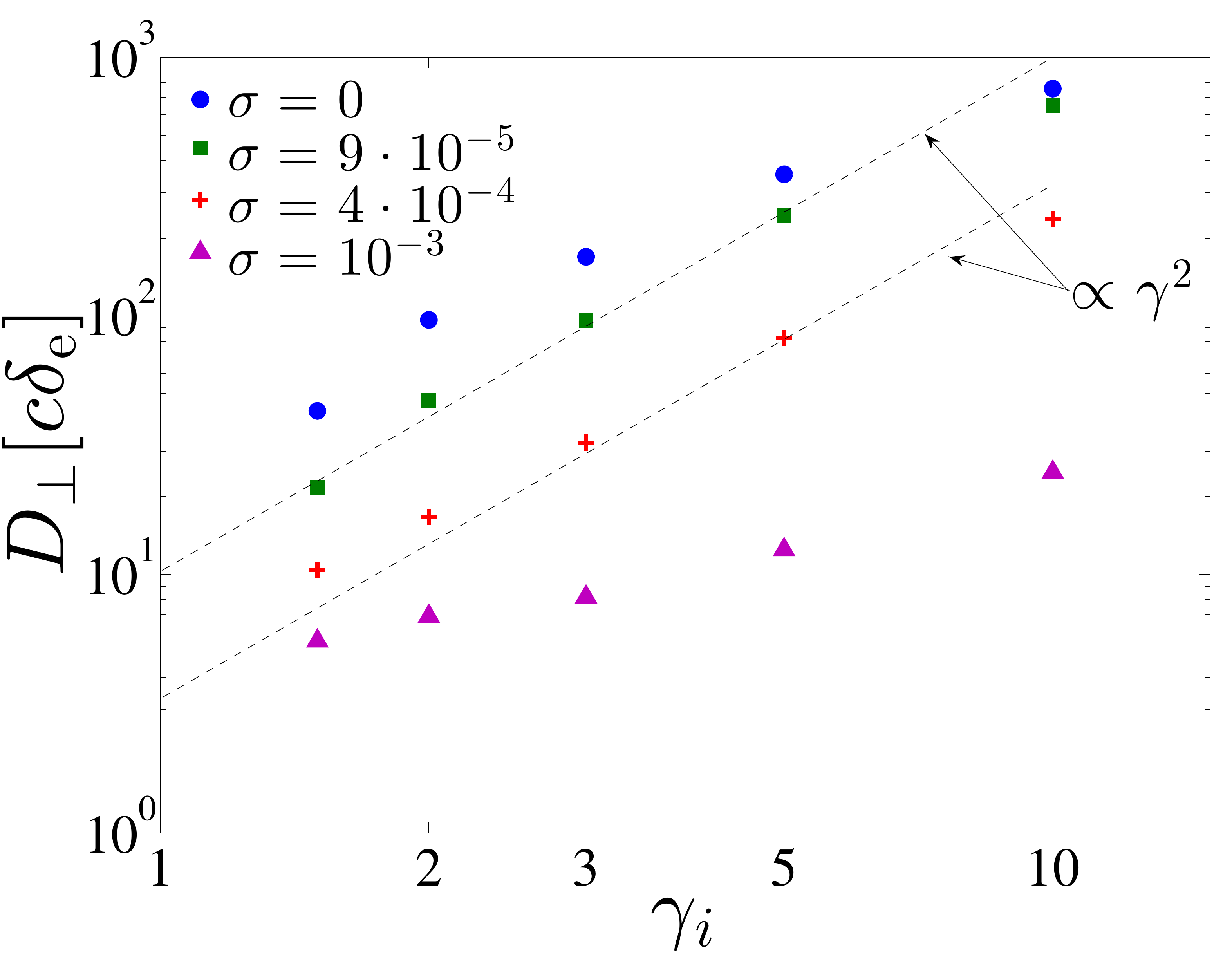}
\caption{Final values of the running diffusion coefficient $D_\perp$ as function of the initial particle energy $\gamma_i$ for the same magnetizations than in Figure~\ref{fig:dxx_time}. The \emph{dashed lines} show the $D_\perp \propto \gamma^{2}$ scaling.}
\label{fig:dxx_energies}%
\end{figure}

Figure~\ref{fig:dxx_energies} presents the final values of $D_\perp \equiv {\langle \Delta x^2 \rangle / 2\Delta t}$ as a function of the particle energy $\gamma_i$ for the four values of $\sigma$ introduced above. 
The dashed guide-lines indicate the $D_\perp \propto \gamma_i^2$ scaling. 
This scaling describes well the behavior of weakly magnetized shocks $\sigma \ll 10^{-3}$, while $D_\perp$ is almost constant for $\sigma=10^{-3}$. 
This is a natural consequence of the increasing strength of $B_0$, when it becomes comparable to $\delta B$. 
In this case $D_\perp$ enters directly in the regime $\omega_{L,0} > \nu_{\rm s}$ where its value is roughly constant. 
These findings confirm the theoretical prediction for the transverse spatial diffusion law given by Eq.~\ref{eq:dperp}.

\begin{figure*}
\centering
\includegraphics[width=0.75\textwidth]{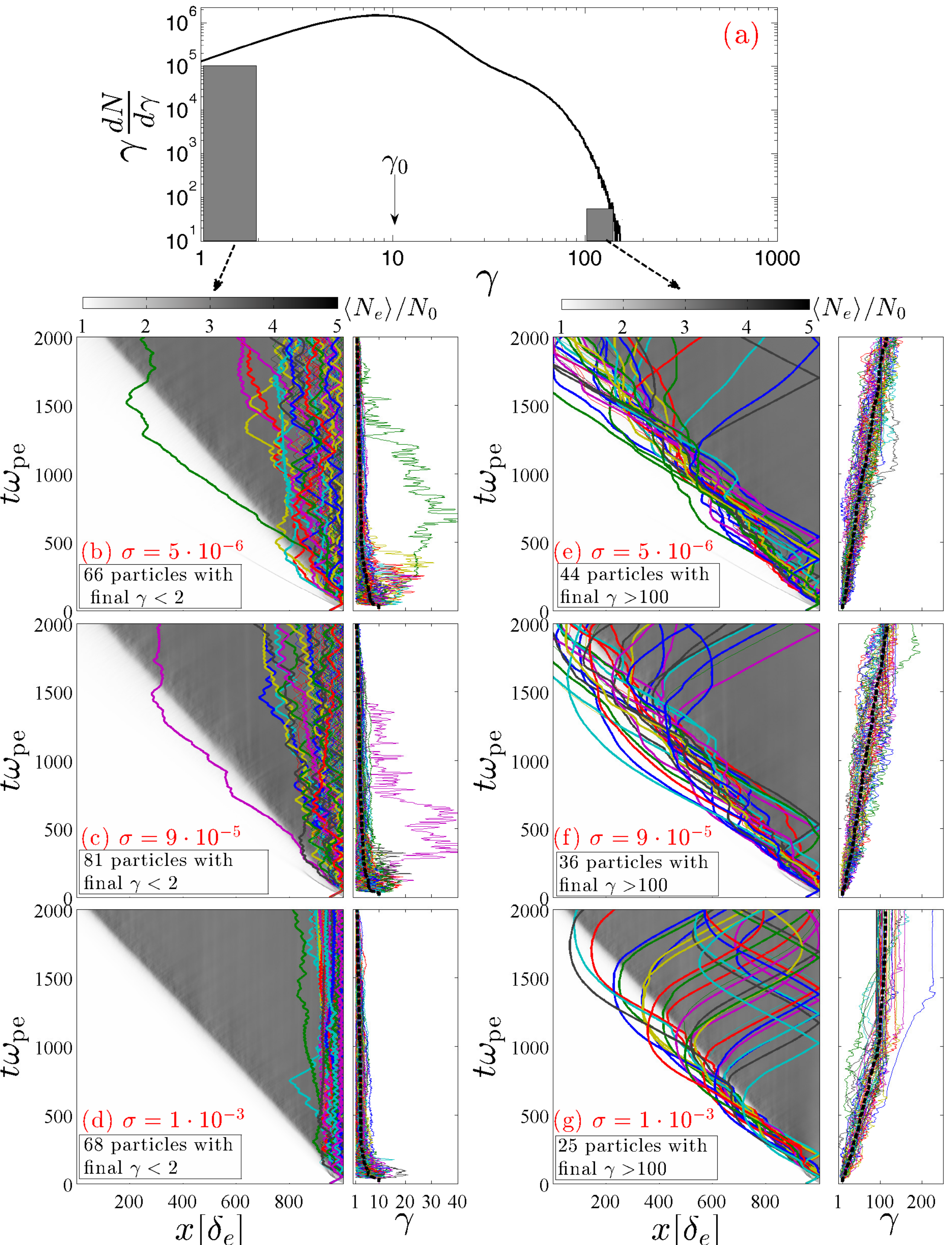}
\caption{Trajectories and temporal evolution of the energy of particles selected from different parts of the distribution function. 
(a) Typical electron (or positron) energy spectrum downstream. The \emph{grey shaded rectangles} illustrate the selections (at low and high energy) 
on the particles whose trajectories and energy evolution are plotted in the lower panels.  
\emph{Left column} (panels b,c,d): low energy particles with final Lorentz factor $\gamma<2$.
\emph{Right column} (panels e,f,g): highest energy particles. 
In each column, the \emph{main set} presents the trajectories of different particles (shown with different colors).
The electron density averaged along the transverse $y$-direction is plotted in gray scale in order to highlight the position of the shock front with time.
The grey and white regions corresponds to the downstream and upstream regions, respectively.
The \emph{subsets} show the temporal evolution of the selected particle energy (Lorentz factor). The \emph{thick dashed black lines} show the temporal evolution of the mean energy $\langle \gamma (t)\rangle$.
Three representative magnetizations are reported: $\sigma=5\times 10^{-6}$ in panels (b) and (e), $\sigma=9\times10^{-5}$ in panels (c) and (f), $\sigma=1\times10^{-3}$ in panels (d) and (g).}
\label{fig:dens_stack_traj}%
\end{figure*}

\subsection{Accelerated particles}
\label{sect:track_particles_accel}

We now look at the dynamics of the particles accelerated at the shock front, as well as of the particles that lye in the low energy part of the downstream distribution function at the end of the simulation. 
We consider population 1 that follows the shock front during its formation and evolution. 
A large part of the considered particles contribute to the thermal distribution of the downstream medium. 
A small fraction of them however can reach energies above 10$\gamma_0$, if the initial magnetization allows for particle acceleration. 

Figure~\ref{fig:dens_stack_traj}(a) shows the typical electron downstream spectrum for $\sigma \leq 10^{-3}$. 
It is composed of a 2D Maxwell-Jüttner contribution for energies $\gamma<5\gamma_0$ and of a power-low component at larger energies. 
We illustrate the trajectories of the low-energy particles having final $\gamma<2$ (left column panels b,c,d) 
and of the high-energy particles having final Lorentz factor $\gamma>10\gamma_0$ (right column panels e,f,g), 
for three representative magnetizations $\sigma=5\times 10^{-6}$, $\sigma=9 \times 10^{-5}$, and $\sigma=10^{-3}$.
Each line represents the trajectory $x(t)$ of a single particle. 
In the background we highlight the evolution in time of the transverse average electron density. 
The white region corresponds to the upstream while grey region to the downstream. 
The shock front position is located at the interface between the two regions and propagates from right to left. 
In panels (b-g), the accompanying right-side plots present the energy $\gamma(t)$ temporal evolution for each particle (using the same color-coding than in the main panels, b-g).
The energy averaged over all the considered particles is presented as a thick dashed line. 

Let us now have a closer look at the low-energy particles (left panels).
Except for one particle in the (b-c) panels, all low-energy particles remain downstream after the shock is formed. 
Their spread in the $x$-direction is reduced with increasing $\sigma$. 
This reflects the decrease of the spatial diffusion coefficient $D_{\perp}$ when $\sigma$ increases (see Figure~\ref{fig:dxx_energies}). 
The average energy decreases from the initial value $\gamma=\gamma_0$ to $\langle \gamma \rangle \simeq 2$ at end of the simulation.

Some rare particles that initially escape upstream of the shock gain a non-negligible amount of energy, but lose it when are advected back downstream. 
This energy decrease is due to an unfavorable shock-crossing condition that occur when a particle transfer part of its energy to the wave turbulence in the precursor before reaching the downstream. 
Those are the "unlucky" particles that are unable to perform more than one Fermi-cycle. 
Indeed, the Fermi mechanism involves a statistical energy gain and does not exclude an energy lost during a cycle for a single particle.

The picture is different for the high-energy particles with final $\gamma > 10\gamma_0$ as highlighted by the right panels (e,f,g) of Figure~\ref{fig:dens_stack_traj}. 
These particles remain close to the shock front and keep gaining energy, as shown in the right-side $\gamma (t)$ panels. 
The stochastic nature of the particles' trajectories is clearly observed in panel (e) for the case of low magnetization ($\sigma=5 \times 10^{-6}$).
For larger magnetizations, the kinematic importance of the ordered magnetic field is no more negligible, and leads to more ordered trajectories, as visible in panel (g) 
for the mildly magnetized case ($\sigma=10^{-3}$).

This confirms that the standard first-order Fermi mechanism operates in the unmagnetized and weakly magnetized ($\sigma \ll 10^{-3}$) relativistic $e^{\pm}$ shocks, 
consistent with the results of \citep{2008ApJ...682L...5S, 2009ApJ...695L.189M} and \citep{SSA13}.
In addition, for $\sigma=10^{-3}$ (panel f) we observe a regular gain in energy when particles are in the upstream region. 
This points towards a non negligible contribution of the Shock Drift Acceleration process. 
Also this acceleration seems to be limited to not very high maximal energy here $\gamma_{\rm max} \sim 10\gamma_0$ and the average particle energy saturates after $t\omega_{\rm pe}=1000$ for this population of particles.

\begin{figure}
\centering
\includegraphics[width=0.49\textwidth]{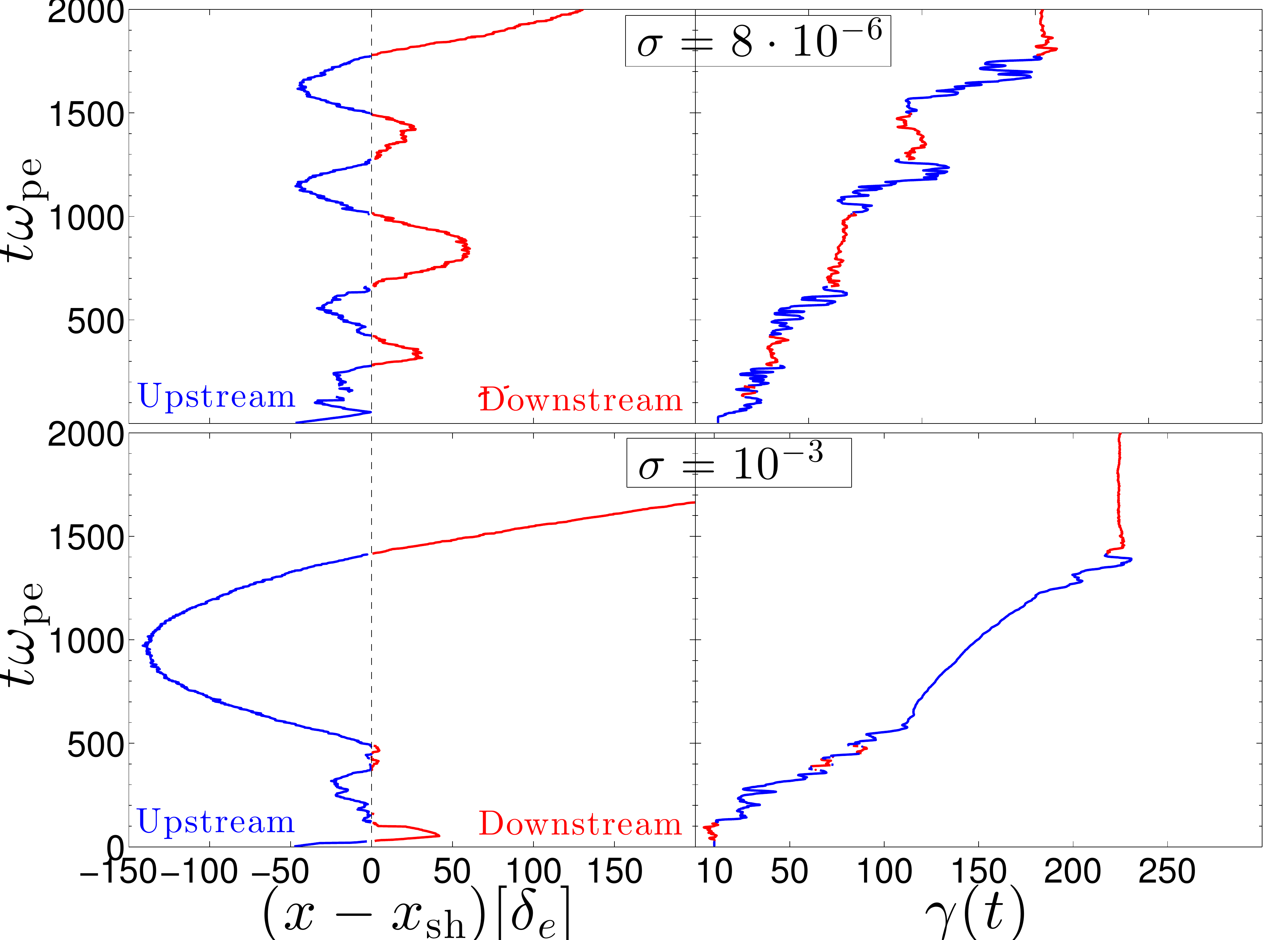}
\caption{Trajectories and temporal evolution of the energy of the highest energy particle obtained in the simulations: for $\sigma=8\times 10^{-6}$ (upper panels) and $\sigma=1\times 10^{-3}$ (lower panels). The \emph{left column} shows the particle distance to the shock front  $x(t)-x_{\rm sh}(t)$ versus time. The \emph{right column} shows the corresponding particle Lorentz factor versus time. The \emph{blue color} corresponds to times when the particle is in the upstream region, and the \emph{red color} to times when the particle is in the downstream region.}
\label{fig:2traj}%
\end{figure}

We further highlight the differences between the two acceleration mechanisms in Figure~\ref{fig:2traj} for $\sigma=5\times 10^{-6}$ (top panels) and $\sigma=10^{-3}$ (bottom panels). The particle with the highest final energy was selected from each simulation. We plot the time evolution of its position $x(t)-x_{\rm sh}(t)$ (left side panels) and of its energy $\gamma(t)$ (right side panels).

In the low-magnetization case the particle scatters on both sides of the shock, gaining energy through each downstream$\to$ upstream $\to$ downstream cycle. 
The energy gain per cycle is $\Delta\gamma \simeq \gamma$ as expected for a typical Fermi process in relativistic shocks \citep{1999MNRAS.305L...6G}, 
and the upstream and downstream residence times are of the same order. 

In the $\sigma= 10^{-3}$ case, we observe the same behavior at early times ($t\omega_{\rm pe}<500$), 
when the particle energy increases from $\gamma=10$ to $\gamma \simeq 90$ through the Fermi-like process. 
For larger times $500<t\omega_{\rm pe}<1500$, the particle located in the upstream sees its energy increasing nearly linearly with time. 
A linear fit gives $\Delta \gamma \simeq 0.14 \Delta t \omega_{\rm pe}$ in good agreement with the energy gain due to the upstream electric field: $e E_{y,0}/( mc \omega_{\rm pe}) \sim 0.143$, 
as expected in the SDA process. 
Once the particle is advected downstream, it can not be scattered back toward the upstream since the self-generated turbulence is not strong enough to turn it back toward the shock front.
This demonstrates a significant contribution from the SDA process in mildly magnetized relativistic shocks, 
similarly to what was found by \citet{2009ApJ...698.1523S} for higher magnetization ($\sigma=0.1$) oblique shocks. 
Our simulations show that, in perpendicular mildly magnetized shocks, SDA can rapidly accelerate electrons up to energies larger than 20 times the typical downstream thermal energy.
This corresponds to electron Lorentz factors above $20 \gamma_0$ in the downstream rest frame (and to $\gamma^{\vert u}_{\rm max} \simeq 20\gamma_0^2=20 \gamma^{\vert u \, 2}_{sh} (2/\Gamma_{\rm ad}-1)$ in the upstream rest frame, where $\Gamma_{\rm ad}$ is the adiabatic index of the downstream) for $\sigma=10^{-3}$. Here we assumed that the process efficiency scales linearly with $\gamma_0$ in the limit $\sigma \ll 1$ and $\gamma_0 \gg 1$.

%% file: 08_discussion.tex
\section{Discussion and conclusions} \label{sect:discuss}

\subsection{Numerical considerations}
We start this discussion with some considerations on the numerical limitations of the present study, 
commenting first on the possible differences in between 2D and 3D simulations, 
second on the need of long term simulations in particular when considering weakly magnetized shocks, 
and finally on the differences between FDTD and spectral codes.

We demonstrate that, even within a 2D configuration, the dominant physical processes behind the shock formation and evolution are captured. 
Nevertheless some differences may appear when considering 3D simulations. 

First, the density jump at the shock would be different since for a relativistic downstream plasma the adiabatic index in 3D is $\Gamma_{\rm ad}= 4/3$ instead of $\Gamma_{\rm ad}= 3/2$ in 2D. Hence, a stronger compression is expected in 3D $N_d/N_0 \simeq 4$ in the limit $\sigma \ll 1$ while in 2D $N_d/N_0 \simeq 3$. As a consequence, the shock front velocity would be $\upsilon_{\rm sh} \simeq 0.3 c$ instead of $\upsilon_{\rm sh}\simeq 0.45 c$ obtained in this study.

More important, the 3D geometry opens a new degree of freedom for plasma fluctuations in the direction of the external field $B_0$ 
(in 2D the $k_z$ wavevector is 0 by definition). 
This could have important consequences on the hierarchy of plasma instabilities in the mildly magnetized regime, while no fundamental variations are expected in the weakly and strongly magnetized limit, see \citet{SSA13} and \citet{2014MNRAS.440.1365L} .
A hint for this is given in Fig.~5 of \citet{SSA13} where 3D simulations for $\sigma=0$ and $\sigma=10^{-3}$ (and $\gamma_0=15$) were presented. 
Clear variations along $z$-direction are seen that implies an important contribution from the $k_z$ component when linear stability analysis of the precursor is done. 
\citet{2014MNRAS.440.1365L} found from a linear instability analysis that in some cases the contribution from this component may be as important as from the$k_x$ and $k_y$ related modes.
Therefore, accounting for the third dimension may change somewhat the value of $\sigma_{\rm crit}$ at which occurs the transition between Weibel-mediated shocks and magnetic reflection-shaped shocks.

Another point to investigate further concerns the long-term evolution of the shock. The simulations presented here could not exceed $t \omega_{\rm pe}=2000$. Imposed by both limited computer resources and the efficiency of the Cherenkov mitigation techniques, this time limit was the main limiting factor of the present study: the saturation energy is not reached for $\sigma < 10^{-3}$ and the diffusion coefficients are not fully stabilized by the end of the simulations. The implementation of alternative filtering techniques in SMILEI are currently under study which may allow to run longer simulations and thus strengthen our findings.

Running long-term simulations will also be of particular importance to reach a better understanding of the behavior of the weak magnetization shocks $\sigma \ll 10^{-4}$. 
The main open questions concern the asymptotic shock structure, such as the length of the particle precursor,  the turbulence properties, and the maximal particle energy. 
This is numerically challenging since it requires long simulation times, beyond $10^5 \omega_{\rm pe}^{-1}$, while up to now the longest simulations in the unmagnetized and weakly magnetized regimes \citep{2008ApJ...682L...5S, 2009ApJ...693L.127K, SSA13} reach $\simeq 8\times 10^4\omega_{\rm pe}^{-1}$. 
Even these time scales seem to be not enough to bring a definitive answer regarding inverse cascades (generation of large wavelength modes) in the particle precursor or to quantitatively investigate the differences between purely Weibel-mediated shocks $\sigma < \xi_{\tiny CR}/\gamma_{\rm sh}^{|\, u~ 2}$ and shocks in the mildly magnetized regime ($\sigma \sim 10^{-4}$). 
As previously stated, the shock steady state is unreachable with  current state-of-art PIC simulations for two main reasons: the available computational resources and the limit imposed by the spurious numerical Cherenkov instability at large times.

The filtering solution employed to deal with the Cherenkov instability is different in the \smilei (FDTD) and \shockapic (spectral) codes. 
Since it entails the cleaning (or damping) of some wavelengths in the light-wave branch, it is unavoidable to impact small-scale physical effects that involve particle-wave interaction on this branch. 
An example is provided in Appendix~\ref{appendix:comparison_codes}. 
By comparing directly the results from \smilei and \shockapic simulations, we show that the temporal Friedman filter used in \smilei damps the electromagnetic precursor waves emitted from the shock front in the strongly magnetized regime. 
This has no important effect on the shock dynamics or on the particle acceleration, but it should be taken into account for a detailed analysis of the precursor wave properties (such as in \citet{1992ApJ...391...73G}).
Concerning the spectral code, we adopted the solution that `cleans' large $k_x$ modes. This suppresses efficiently the numerical instability, but also any physical processes at wavelengths $\lambda<\delta_e$. This does not affect the processes involved in the relativistic shock dynamics, but it could be important in other systems.

Finally, concerning the spectral codes, one could be concerned with causality issue as they rely on a large numerical stencil. For the study at hand, however, this issue seems not to be important as both the FDTD and spectral codes give very similar results. The spectral method however requires global communications over the whole simulation box, which makes its optimization quite challenging for massively parallel, high-performance computing (HPC).
We would like to point out that some HPC-relevant strategies for spectral methods start to emerge in the literature \citep[e.g.,][]{2013JCoPh.243..260V}.

\subsection{Maximal energy of particles}

Much attention has been paid in this work to the study of particle acceleration and transport in the downstream turbulence.
Our simulations have shown (i) that particle gain energy when located in the upstream region, and (ii) that, overall, the particle maximal energy increases with the square-root of time, supporting the idea that particle acceleration proceeds in the small-angle scattering regime. 
The particle maximal energy will thus be limited by the particle ability to return from the downstream, back into the upstream. 

For (weak yet finite\footnote{We here consider planar shocks and will thus not appeal to any geometrical considerations, e.g. shock front corrugation, that may further limit the maximal particle energy.}) magnetizations in the range $10^{-4} < \sigma \lesssim 10^{-3}$, \citet{SSA13} found a scaling $\gamma_{\rm max} \propto \sigma^{-1/4}$.
In that case indeed, the particle diffusion length in the downstream turbulence scales with the square of the particle energy $\lambda_{\perp} = D_{\perp}/c \propto \gamma^2$, and the main limitation on the particle maximal energy is given by the width of the downstream turbulence $L_{B,{\rm sat}}$. While the width of the downstream turbulence is still a matter of debate \citep{2008ApJ...674..378C, 2009ApJ...693L.127K, 2015JPlPh..81a4501L}, \citet{SSA13} estimate $L_{B,{\rm sat}}$ to be of the order of the precursor (upstream) length $L_p$, consistent with simulations. These authors also reported that, for such magnetizations, the precursor length is limited by the injected particle Larmor radius in the upstream magnetic field, yielding $L_{B,{\rm sat}} \sim L_p \propto \eta_{\rm inj}/\sqrt{\sigma}$, where $\eta_{\rm inj}=\gamma_{\rm inj}/\gamma_0$ is found not to depend on the upstream magnetization.
The maximal energy of the particle will keep increasing as long as the diffusion length in the downstream turbulence remains smaller than this characteristic width of the downstream turbulence, that is as long as $\lambda_{\perp} < L_{B,{\rm sat}}$. Equating the two lengths thus provides the scaling for the particle maximal energy $\gamma_{\rm max} \propto \sigma^{-1/4}$.

The simulations presented in this study were however not long enough to extract the maximal particle energy for such weak magnetizations.
Nevertheless, we found that, for intermediate magnetizations $10^{-3} < \sigma < 10^{-2}$, the maximal particle energy
scales with $\sigma^{-1/2}$ with seemingly better confidence than the $\sigma^{-1/4}$ scaling observed by \citet{SSA13} (even though it spans over a single decade). 
This can be explained by the fact that, for these intermediate magnetizations, the role of the ordered magnetic field on the particle diffusion in the downstream is important.
In this case particles can gain energy as long as their scattering frequency in the downstream turbulence $\nu_s \propto 1/\gamma^2$ 
remains larger than their relativistic gyration frequency in the regular field $\omega_{L,{\rm d}}=e B_{0,{\rm d}}/(\gamma m c)$, that is as long as $\nu_s > \omega_{L,{\rm d}}$ \citep[see, e.g.,][]{2009MNRAS.393..587P, 2010MNRAS.402..321L}.  As soon as this inequality is not satisfied, the particle would effectively freeze within the downstream fluid and see the shock front receding at a speed $v_{\rm sh} \simeq c/2$ (in 2D). Equating the two frequencies then provides the scaling for the particle maximal energy $\gamma_{\rm max} \propto \sigma^{-1/2}$. 

We finally stress that for very weak magnetizations, such as encountered in the ISM ($\sigma < 10^{-8}$), the question of the maximal particle energy dependence on the upstream magnetization remains open.
While one could expect the scaling $\propto \sigma^{-1/4}$ obtained by \citet{SSA13} at $10^{-4} < \sigma \lesssim 10^{-3}$, the $\sigma^{-1/2}$-scaling cannot be ruled out~\citep[e.g.][]{2010MNRAS.402..321L} and has been used in several studies \citep[e.g., ][]{2014MNRAS.439.2050R, araudo15}.
 
\subsection{Astrophysical implications}

We saw that the shock structure changes gradually between $\sigma=10^{-3}$ and $\sigma=10^{-2}$ from Weibel-mediated to magnetic reflection-shaped shocks. The critical magnetization at which this transition occurs seems to be independent from the shock Lorentz factor in the ultra-relativistic limit, see \citet{SSA13}. 
However, the range of magnetization for which the Weibel-filamentation instability can fully develop in the shock precursor is $\sigma \leq 10^{-4}$.
This discrepancy can be explained, as proposed by \citet{2014MNRAS.440.1365L, 2014EL....10655001L}, by the instability triggered by the perpendicular current in the precursor in mildly magnetized shocks. 
The effect of this instability modifies the motion of the center-of-mass in the shock precursor. It also explains how Fermi-like acceleration processes can be sustained up to $\sigma \sim \xi_{\tiny CR}^2 \simeq 10^{-2}$, assuming $\xi_{\tiny CR}\sim 10^{-1}$. 
In the present study, we demonstrated that the beam of returning/accelerated particles carries a measurable perpendicular current and that it gives its maximum contribution for $\sigma \sim 10^{-3}$ (Sect~\ref{subsect:perp_current}). We also observed that particle acceleration efficiency drops rapidly between $\sigma=10^{-3}$ and $\sigma=10^{-2}$ (Sect.~\ref{sect:fdist}) in good agreement with the scaling provided by \citet{2014EL....10655001L}.

As we have shown, weakly magnetized shocks are efficient particle accelerators.
However an isolated ultra-relativistic shock can not accelerate electrons above $\gamma=10^6$ due to the synchrotron radiation limit \citep{2010ApJ...710L..16K, SSA13} and protons above $\gamma=10^8$ due to the pre-shock magnetization in the ISM and to the particle scattering properties \citep{1999MNRAS.305L...6G, 2010MNRAS.402..321L, 2014MNRAS.439.2050R}. 
It is required to consider some alternatives, such as supplementary downstream turbulence on large scales in order to unlock the phase space and decrease the scattering timescale, or multi-shock front interaction such as in internal shocks of AGN or GRB jets.
A possible alternative is the injection of a pre-existent energetic particle population in the shock, so that it can boost the particles energy by $\gamma_{\rm sh}^2$ factor in one Fermi cycle, as seen in the upstream rest frame \citep[e.g.,][]{1999MNRAS.305L...6G, 2015ApJ...811L..38C}.

In the case of the termination shock in the Pulsar Wind Nebulae (PWNe) the magnetization is always lager than $\sigma =10^{-2}$ and the shock is perpendicular, leading to an inefficient particle acceleration. 
Instead, one expects an intense semi-coherent electromagnetic waves emission towards the internal part of the pulsar magnetosphere through the Maser Synchrotron Instability at the shock front. Our simulations show that this emission occurs for $\sigma\gtrsim 10^{-3}$ and its intensity peaks around $\sigma \sim 0.1$. 
For larger magnetizations the fraction of the incoming energy channeled to the MSI decreases rapidly. A detailed investigation of the MSI emitted waves by relativistic shocks will be presented in a dedicated study  \citep[in preparation; see also][]{2017ApJ...840...52I}.

A particular region of the PWNe is the equatorial plane (for an aligned rotator) where the presence of the current sheet decreases the local magnetization to very low values as result of efficient channeling of magnetic energy into kinetic energy of the background plasma \citep{2015MNRAS.448..606C, 2014ApJ...780....3U}. Such regions are known to provide enhanced levels of wave turbulence and increased background plasma density, similarly to the Heliospheric Current Sheet where in-situ measurements are available. Yet, particles in this region are already significantly energized through magnetic reconnection and the acceleration by the terminal shock should account for different upstream conditions than assumed in the present study.

\subsection{Summary and conclusion}
The present study presents a systematic description of magnetized relativistic pair shocks by means of 2D Particle-In-Cell simulations. 
It provides a discussion of the shock structure, jump conditions, formation timescales, particle acceleration efficiency and mechanisms as well as particle transport properties from weakly magnetized $\sigma =10^{-5}$ to highly magnetized $\sigma \geq 1$ shocks. 

It was found, in agreement with previous studies, that the shock structure evolves from Weibel-dominated to magnetic-reflection-shaped shocks. 
The transition occurs between $\sigma=10^{-3}$ and $\sigma=10^{-2}$ beyond which the particle populated precursor disappears leaving place to a large amplitude electromagnetic wave emission from the shock front through Maser Synchrotron Instability, in agreement with \citet{SSA13}. 
The  shock jump conditions are derived systematically and found to be in good agreement with the ideal MHD prediction except in the range $10^{-4} < \sigma <10^{-2}$ where the density compression ratio is $\simeq 10\%$ lower and the front speed is $10\%$ faster than predicted. We argued that including the supplementary downstream pressure from magnetic wave turbulence, with intensity deduced from simulations, into the conservation equations can account for this discrepancy.
We also demonstrated the presence of a perpendicular current in the precursor for $\sigma \sim 10^{-3}$ as predicted theoretically by \citet{2014MNRAS.440.1365L}. 
This net current triggers the current filamentation instability that plays an important role in shaping the shock precursor and generating magnetic turbulence at these intermediate magnetizations. 

We studied in detail the particle acceleration efficiency and spatial transport in the downstream micro-turbulence. 
Particle acceleration is efficient in weakly magnetized perpendicular shocks. 
The maximal particle energy increases in time as $E_{\rm max} \propto \sqrt{t}$ for $\sigma<10^{-2}$ and saturates in simulations with $\sigma>10^{-3}$. 
For this magnetizations, the saturation energy obeys the scaling $E_{\rm sat} \propto \sigma^{-1/2}$.
Our simulations were however too short to reach the saturation energy in shocks with lower upstream magnetization. Hence, we do not exclude the
scaling $E_{\rm sat} \propto \sigma^{-1/4}$ observed by \citet{SSA13} for $\sigma<10^{-3}$ shocks. 
The non-thermal tail disappears above the transition magnetization beyond which ($\sigma >10^{-2}$) the particle energy spectrum is that expected for a Maxwell-J\"uttner distribution. 
It is found that Diffusive Shock Acceleration is efficient only in weakly magnetized shocks, while a dominant contribution of Shock Drift Acceleration is evidenced at intermediate magnetizations.
The diffusion coefficients are extracted from the simulations and found to evolve with the square of the particle energy in weakly magnetized shocks, as expected in the small-angle scattering regime. They reach a constant value when the external and irregular magnetic field are of comparable strength downstream of the shock, i.e. when the scattering frequency on the microturbulence becomes comparable with the regular gyro-frequency.

We conclude by a remark that further developments could shed some light on plasma physics of particle acceleration in astrophysical environments by studying the accurate long-term behavior of $\sigma <10^{-5}$ shocks, multi-front interaction, and relativistic shocks in a turbulent upstream medium. Enhanced particle acceleration is expected in these environments.

%% file: appendix_codes_compare.tex
\section{Comparison of the finite-difference and spectral codes}
\label{appendix:comparison_codes}

To compare the two codes we chose the mildly magnetized case $\sigma=2\cdot 10^{-3}$ as the returning particle beam-filamentation is present in the precursor along with the electromagnetic wave emission from the shock front. 
The numerical set-up is the same for the two codes: upstream plasma is injected with $\gamma_0=10$, the transverse box size in the $y$-direction is $42\delta_e$ wide, there are 2 particles/cell/species and the simulation time is $t\omega_{\rm pe}=1200$. 

In addition to the difference in the way Maxwell's equations are solved in the two codes, there are differences in the current deposition algorithms on the grid 
as well as in the filtering techniques adopted to avoid the spurious Cherenkov instability. 
In the FDTD code \smilei we used a 3-pass digital filter on currents and the temporal Friedmann filter on the electric field \citep{Greenwood04}, 
with the control parameter $\theta=0.1$ or $0.3$. 
In the case of the spectral code \shockapic, we used a 2-pass digital filter and a spectral cut of large $k_x$ modes on all field components.

\begin{figure}
\centering
\includegraphics[width=0.49\textwidth]{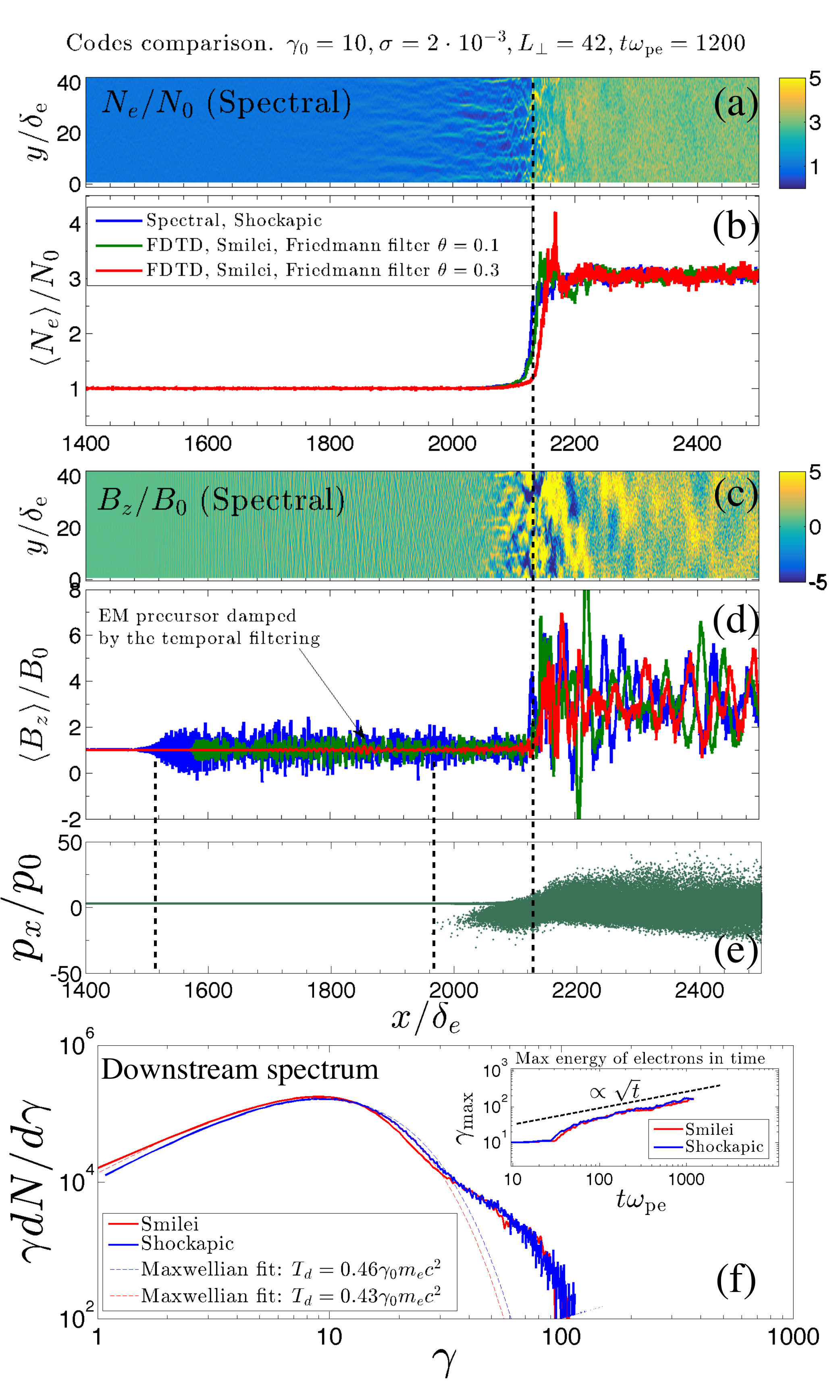}
\caption{Comparison of the two PIC code results: 
spectral 2D code \shockapic and FDTD 2D code \smilei using two different parameters ($\theta=0.1$ and $\theta=0.3$) for the Friedmann filter. 
Simulation parameters are set to $\gamma_0=10$, $\sigma=2\cdot 10^{-3}$, $L_y=42\delta_e$, $N_{\rm ppc}=2$.
The comparison is made at time $t\omega_{\rm pe}=1200$. 
(a)~Electron density map $N_e/N_0$ in the simulation plane from \shockapic.
(b)~Transversely averaged electron density profiles from the three simulations.  
(c)~$B_z$ component of the magnetic field in the simulation plane from \shockapic.
(d)~Transversely averaged $B_z$ from the three simulations.  
(e)~$x-p_x$ electron phase-space distribution from \shockapic.  
(f)~Downstream electron energy spectra from the two codes (\emph{blue line}: \shockapic, \emph{red line}: \smilei with $\theta = 0.3$). 
The \emph{dashed lines} present the respective fits of the two spectra using the energy dispersion expected for a 2D Maxwell-J\"uttner distribution. 
\emph{Inset} Temporal evolution of the maximal energy $\gamma_{\rm max}(t)$.
} 
\label{fig:codes_comparison}%
\end{figure}

As previously, the shock is triggered by reflecting the incident flow on the right-side wall (see Figure~\ref{fig:scheme}). 
The shock is formed after $t\omega_{\rm pe} \simeq 140$ (see Section~\ref{sect:formation}). 
The formation is slightly retarded in the FDTD simulations, using a strong filtering on the electric field $\theta=0.3$. 
At longer times the shock structure sets up with the precursor populated with hot, energetic particles. 
An electromagnetic wave is also emitted from the shock front and propagates at the speed of light in the $-x$ direction. 
In Figure~\ref{fig:codes_comparison} we present the state of the simulations at $t\omega_{\rm pe}=1200$. 

Panel (a) presents the electron density map in the simulation plane derived from the \shockapic code. The shock front position is indicated by the vertical dashed line at $x \simeq 2125 \delta_e$. On the right side the plasma is compressed towards $N_d/N_0 \simeq 3.1$ (see Figure~\ref{fig:jump_dens} for the compression ratio dependence on $\sigma$). On the left side the precursor exhibits two types of filamentary structures. The first type of filamentary structures extends from $x=1800\delta_e$ to the shock front. It is elongated in the shock normal direction. This filamentation is produced in reaction to the strong electromagnetic wave emitted from the front. The second type of filamentation is seen more closely to the front, between $x \simeq 2050 \delta_e$ and the shock front. It is oblique and is produced by the returning beam of the hot shock reflected/accelerated particles. 
Panel (b) presents the transversely averaged density profile from the three simulations: blue line is for the \shockapic code, green line for the \smilei code with $\theta=0.1$ and red line the \smilei code with $\theta=0.3$. The shock front is slightly retarded in the \smilei simulation with $\theta=0.3$ as the shock formation is slightly delayed. Still the shock front speed and density compression ratio are identical in the three simulations. 

In panels (c) and (d) we present the magnetic field $B_z$ in the simulation plane derived form the \shockapic simulation (c) and the transversely averaged $\langle B_z \rangle$ from the three simulations (d). The precursor ($x<x_{\rm sh}$) is composed of the electromagnetic X-mode wave between $x=1500\delta_e$ and the front and of the filamentary oblique structures closely to the shock front. The magnetic field is increased towards its downstream shock compressed value in $x>2125\delta_e$ region. Large amplitude fluctuations can be observed as the plasma has not yet fully relaxed.
Vertical dashed lines, from left to right, delimit the extent of the electromagnetic wave precursor, the extent of the particle precursor, and the shock front position, respectively. Inspecting panel~(d) we remark that the electromagnetic wave in the precursor is of larger amplitude when derived from the spectral code (blue line) than in $\theta=0.1$ case derived from the FDTD code (green line). It is considerably damped in the $\theta=0.3$ FDTD simulation (red line).

Panel (e) presents the phase-space projected on the longitudinal $x-p_x$ momentum (from \shockapic simulation), where $p_x=\gamma \beta_x$. A cold upstream flow is seen at the left part of the box, $p_x/p_0=1$. The particle precursor with hot returning beam fills the $1965<x/\delta_e<2125$ region. The shock front is at $x_{\rm sh} \simeq 2125 \delta_e$ and the downstream hot plasma fills the $x>x_{\rm sh}$ medium.

Finally, panel (f) presents the downstream electron energy spectrum as derived from the two codes ($\theta=0.3$ for the \smilei simulation presented here). 
Both are in general agreement with the downstream distribution function composed of the Maxwellian part and non-thermal part for $\gamma>40$. 
The downstream `temperature' is slightly lower for \smilei ($T_d=0.43 \gamma_0 m_e c^2$) than for \shockapic ($0.46 \gamma_0 m_e c^2$). 
We remark that the maximal energy evolution $\gamma_{\rm max}(t)$ is in very good agreement for the two codes (panel inset) except that the shock forms slightly earlier in the \shockapic simulation.

In conclusion, both codes provide very similar results. It is shown that the usage of the temporal Friedmann filter produces some damping of the electromagnetic wave precursor. This damping is negligible for $\theta=0.1$, but becomes important for stronger filtering $\theta=0.3$. 
The filter acts mainly on the small wavelength in the light-wave branch of the plasma dispersion in order to damp efficiently the spurious Cherenkov modes. 
It is therefore logical that it damps also the physical modes that involve resonance or wave emission on the same branch. 
We also remarke that, while it has an impact on the amplitude of the emitted wave, it has no impact on the particle acceleration efficiency that is governed by the beam-type instabilities in the shock foot.

%% file: appendix_jumpMHD.tex
\section{Derivation of the ideal MHD jump conditions}
\label{appendix:jump}
Here we follow the derivation of jump conditions
as stated in \citet{1992ApJ...391...73G} or in \citet{2016ApJ...827...44L}. The difference is
that we need to alleviate here some approximations that they
adopted (ultra-relativistic or 3D ideal gas adiabatic index
downstream) and derive the shock front speed and density compression ratio in a
most general form. We note that a similar derivation to the one presented here was 
carried out recently by \citet{2016NJPh...18j5002S}.

If the upstream magnetic field is perpendicular to the
shock normal ($\theta_{\rm Bn}=90^{\circ}$), the ideal MHD conservation equations of the particle number density, magnetic flux,
total energy and total momentum as written in the downstream rest frame take the following form:
\begin{eqnarray}
\gamma_0 \left( \beta_0+\beta_{\rm sh} \right) n_0 &=& \beta_{\rm sh} n_d  \label{eq:consMHD1}\\
\gamma_0 \left( \beta_0+\beta_{\rm sh} \right)b_0 &=& \beta_{\rm sh} b_d  \label{eq:consMHD2}\\
\gamma_0^2 \left( \beta_0 +\beta_{\rm sh} \right) \left( w_0+{b_0^2 \over 4 \pi}\right) - \beta_{\rm sh} \left( p_0 + {b_0^2 \over 8 \pi} \right) &=&  \nonumber \\
\beta_{\rm sh} \left( w_d - p_d +{b_d^2 \over 8 \pi}\right) \label{eq:consMHD3}\\
\gamma_0^2 \beta_0 \left(\beta_0 +\beta_{\rm sh} \right) \left( w_0+{b_0^2 \over 4 \pi}\right) +  \left( p_0 + {b_0^2 \over 8 \pi} \right) &=& p_d +{b_d^2 \over 8 \pi} \label{eq:consMHD4}
\end{eqnarray}
where $n_0$ and $n_d$ are the proper densities of upstream and downstream fluids, respectively. They are related
to the apparent density as $N_i=\gamma_i n_i$, where $\gamma_i$ is the bulk Lorentz factor of the flow. The upstream bulk flow is defined by its speed $\beta_0$ in units of $c$ and Lorentz factor $\gamma_0$, $\beta_{\rm sh}$ is the absolute value of the shock speed as seen from the downstream frame, $b_i$ is the perpendicular magnetic field, $w_i$ is the fluid enthalpy and $p_i$ is the kinetic pressure. The subscript $i$ can take values $0$ and $d$ for upstream and downstream, respectively. The enthalpy is related to the rest mass energy and pressure as $w_i=n_i m c^2 + (\Gamma_{\rm ad}/(\Gamma_{\rm ad}-1))p_i$, where $m$ is the total mass of the fluid components and $\Gamma_{\rm ad}$ is the gas adiabatic index as defined using Synge equation of state.  We note that the set of equations~\ref{eq:consMHD1}-\ref{eq:consMHD4} can be obtained by writing the conservation laws in the shock front rest frame and then Lorentz transforming them into the downstream frame or by directly writing the set in the downstream frame, as done in \citet{2016ApJ...827...44L}.

The only approximation we need to make is the strong shock limit ($w_0 \simeq n_0 m c^2$), but still not ultra-relativistic, i.e. $w_d=n_d m c^2 + (\Gamma_{\rm ad}/(\Gamma_{\rm ad}-1))p_d$. In that case the second terms on the left hand side of the equations \ref{eq:consMHD3} and \ref{eq:consMHD4}, i.e. upstream total pressure terms, are negligible. 

Defining the upstream magnetisation as $\sigma=b_0^2/(4\pi w_0)$ \citep{KirkDuffy} and inserting $b_d$ as deduced from the equation \ref{eq:consMHD2} into the equation \ref{eq:consMHD4} the downstream kinetic pressure can be written as
\begin{eqnarray}
{p_d \over w_0} &=& \gamma_0^2 \beta_0 (\beta_0+\beta_{\rm sh})(1+\sigma) - \gamma_0^2 {(\beta_0+\beta_{\rm sh})^2\over \beta_{\rm sh}^2}  {\sigma \over 2} \, .
\end{eqnarray}
Inserting this expression into the equation~\ref{eq:consMHD3} and using the equation~\ref{eq:consMHD1} for $n_d/n_0$, results in a second order equation for $\beta_{\rm sh}$:
\begin{eqnarray}
2\gamma_0 \beta_0 (1 + \sigma) \beta_{\rm sh}^2 - \Large[ 2(\gamma_0-1) \left(\Gamma_{\rm ad}-1\right) + \nonumber  \\
\gamma_0 \Gamma_{\rm ad}  \sigma \Large]\beta_{\rm sh} +\gamma_0 \beta_0 \left(2-\Gamma_{\rm ad}\right) \sigma &=& 0 \, .
\end{eqnarray}
The physical solution of this equation in its complete form is
\begin{eqnarray}\label{eq:beta_sh}
  \beta_{\rm sh}  &=&  \frac{A+\sqrt{8 \gamma_0^2 \beta_0^2 \sigma (1+\sigma )(2-\Gamma_{\rm ad}) + A^2}}{4 (1+\sigma)\gamma_0 \beta_0} \, ,
\end{eqnarray}
where
\begin{eqnarray}
A & =& 2(\gamma_0-1)(\Gamma_{\rm ad}-1) + \Gamma_{\rm ad}  \gamma_0 \sigma \, .
\end{eqnarray}
This is a general solution for a strong perpendicular shock. 
In the unmagnetized limit, $\sigma=0$, the shock speed is equal to 
$(\Gamma_{\rm ad}-1)(\gamma_0-1)/(\gamma_0 \beta_0)$ and $N_d/N_0=1+(\gamma_0+1)/\left[(\Gamma_{\rm ad}-1)\gamma_0\right]$,
in agreement with \citep{1976PhFl...19.1130B, KirkDuffy,2008ApJ...682L...5S}.
In a strongly magnetised case $\sigma \gg 1$, the shock speed tends to $c$ and the compression ratio tends to 2, in agreement with \citet{1984ApJ...283..694K} and \citet{2016ApJ...827...44L}.
Equation~\ref{eq:beta_sh} is also relevant for non-relativisic shocks if the appropriate $\Gamma_{\rm ad}$ is chosen. 
Considering additional approximations allows to recover \citet{1992ApJ...391...73G} (ultra-relativistic limit) 
or \citet{1984ApJ...283..694K} (ultra-relativistic and $\Gamma_{\rm ad}=4/3$). 

The compression ratio can be deduced using equation~\ref{eq:consMHD1}:
\begin{eqnarray}
\label{eq:jump}
{N_d \over N_0} &=& {n_d \over \gamma_0 n_0} = 1 + {\beta_0 \over \beta_{\rm sh}} \,,
\end{eqnarray}
and the jump in magnetic field is the same in virtue of equation~\ref{eq:consMHD2}.
We recover the classical results for a perfect 2D relativistic gas ($\Gamma_{\rm ad}=3/2$) and ultra-relativistic shock ($\gamma_0 \gg 1$) the density jump is equal to $\Gamma_{\rm ad}/(\Gamma_{\rm ad}-1)=3$.

Complementarily, using $p_d=N_d k_B T_d$, one can derive the expected downstream temperature:
\begin{eqnarray}
\label{eq:Tdownstream}
{k_B T_d \over \gamma_0 m c^2} &=&
\beta_{\rm sh} \beta_0 \left( 1+\sigma \right) - {\beta_0+\beta_{\rm sh} \over \beta_{\rm sh}}{\sigma \over 2} \, .
\end{eqnarray}
Its unmagnetized limit is $k_B T_d=(\Gamma_{\rm ad}-1)(\gamma_0-1) m c^2$.

Finally, it is important to note that the strong shock approximation is valid as far as $\sigma<\gamma_0^2$, i.e., as long as the alfvenic Mach number of the upstream flow remains larger than unity \citep{AlsopArons88}.
Indeed, Equation~\ref{eq:beta_sh} gives nonphysical shock front speed, $\beta_{\rm sh}>1$, for $\sigma>\gamma_0^2$. 
This is because for such high magnetizations the magnetic pressure of the upstream flow can not be neglected anymore and additional pressure terms in the left hand side of Equations \ref{eq:consMHD3} and \ref{eq:consMHD4} need to be accounted for. 
Including these terms in the system of conservation laws yields to a third order equation on $\beta_{\rm sh}$ that ensures that the shock front is always subluminal for any $\sigma \gg 1$.
The asymptotic behaviors of the shock front speed and Lorentz factor for $\sigma > \gamma_0^2 \gg 1$ are then $\beta_{\rm sh} \to \sqrt{1-\sigma^{-2}}$ and $\gamma_{\rm sh} \to \sqrt{\sigma}$, respectively.

%% file: article_main.bbl
\begin{thebibliography}{}
\makeatletter
\relax
\def\mn@urlcharsother{\let\do\@makeother \do\$\do\&\do\#\do\^\do\_\do\%\do\~}
\def\mn@doi{\begingroup\mn@urlcharsother \@ifnextchar [ {\mn@doi@}
  {\mn@doi@[]}}
\def\mn@doi@[#1]#2{\def\@tempa{#1}\ifx\@tempa\@empty \href
  {http://dx.doi.org/#2} {doi:#2}\else \href {http://dx.doi.org/#2} {#1}\fi
  \endgroup}
\def\mn@eprint#1#2{\mn@eprint@#1:#2::\@nil}
\def\mn@eprint@arXiv#1{\href {http://arxiv.org/abs/#1} {{\tt arXiv:#1}}}
\def\mn@eprint@dblp#1{\href {http://dblp.uni-trier.de/rec/bibtex/#1.xml}
  {dblp:#1}}
\def\mn@eprint@#1:#2:#3:#4\@nil{\def\@tempa {#1}\def\@tempb {#2}\def\@tempc
  {#3}\ifx \@tempc \@empty \let \@tempc \@tempb \let \@tempb \@tempa \fi \ifx
  \@tempb \@empty \def\@tempb {arXiv}\fi \@ifundefined
  {mn@eprint@\@tempb}{\@tempb:\@tempc}{\expandafter \expandafter \csname
  mn@eprint@\@tempb\endcsname \expandafter{\@tempc}}}

\bibitem[\protect\citeauthoryear{{Achterberg}, {Gallant}, {Kirk}  \&
  {Guthmann}}{{Achterberg} et~al.}{2001}]{2001MNRAS.328..393A}
{Achterberg} A.,  {Gallant} Y.~A.,  {Kirk} J.~G.,   {Guthmann} A.~W.,  2001,
  \mn@doi [\mnras] {10.1046/j.1365-8711.2001.04851.x}, \href
  {http://adsabs.harvard.edu/abs/2001MNRAS.328..393A} {328, 393}

\bibitem[\protect\citeauthoryear{Aloisio \& Berezinsky}{Aloisio \&
  Berezinsky}{2004}]{Aloisio04}
Aloisio R.,  Berezinsky V.,  2004, \apj, 612, 900

\bibitem[\protect\citeauthoryear{{Alsop} \& {Arons}}{{Alsop} \&
  {Arons}}{1988}]{AlsopArons88}
{Alsop} D.,  {Arons} J.,  1988, \mn@doi [Physics of Fluids] {10.1063/1.866765},
  \href {http://cdsads.u-strasbg.fr/abs/1988PhFl...31..839A} {31, 839}

\bibitem[\protect\citeauthoryear{{Araudo}, {Bell}  \& {Blundell}}{{Araudo}
  et~al.}{2015}]{araudo15}
{Araudo} A.~T.,  {Bell} A.~R.,   {Blundell} K.~M.,  2015, \mn@doi [\apj]
  {10.1088/0004-637X/806/2/243}, \href
  {http://adsabs.harvard.edu/abs/2015ApJ...806..243A} {806, 243}

\bibitem[\protect\citeauthoryear{{Begelman} \& {Kirk}}{{Begelman} \&
  {Kirk}}{1990}]{1990ApJ...353...66B}
{Begelman} M.~C.,  {Kirk} J.~G.,  1990, \mn@doi [\apj] {10.1086/168590}, \href
  {http://adsabs.harvard.edu/abs/1990ApJ...353...66B} {353, 66}

\bibitem[\protect\citeauthoryear{{Bell}}{{Bell}}{1978}]{Bell78}
{Bell} A.~R.,  1978, MNRAS, \href
  {http://cdsads.u-strasbg.fr/abs/1978MNRAS.182..147B} {182, 147}

\bibitem[\protect\citeauthoryear{{Birdsall} \& {Langdon}}{{Birdsall} \&
  {Langdon}}{1991}]{1991ppcs.book.....B}
{Birdsall} C.~K.,  {Langdon} A.~B.,  1991, {Plasma Physics via Computer
  Simulation}

\bibitem[\protect\citeauthoryear{{Blandford} \& {McKee}}{{Blandford} \&
  {McKee}}{1976}]{1976PhFl...19.1130B}
{Blandford} R.~D.,  {McKee} C.~F.,  1976, \mn@doi [Physics of Fluids]
  {10.1063/1.861619}, \href {http://adsabs.harvard.edu/abs/1976PhFl...19.1130B}
  {19, 1130}

\bibitem[\protect\citeauthoryear{{Bret}}{{Bret}}{2009}]{2009ApJ...699..990B}
{Bret} A.,  2009, \mn@doi [\apj] {10.1088/0004-637X/699/2/990}, \href
  {http://adsabs.harvard.edu/abs/2009ApJ...699..990B} {699, 990}

\bibitem[\protect\citeauthoryear{{Bret}, {Stockem}, {Narayan}  \&
  {Silva}}{{Bret} et~al.}{2014}]{2014PhPl...21g2301B}
{Bret} A.,  {Stockem} A.,  {Narayan} R.,   {Silva} L.~O.,  2014, \mn@doi
  [Physics of Plasmas] {10.1063/1.4886121}, \href
  {http://adsabs.harvard.edu/abs/2014PhPl...21g2301B} {21, 072301}

\bibitem[\protect\citeauthoryear{{Caprioli}}{{Caprioli}}{2015}]{2015ApJ...811L..38C}
{Caprioli} D.,  2015, \mn@doi [\apjl] {10.1088/2041-8205/811/2/L38}, \href
  {http://adsabs.harvard.edu/abs/2015ApJ...811L..38C} {811, L38}

\bibitem[\protect\citeauthoryear{{Cerutti}, {Philippov}, {Parfrey}  \&
  {Spitkovsky}}{{Cerutti} et~al.}{2015}]{2015MNRAS.448..606C}
{Cerutti} B.,  {Philippov} A.,  {Parfrey} K.,   {Spitkovsky} A.,  2015, \mn@doi
  [\mnras] {10.1093/mnras/stv042}, \href
  {http://adsabs.harvard.edu/abs/2015MNRAS.448..606C} {448, 606}

\bibitem[\protect\citeauthoryear{{Chang}, {Spitkovsky}  \& {Arons}}{{Chang}
  et~al.}{2008}]{2008ApJ...674..378C}
{Chang} P.,  {Spitkovsky} A.,   {Arons} J.,  2008, \mn@doi [\apj]
  {10.1086/524764}, \href {http://adsabs.harvard.edu/abs/2008ApJ...674..378C}
  {674, 378}

\bibitem[\protect\citeauthoryear{Chen et~al.,}{Chen et~al.}{2015}]{chen2015}
Chen H.,  et~al., 2015, \mn@doi [Phys. Rev. Lett.]
  {10.1103/PhysRevLett.114.215001}, 114, 215001

\bibitem[\protect\citeauthoryear{{Derouillat} et~al.,}{{Derouillat}
  et~al.}{2018}]{derouillat2018}
{Derouillat} J.,  et~al., 2018, \mn@doi [Computer Physics Communications]
  {10.1016/j.cpc.2017.09.024}, \href
  {http://adsabs.harvard.edu/abs/2018CoPhC.222..351D} {222, 351}

\bibitem[\protect\citeauthoryear{{Ellison}, {Warren}  \& {Bykov}}{{Ellison}
  et~al.}{2016}]{2016MNRAS.456.3090E}
{Ellison} D.~C.,  {Warren} D.~C.,   {Bykov} A.~M.,  2016, \mn@doi [\mnras]
  {10.1093/mnras/stv2912}, \href
  {http://adsabs.harvard.edu/abs/2016MNRAS.456.3090E} {456, 3090}

\bibitem[\protect\citeauthoryear{{Esirkepov}}{{Esirkepov}}{2001}]{Esirkepov01}
{Esirkepov} T.~Z.,  2001, \mn@doi [Computer Physics Communications]
  {10.1016/S0010-4655(00)00228-9}, \href
  {http://cdsads.u-strasbg.fr/abs/2001CoPhC.135..144E} {135, 144}

\bibitem[\protect\citeauthoryear{{Fried}}{{Fried}}{1959}]{Fried59}
{Fried} B.~D.,  1959, \mn@doi [Physics of Fluids] {10.1063/1.1705933}, \href
  {http://cdsads.u-strasbg.fr/abs/1959PhFl....2..337F} {2, 337}

\bibitem[\protect\citeauthoryear{{Gallant} \& {Achterberg}}{{Gallant} \&
  {Achterberg}}{1999}]{1999MNRAS.305L...6G}
{Gallant} Y.~A.,  {Achterberg} A.,  1999, \mn@doi [\mnras]
  {10.1046/j.1365-8711.1999.02566.x}, \href
  {http://adsabs.harvard.edu/abs/1999MNRAS.305L...6G} {305, L6}

\bibitem[\protect\citeauthoryear{{Gallant}, {Hoshino}, {Langdon}, {Arons}  \&
  {Max}}{{Gallant} et~al.}{1992}]{1992ApJ...391...73G}
{Gallant} Y.~A.,  {Hoshino} M.,  {Langdon} A.~B.,  {Arons} J.,   {Max} C.~E.,
  1992, \mn@doi [\apj] {10.1086/171326}, \href
  {http://adsabs.harvard.edu/abs/1992ApJ...391...73G} {391, 73}

\bibitem[\protect\citeauthoryear{Grassi, Grech, Amiranoff, Pegoraro, Macchi  \&
  Riconda}{Grassi et~al.}{2017}]{grassi2017}
Grassi A.,  Grech M.,  Amiranoff F.,  Pegoraro F.,  Macchi A.,   Riconda C.,
  2017, \mn@doi [Phys. Rev. E] {10.1103/PhysRevE.95.023203}, 95, 023203

\bibitem[\protect\citeauthoryear{{Greenwood}, {Cartwright}, {Luginsland}  \&
  {Baca}}{{Greenwood} et~al.}{2004}]{Greenwood04}
{Greenwood} A.~D.,  {Cartwright} K.~L.,  {Luginsland} J.~W.,   {Baca} E.~A.,
  2004, \mn@doi [Journal of Computational Physics] {10.1016/j.jcp.2004.06.021},
  \href {http://cdsads.u-strasbg.fr/abs/2004JCoPh.201..665G} {201, 665}

\bibitem[\protect\citeauthoryear{{Haugb{\o}lle}}{{Haugb{\o}lle}}{2011}]{2011ApJ...739L..42H}
{Haugb{\o}lle} T.,  2011, \mn@doi [\apjl] {10.1088/2041-8205/739/2/L42}, \href
  {http://adsabs.harvard.edu/abs/2011ApJ...739L..42H} {739, L42}

\bibitem[\protect\citeauthoryear{{Hoshino} \& {Arons}}{{Hoshino} \&
  {Arons}}{1991}]{1991PhFlB...3..818H}
{Hoshino} M.,  {Arons} J.,  1991, \mn@doi [Physics of Fluids B]
  {10.1063/1.859877}, \href {http://adsabs.harvard.edu/abs/1991PhFlB...3..818H}
  {3, 818}

\bibitem[\protect\citeauthoryear{{Huntington} et~al.,}{{Huntington}
  et~al.}{2015}]{2015NatPh..11..173H}
{Huntington} C.~M.,  et~al., 2015, \mn@doi [Nature Physics]
  {10.1038/nphys3178}, \href
  {http://adsabs.harvard.edu/abs/2015NatPh..11..173H} {11, 173}

\bibitem[\protect\citeauthoryear{{Iwamoto}, {Amano}, {Hoshino}  \&
  {Matsumoto}}{{Iwamoto} et~al.}{2017}]{2017ApJ...840...52I}
{Iwamoto} M.,  {Amano} T.,  {Hoshino} M.,   {Matsumoto} Y.,  2017, \mn@doi
  [\apj] {10.3847/1538-4357/aa6d6f}, \href
  {http://adsabs.harvard.edu/abs/2017ApJ...840...52I} {840, 52}

\bibitem[\protect\citeauthoryear{{Kato}}{{Kato}}{2007}]{Kato07}
{Kato} T.~N.,  2007, \mn@doi [\apj] {10.1086/521297}, \href
  {http://cdsads.u-strasbg.fr/abs/2007ApJ...668..974K} {668, 974}

\bibitem[\protect\citeauthoryear{{Kennel} \& {Coroniti}}{{Kennel} \&
  {Coroniti}}{1984}]{1984ApJ...283..694K}
{Kennel} C.~F.,  {Coroniti} F.~V.,  1984, \mn@doi [\apj] {10.1086/162356},
  \href {http://adsabs.harvard.edu/abs/1984ApJ...283..694K} {283, 694}

\bibitem[\protect\citeauthoryear{{Keshet} \& {Waxman}}{{Keshet} \&
  {Waxman}}{2005}]{2005PhRvL..94k1102K}
{Keshet} U.,  {Waxman} E.,  2005, \mn@doi [Physical Review Letters]
  {10.1103/PhysRevLett.94.111102}, \href
  {http://adsabs.harvard.edu/abs/2005PhRvL..94k1102K} {94, 111102}

\bibitem[\protect\citeauthoryear{{Keshet}, {Katz}, {Spitkovsky}  \&
  {Waxman}}{{Keshet} et~al.}{2009}]{2009ApJ...693L.127K}
{Keshet} U.,  {Katz} B.,  {Spitkovsky} A.,   {Waxman} E.,  2009, \mn@doi
  [\apjl] {10.1088/0004-637X/693/2/L127}, \href
  {http://adsabs.harvard.edu/abs/2009ApJ...693L.127K} {693, L127}

\bibitem[\protect\citeauthoryear{{Kirk} \& {Duffy}}{{Kirk} \&
  {Duffy}}{1999}]{KirkDuffy}
{Kirk} J.,  {Duffy} P.,  1999, Journal of Physics G: Nuclear and Particle
  Physics, 25, R163

\bibitem[\protect\citeauthoryear{{Kirk} \& {Reville}}{{Kirk} \&
  {Reville}}{2010}]{2010ApJ...710L..16K}
{Kirk} J.~G.,  {Reville} B.,  2010, \mn@doi [\apjl]
  {10.1088/2041-8205/710/1/L16}, \href
  {http://adsabs.harvard.edu/abs/2010ApJ...710L..16K} {710, L16}

\bibitem[\protect\citeauthoryear{{Kirk}, {Guthmann}, {Gallant}  \&
  {Achterberg}}{{Kirk} et~al.}{2000}]{2000ApJ...542..235K}
{Kirk} J.~G.,  {Guthmann} A.~W.,  {Gallant} Y.~A.,   {Achterberg} A.,  2000,
  \mn@doi [\apj] {10.1086/309533}, \href
  {http://adsabs.harvard.edu/abs/2000ApJ...542..235K} {542, 235}

\bibitem[\protect\citeauthoryear{{Krymskii}}{{Krymskii}}{1977}]{1977DoSSR.234.1306K}
{Krymskii} G.~F.,  1977, Akademiia Nauk SSSR Doklady, \href
  {http://adsabs.harvard.edu/abs/1977DoSSR.234.1306K} {234, 1306}

\bibitem[\protect\citeauthoryear{{Langdon}, {Arons}  \& {Max}}{{Langdon}
  et~al.}{1988}]{Langdon88}
{Langdon} A.~B.,  {Arons} J.,   {Max} C.~E.,  1988, \mn@doi [Physical Review
  Letters] {10.1103/PhysRevLett.61.779}, \href
  {http://cdsads.u-strasbg.fr/abs/1988PhRvL..61..779L} {61, 779}

\bibitem[\protect\citeauthoryear{{Lemoine}}{{Lemoine}}{2015}]{2015JPlPh..81a4501L}
{Lemoine} M.,  2015, \mn@doi [Journal of Plasma Physics]
  {10.1017/S0022377814000920}, \href
  {http://adsabs.harvard.edu/abs/2015JPlPh..81a4501L} {81, 455810101}

\bibitem[\protect\citeauthoryear{{Lemoine} \& {Pelletier}}{{Lemoine} \&
  {Pelletier}}{2003}]{2003ApJ...589L..73L}
{Lemoine} M.,  {Pelletier} G.,  2003, \mn@doi [\apjl] {10.1086/376353}, \href
  {http://adsabs.harvard.edu/abs/2003ApJ...589L..73L} {589, L73}

\bibitem[\protect\citeauthoryear{{Lemoine} \& {Pelletier}}{{Lemoine} \&
  {Pelletier}}{2010}]{2010MNRAS.402..321L}
{Lemoine} M.,  {Pelletier} G.,  2010, \mn@doi [\mnras]
  {10.1111/j.1365-2966.2009.15869.x}, \href
  {http://adsabs.harvard.edu/abs/2010MNRAS.402..321L} {402, 321}

\bibitem[\protect\citeauthoryear{{Lemoine} \& {Pelletier}}{{Lemoine} \&
  {Pelletier}}{2011}]{2011MNRAS.417.1148L}
{Lemoine} M.,  {Pelletier} G.,  2011, \mn@doi [\mnras]
  {10.1111/j.1365-2966.2011.19331.x}, \href
  {http://adsabs.harvard.edu/abs/2011MNRAS.417.1148L} {417, 1148}

\bibitem[\protect\citeauthoryear{{Lemoine}, {Pelletier}  \& {Revenu}}{{Lemoine}
  et~al.}{2006}]{2006ApJ...645L.129L}
{Lemoine} M.,  {Pelletier} G.,   {Revenu} B.,  2006, \mn@doi [\apjl]
  {10.1086/506322}, \href {http://adsabs.harvard.edu/abs/2006ApJ...645L.129L}
  {645, L129}

\bibitem[\protect\citeauthoryear{{Lemoine}, {Pelletier}, {Gremillet}  \&
  {Plotnikov}}{{Lemoine} et~al.}{2014a}]{2014EL....10655001L}
{Lemoine} M.,  {Pelletier} G.,  {Gremillet} L.,   {Plotnikov} I.,  2014a,
  \mn@doi [EPL (Europhysics Letters)] {10.1209/0295-5075/106/55001}, \href
  {http://adsabs.harvard.edu/abs/2014EL....10655001L} {106, 55001}

\bibitem[\protect\citeauthoryear{{Lemoine}, {Pelletier}, {Gremillet}  \&
  {Plotnikov}}{{Lemoine} et~al.}{2014b}]{2014MNRAS.440.1365L}
{Lemoine} M.,  {Pelletier} G.,  {Gremillet} L.,   {Plotnikov} I.,  2014b,
  \mn@doi [\mnras] {10.1093/mnras/stu213}, \href
  {http://adsabs.harvard.edu/abs/2014MNRAS.440.1365L} {440, 1365}

\bibitem[\protect\citeauthoryear{{Lemoine}, {Ramos}  \& {Gremillet}}{{Lemoine}
  et~al.}{2016}]{2016ApJ...827...44L}
{Lemoine} M.,  {Ramos} O.,   {Gremillet} L.,  2016, \mn@doi [\apj]
  {10.3847/0004-637X/827/1/44}, \href
  {http://adsabs.harvard.edu/abs/2016ApJ...827...44L} {827, 44}

\bibitem[\protect\citeauthoryear{Liu}{Liu}{1997}]{Liu97PSTD}
Liu Q.~H.,  1997, \mn@doi [Microwave and Optical Technology Letters]
  {10.1002/(SICI)1098-2760(19970620)15:3<158::AID-MOP11>3.0.CO;2-3}, 15, 158

\bibitem[\protect\citeauthoryear{Lobet, Ruyer, Debayle, d'Humi\`eres, Grech,
  Lemoine  \& Gremillet}{Lobet et~al.}{2015}]{lobet2015}
Lobet M.,  Ruyer C.,  Debayle A.,  d'Humi\`eres E.,  Grech M.,  Lemoine M.,
  Gremillet L.,  2015, \mn@doi [Phys. Rev. Lett.]
  {10.1103/PhysRevLett.115.215003}, 115, 215003

\bibitem[\protect\citeauthoryear{{Martins}, {Fonseca}, {Silva}  \&
  {Mori}}{{Martins} et~al.}{2009}]{2009ApJ...695L.189M}
{Martins} S.~F.,  {Fonseca} R.~A.,  {Silva} L.~O.,   {Mori} W.~B.,  2009,
  \mn@doi [\apjl] {10.1088/0004-637X/695/2/L189}, \href
  {http://adsabs.harvard.edu/abs/2009ApJ...695L.189M} {695, L189}

\bibitem[\protect\citeauthoryear{{Max}, {Arons}  \& {Langdon}}{{Max}
  et~al.}{1974}]{1974PhRvL..33..209M}
{Max} C.~E.,  {Arons} J.,   {Langdon} A.~B.,  1974, \mn@doi [Physical Review
  Letters] {10.1103/PhysRevLett.33.209}, \href
  {http://adsabs.harvard.edu/abs/1974PhRvL..33..209M} {33, 209}

\bibitem[\protect\citeauthoryear{{Medvedev} \& {Loeb}}{{Medvedev} \&
  {Loeb}}{1999}]{1999ApJ...526..697M}
{Medvedev} M.~V.,  {Loeb} A.,  1999, \mn@doi [\apj] {10.1086/308038}, \href
  {http://adsabs.harvard.edu/abs/1999ApJ...526..697M} {526, 697}

\bibitem[\protect\citeauthoryear{{Moiseev} \& {Sagdeev}}{{Moiseev} \&
  {Sagdeev}}{1963}]{1963JNuE....5...43M}
{Moiseev} S.~S.,  {Sagdeev} R.~Z.,  1963, \mn@doi [Journal of Nuclear Energy]
  {10.1088/0368-3281/5/1/309}, \href
  {http://adsabs.harvard.edu/abs/1963JNuE....5...43M} {5, 43}

\bibitem[\protect\citeauthoryear{{Niemiec}, {Ostrowski}  \& {Pohl}}{{Niemiec}
  et~al.}{2006}]{2006ApJ...650.1020N}
{Niemiec} J.,  {Ostrowski} M.,   {Pohl} M.,  2006, \mn@doi [\apj]
  {10.1086/506901}, \href {http://adsabs.harvard.edu/abs/2006ApJ...650.1020N}
  {650, 1020}

\bibitem[\protect\citeauthoryear{{Nishikawa}, {Hardee}, {Richardson}, {Preece},
  {Sol}  \& {Fishman}}{{Nishikawa} et~al.}{2005}]{Nishikawa05}
{Nishikawa} K.-I.,  {Hardee} P.,  {Richardson} G.,  {Preece} R.,  {Sol} H.,
  {Fishman} G.~J.,  2005, \mn@doi [\apj] {10.1086/428394}, \href
  {http://cdsads.u-strasbg.fr/abs/2005ApJ...622..927N} {622, 927}

\bibitem[\protect\citeauthoryear{Nuter, Grech, Gonzalez~de Alaiza~Martinez,
  Bonnaud  \& d'Humi{\`e}res}{Nuter et~al.}{2014}]{nuter2014}
Nuter R.,  Grech M.,  Gonzalez~de Alaiza~Martinez P.,  Bonnaud G.,
  d'Humi{\`e}res E.,  2014, \mn@doi [The European Physical Journal D]
  {10.1140/epjd/e2014-50162-y}, 68, 177

\bibitem[\protect\citeauthoryear{{Peacock}}{{Peacock}}{1981}]{Peacock81}
{Peacock} J.~A.,  1981, MNRAS, \href
  {http://cdsads.u-strasbg.fr/abs/1981MNRAS.196..135P} {196, 135}

\bibitem[\protect\citeauthoryear{{Pelletier}, {Lemoine}  \&
  {Marcowith}}{{Pelletier} et~al.}{2009}]{2009MNRAS.393..587P}
{Pelletier} G.,  {Lemoine} M.,   {Marcowith} A.,  2009, \mn@doi [\mnras]
  {10.1111/j.1365-2966.2008.14219.x}, \href
  {http://adsabs.harvard.edu/abs/2009MNRAS.393..587P} {393, 587}

\bibitem[\protect\citeauthoryear{{Pelletier}, {Lemoine}, {Gremillet}  \&
  {Plotnikov}}{{Pelletier} et~al.}{2014}]{2014IJMPS..2860193P}
{Pelletier} G.,  {Lemoine} M.,  {Gremillet} L.,   {Plotnikov} I.,  2014, in
  International Journal of Modern Physics Conference Series. p. 1460193
  (\mn@eprint {arXiv} {1401.7210}), \mn@doi{10.1142/S2010194514601938}

\bibitem[\protect\citeauthoryear{{Plotnikov}, {Pelletier}  \&
  {Lemoine}}{{Plotnikov} et~al.}{2011}]{2011A&A...532A..68P}
{Plotnikov} I.,  {Pelletier} G.,   {Lemoine} M.,  2011, \mn@doi [\aap]
  {10.1051/0004-6361/201117182}, \href
  {http://adsabs.harvard.edu/abs/2011A%26A...532A..68P} {532, A68}

\bibitem[\protect\citeauthoryear{{Reville} \& {Bell}}{{Reville} \&
  {Bell}}{2014}]{2014MNRAS.439.2050R}
{Reville} B.,  {Bell} A.~R.,  2014, \mn@doi [\mnras] {10.1093/mnras/stu088},
  \href {http://adsabs.harvard.edu/abs/2014MNRAS.439.2050R} {439, 2050}

\bibitem[\protect\citeauthoryear{Ross et~al.,}{Ross et~al.}{2017}]{ross2017}
Ross J.~S.,  et~al., 2017, \mn@doi [Phys. Rev. Lett.]
  {10.1103/PhysRevLett.118.185003}, 118, 185003

\bibitem[\protect\citeauthoryear{{Sagdeev}}{{Sagdeev}}{1966}]{1966RvPP....4...23S}
{Sagdeev} R.~Z.,  1966, Reviews of Plasma Physics, \href
  {http://adsabs.harvard.edu/abs/1966RvPP....4...23S} {4, 23}

\bibitem[\protect\citeauthoryear{{Silva}, {Fonseca}, {Tonge}, {Dawson}, {Mori}
  \& {Medvedev}}{{Silva} et~al.}{2003}]{Silva03}
{Silva} L.~O.,  {Fonseca} R.~A.,  {Tonge} J.~W.,  {Dawson} J.~M.,  {Mori}
  W.~B.,   {Medvedev} M.~V.,  2003, \mn@doi [\apjl] {10.1086/379156}, \href
  {http://cdsads.u-strasbg.fr/abs/2003ApJ...596L.121S} {596, L121}

\bibitem[\protect\citeauthoryear{{Sironi} \& {Spitkovsky}}{{Sironi} \&
  {Spitkovsky}}{2009}]{2009ApJ...698.1523S}
{Sironi} L.,  {Spitkovsky} A.,  2009, \mn@doi [\apj]
  {10.1088/0004-637X/698/2/1523}, \href
  {http://adsabs.harvard.edu/abs/2009ApJ...698.1523S} {698, 1523}

\bibitem[\protect\citeauthoryear{{Sironi}, {Spitkovsky}  \& {Arons}}{{Sironi}
  et~al.}{2013}]{SSA13}
{Sironi} L.,  {Spitkovsky} A.,   {Arons} J.,  2013, \mn@doi [\apj]
  {10.1088/0004-637X/771/1/54}, \href
  {http://adsabs.harvard.edu/abs/2013ApJ...771...54S} {771, 54}

\bibitem[\protect\citeauthoryear{{Spitkovsky}}{{Spitkovsky}}{2008}]{2008ApJ...682L...5S}
{Spitkovsky} A.,  2008, \mn@doi [\apjl] {10.1086/590248}, \href
  {http://adsabs.harvard.edu/abs/2008ApJ...682L...5S} {682, L5}

\bibitem[\protect\citeauthoryear{{Stockem Novo}, {Bret}  \& {Sinha}}{{Stockem
  Novo} et~al.}{2016}]{2016NJPh...18j5002S}
{Stockem Novo} A.,  {Bret} A.,   {Sinha} U.,  2016, \mn@doi [New Journal of
  Physics] {10.1088/1367-2630/18/10/105002}, \href
  {http://adsabs.harvard.edu/abs/2016NJPh...18j5002S} {18, 105002}

\bibitem[\protect\citeauthoryear{{Stockem}, {Fi{\'u}za}, {Fonseca}  \&
  {Silva}}{{Stockem} et~al.}{2012}]{2012ApJ...755...68S}
{Stockem} A.,  {Fi{\'u}za} F.,  {Fonseca} R.~A.,   {Silva} L.~O.,  2012,
  \mn@doi [\apj] {10.1088/0004-637X/755/1/68}, \href
  {http://adsabs.harvard.edu/abs/2012ApJ...755...68S} {755, 68}

\bibitem[\protect\citeauthoryear{Taflove \& Hagness}{Taflove \&
  Hagness}{2005}]{taflove2005}
Taflove A.,  Hagness S.,  2005, Computational Electrodynamics: The
  Finite-Difference Time-Domain Method, 3rd Ed..
Artech House, Norwood

\bibitem[\protect\citeauthoryear{{Uzdensky} \& {Spitkovsky}}{{Uzdensky} \&
  {Spitkovsky}}{2014}]{2014ApJ...780....3U}
{Uzdensky} D.~A.,  {Spitkovsky} A.,  2014, \mn@doi [\apj]
  {10.1088/0004-637X/780/1/3}, \href
  {http://adsabs.harvard.edu/abs/2014ApJ...780....3U} {780, 3}

\bibitem[\protect\citeauthoryear{{Vay}, {Haber}  \& {Godfrey}}{{Vay}
  et~al.}{2013}]{2013JCoPh.243..260V}
{Vay} J.-L.,  {Haber} I.,   {Godfrey} B.~B.,  2013, \mn@doi [Journal of
  Computational Physics] {10.1016/j.jcp.2013.03.010}, \href
  {http://adsabs.harvard.edu/abs/2013JCoPh.243..260V} {243, 260}

\bibitem[\protect\citeauthoryear{{Vietri}}{{Vietri}}{1995}]{Vietri95}
{Vietri} M.,  1995, \mn@doi [\apj] {10.1086/176448}, \href
  {http://cdsads.u-strasbg.fr/abs/1995ApJ...453..883V} {453, 883}

\bibitem[\protect\citeauthoryear{{Weibel}}{{Weibel}}{1959}]{Weibel59}
{Weibel} E.~S.,  1959, \mn@doi [Physical Review Letters]
  {10.1103/PhysRevLett.2.83}, \href
  {http://cdsads.u-strasbg.fr/abs/1959PhRvL...2...83W} {2, 83}

\makeatother
\end{thebibliography}
